\documentclass[12pt]{article}
\usepackage{amssymb,amsfonts}
\usepackage{graphicx}

\newtheorem{theorem}{Theorem}[section]

\newtheorem{lemma}{Lemma}[section]

\newtheorem{proposition}{Proposition}[section]

\newtheorem{definition}{Definition}[section]

\newtheorem{corollary}{Corollary}[section]

\newtheorem{example}{Example}[section]

\begin{document}

\title{Singularities of Transition Processes \\ in Dynamical Systems: \\ Qualitative Theory of Critical Delays}

\author{Alexander N. Gorban\thanks{agorban$@$mat.ethz.ch}\\
ETH-Zentrum, Department of Materials, \\ Institute of Polymers, Polymer Physics \\ Sonneggstr. 3,
CH-8092 Z{\"u}rich, Switzerland; \\ Institute of Computational Modeling SB RAS, \\ Akademgorodok,
Krasnoyarsk 660036, Russia}

\date{}

\maketitle

\begin{abstract}

The paper gives a systematic analysis of singularities of transition processes in dynamical systems.
General dynamical systems with dependence on parameter are studied. A system of relaxation times is
constructed. Each relaxation time depends on three variables: initial conditions, parameters $k$ of
the system and accuracy $\varepsilon$ of the relaxation. The singularities of relaxation times as
functions of $(x_0,k)$ under fixed $\varepsilon$ are studied. The classification of different
bifurcations (explosions) of limit sets is performed. The relations between the singularities of
relaxation times and bifurcations of limit sets are studied. The peculiarities of dynamics which
entail singularities of transition processes without bifurcations are described as well. The analogue
of the Smale order for general dynamical systems under perturbations is constructed. It is shown that
the perturbations simplify the situation: the interrelations between the singularities of relaxation
times and other peculiarities of dynamics for general dynamical system under small perturbations are
the same as for the Morse-Smale systems.

\end{abstract}

\noindent Math Subject Classifications: 54H20,  58D30, 37B25.

\noindent Key Words: Dynamical system, Transition process, Relaxation time, Bifurcation, Limit set,
Smale order.

 \clearpage

\tableofcontents

\clearpage

\addcontentsline{toc}{section}{Introduction}
\section*{Introduction} \label{S0}

Are there ``white spots" in topological dynamics? Undoubtedly, they exist: transition processes in
dynamical systems are still not very well studied. A a consequence, it is difficult to interpret the
experiments which reveal singularities of transition processes, and in particular, anomalously slow
relaxation. ``Anomalously slow" means here ``unexpectedly slow"; but what can one  expect from a
dynamical system in a general case?

In this paper, the transition processes in general dynamical systems are studied. The approach based
on the topological dynamics is quite general, but one  pays for these generality by the loss of
constructivity. Nevertheless, this stage of a general consideration is needed.

The limiting behaviour (as $t \rightarrow \infty$) of dynamical systems have been studied very
intensively in the XX century \cite{[1],[2],[3],[4],[5],[6]}. New types  of  limit  sets (``strange
attractors") were discovered \cite{[7],[8]}. Fundamental   results concerning the structure of limit
sets were obtained, such as the Kolmogorov-Arnold-Moser  theory \cite{[10],[11]}, the Pugh lemma
\cite{[9]}, the qualitative \cite{Sm,Ku,[4]} and quantitative \cite{Gro,Yo,Ka} Kupka--Smale theorem,
etc. The theory of limit behaviour ``on the average", the ergodic theory \cite{[12]}, was
considerably developed. Theoretical and applied achievements of the bifurcation theory have become
obvious \cite{[13],[14],[15]}. The fundamental textbook on dynamical systems \cite{Katok} and the
introductory review \cite{Katok2} are now available.

The achievements regarding transition processes have been not so impressive, and only relaxations in
linear and  linearized  systems  are well studied. The applications of this elementary theory
received the name the ``relaxation spectroscopy". Development of this discipline with applications in
chemistry and physics was distinguished by Nobel Prize (M.~Eigen \cite{[16]}).

A general theory of transition processes  of essentially non-linear systems does not exist. We
encountered this problem while studying transition processes in catalytic reactions. It was necessary
to give an interpretation on anomalously long  transition processes observed in experiments. To this
point, a discussion arose and even some papers were published. The focus of the discussion was:  do
the slow relaxations arise from slow ``strange processes" (diffusion,  phase transitions, and so on),
or could they have a purely kinetic (that is dynamic) nature?

Since a general theory  of  relaxation  times and their singularities was not available at that time,
we constructed it by ourselves from the very beginning \cite{[17],[18],[19],[20],[21],[22]}. In the
present paper the first, topological part of this theory is presented. It is quite elementary theory,
though rather lengthy $\varepsilon - \delta$ reasonings may require some time and effort. Some
examples of slow relaxation in chemical systems, their theoretical and numerical analysis, and also
an elementary introduction into the theory can be found in the monograph \cite{[23]}.

Two simplest mechanisms of slow relaxations  can  be readily mentioned: the delay of motion near an
unstable fixed  point, and the delay  of  motion  in a domain where a fixed point appears under a
small change of parameters. Let us give some simple examples of motion in  the segment $[-1,1]$.

The delay near an unstable fixed point exists  in  the  system $\dot x = x^2 -1$. There are two fixed
points $x = \pm1$ on the segment $[-1,1]$, the point $x=1$ is unstable and the point $x = -1$ is
stable. The equation is integrable explicitly: $$ x = [(1+x_0) {\rm e}^{-t} - (1-x_0) {\rm e}^t] /
[(1+x_0) {\rm e}^{-t} + (1-x_0) {\rm e}^t], $$ where $x_0 = x(0)$ is initial condition at $t=0$. If
$x_0 \neq 1$ then, after some time, the motion will come  into the $\varepsilon$-neighborhood of the
point $x=-1$, for whatever $\varepsilon > 0$. This process requires the time $$ \tau (\varepsilon,
x_0) = -\frac{1}{2} \ln \frac{\varepsilon}{2-\varepsilon} - \frac{1}{2} \ln \frac{1-x_0}{1+x_0}. $$
It is assumed that $1 > x_0 > \varepsilon-1$. If $\varepsilon$ is fixed then $\tau$  tends to $+
\infty$ as $x_0 \rightarrow 1$ like $-\frac{1}{2} \ln (1-x_0)$. The motion  that begins  near the
point  $x=1$ remains near  this point for a long time ( $ \sim -\frac{1}{2} \ln (1-x_0)$), and then
goes to  the  point  $x = -1$.  In order  to show it more clear, let us compute the time $\tau'$ of
residing in  the  segment $[-1 + \varepsilon,\ 1-\varepsilon]$ of the motion, beginning near the
point $x=1$,  i.e.  the time of its stay outside the $\varepsilon$-neighborhoods of  fixed  points $x
= \pm1$. Assuming $1-x_0 < \varepsilon$, we obtain $$ \tau' (\varepsilon, x_0) = \tau (\varepsilon,
x_0) - \tau (2- \varepsilon, x_0) = -\ln\, \frac{\varepsilon}{2-\varepsilon}. $$ One can see that  if
$1-x_0 <  \varepsilon$ then $\tau' (\varepsilon, x_0)$ does not depend on $x_0$. This is obvious: the
time $\tau'$ is the time of travel from point $1-\varepsilon$   to  point $-1+\varepsilon$.

Let us consider the system $\dot x = (k+x^2)(x^2-1)$ on $[-1,1]$ in order to obtain the example of
delay of motion  in  a domain where a  fixed point  appears under small change  of parameter. If
$k>0$, there are again only two fixed points $x = \pm 1,\ x = -1$ is a stable  point and $x=1$ is an
unstable one. If $k=0$ there appears the third point $x=0$. It is not stable, but ``semistable" in
the following sense:  If  the initial position is $x_0 > 0$ then the motion goes from $x_0$  to $x =
0$. If $x_0 < 0$ then the motion goes from $x_0$ to $x = -1$. If $k<0$ then apart from $x = \pm1$,
there are two other fixed  points $x = \pm \sqrt{|k|}$. The positive point is stable, and the
negative one is unstable. Let us consider the case $k>0$. The time of motion from the point $x_0$ to
the point $x_1$ can be found explicitly $(x_{0,1} \neq \pm1)$: $$ t = \frac{1}{2} \ln
\frac{1-x_1}{1+x_1} - \frac{1}{2} \ln \frac{1-x_0}{1+x_0} - \frac{1}{\sqrt{k}} \left(\arctan
\frac{x_1}{\sqrt{k}} - \arctan \frac{x_0}{\sqrt{k}} \right). $$ If $x_0 > 0,\ x_1 < 0,\ k > 0,\ k
\rightarrow 0$, then $t \rightarrow \infty$ like $\pi / \sqrt{k}$.

The examples presented do not  exhaust  all  the  possibilities: they rather illustrate two common
mechanisms of slow  relaxations appearance.

Below we study parameter-dependent dynamical systems. The point of view of topological dynamics is
adopted (see \cite{[1],[2],[3],[6],[24],[25]}). In the first place this  means  that, as a rule, the
properties associated with the smoothness, analyticy and so  on will be  of  no importance. The phase
space $X$ and the parameter space $K$ are  compact  metric spaces:  for any points  $x_1, x_2$ from
$X$ ($k_1, k_2$ from $K$) the distance $\rho (x_1, x_2)$ ($\rho_K (k_1, k_2)$) is  defined with the
following properties:
\begin{eqnarray*}
& &\rho (x_1, x_2) = \rho (x_2, x_1), \ \
\rho(x_1, x_2) + \rho(x_2, x_3) \geq \rho(x_1, x_3),\\
& &\rho (x_1, x_2) = 0 \ \mbox{if\ and\ only\ if}\ x_1 = x_2
\ (\mbox{similarly\ for}\ \rho_K).
\end{eqnarray*}

The sequence $x_i$ converges to $x^* \ (x_i \rightarrow x^*)$  if $\rho (x_i, x^*) \rightarrow 0$.
The compactness means that from any sequence a convergent subsequence can be chosen.

The states of the system are represented by the points of the phase space $X$. The reader can think
of $X$  and $K$  as closed, bounded subsets of finite-dimensional  Euclidean  spaces,  for example
polyhedrons, and $\rho$  and $\rho_K$ are  the standart  Euclidean distances.

Let  us   define   the phase flow (the transformation {\it ``shift over the time $t$"}). It is a
function $f$ of three arguments: $x \in X$ (of the  initial  condition), $k \in K$   (the parameter
value) and $t \geq 0$, with values in $X$: $f(t,x,k) \in X$. This  function is assumed continuous on
$[0, \infty) \times X \times K $ and satisfying the following conditions:
\begin{itemize}
\item{$f(0,x,k)=x$ (shift over zero time leaves any point in its place);} \item{ $f(t,f (t', x, k), k) =
f(t+t', x, k)$ (the  result of sequentially executed shifts over $t$ and $t'$ is the  shift  over
$t+t'$);}
\item{ if $x \neq x'$, then $f(t,x,k) \neq f(t, x', k)$ (for any $t$ distinct initial points are
shifted in time $t$ into distinct points for. } \end{itemize}

For given value of parameter $k \in K$ and initial  state $x \in X$ the {\it $\omega$-limit set}
$\omega (x,k)$  is the set  of  all limit points of $f(t,x,k)$ for $t \rightarrow \infty$:
\begin{eqnarray*}
& y \in \omega (x,k) \ \mbox{if and only if there exists such a sequence} \ t_i \geq 0 & \\ &
\mbox{that} \ t_i \rightarrow \infty \ \mbox{and} \ f(t_i, x, k) \rightarrow y. &
\end{eqnarray*}
Examples  of $\omega$-limit points are stationary  (fixed)  points, points of limit cycles and so on.

{\it The relaxation} of a system can be understood as its motion to the $\omega$-limit set
corresponding to given initial  state and value of parameter. {\it The relaxation time} can be
defined as the time of this motion. However, there are several possibilities to make this definition
precise.

Let $\varepsilon > 0$. For given  value  of parameter $k$ we denote  by $\tau_1(x, k, \varepsilon)$
the time during which the system will come from the initial state $x$ into the
$\varepsilon$-neighbourhood of $\omega(x,k)$ (for the first time). The $(x,k)$-motion can enter the
$\varepsilon$-neighborhood of the $\omega$-limit set, then this motion can leave it, then reenter it,
and so on it can enter and leave the $\varepsilon$-neighbourhood of $\omega(x,k)$ several times.
After all, the motion will enter this neighbourhood finally, but this may take more time than the
first entry. Therefore, let us introduce for  the $(x,k)$-motion  the time of being outside the
$\varepsilon$-neighborhood of $\omega(x,k)\ (\tau_2)$ and the time of final entry into it $(\tau_3)$.
Thus, we have a system of relaxation times that describes the relaxation of the $(x,k)$-motion to its
$\omega$-limit set $\omega(x,k)$:
\begin{eqnarray*}
& &\tau_1 (x,k,\varepsilon) = \inf \{t>0 \ | \ \rho^* (f(t,x,k),\
\omega(x,k)) < \varepsilon\};\\
& &\tau_2(x,k, \varepsilon) = \mbox{mes} \{t > 0 \ | \ \rho^* (f (t, x, k),\
\omega(x,k)) \geq \varepsilon \};\\
& &\tau_3 (x,k, \varepsilon) = \inf \{ t>0 \ | \ \rho^*(f (t',x,k),\
\omega(x,k)) < \varepsilon\  \mbox{for}\ t'>t\}.
\end{eqnarray*}
Here mes is the Lebesgue measure (on the real line it is length), $\rho^*$ is the distance from the
point to the set: $\rho^*(x, P) = \inf_{y \in P} \rho (x, y)$.

The $\omega$-limit  set depends on an initial state (even under the fixed value of $k$). The limit
behavior of the system can be characterized also by the {\it total limit set} $$ \omega (k) =
\bigcup_{x \in X} \omega (x, k). $$ The set $\omega (k)$ is the union of all $\omega (x, k)$ under
given $k$. Whatever  initial state would be, the system after some time will  be  in the
$\varepsilon$-neighborhood of $\omega (k)$. The relaxation can be also considered as a motion towards
$\omega(k)$. Introduce the corresponding system of relaxation times:
\begin{eqnarray*}
& &\eta_1 (x,k,\varepsilon) = \inf \{t>0 \ | \ \rho^* (f(t,x,k),\
\omega(k)) < \varepsilon\};\\
& &\eta_2(x,k, \varepsilon) = \mbox{mes} \{t > 0 \ | \ \rho^* (f (t, x, k),\
\omega(k)) \geq \varepsilon \};\\
& &\eta_3 (x,k, \varepsilon) = \inf \{ t>0 \ | \ \rho^*(f (t',x,k),\
\omega(k)) < \varepsilon\  \mbox{for}\ t'>t\}.
\end{eqnarray*}

Now we are able to define    a   {\it slow transition process}. There is no distinguished  scale  of
time, which could be compared with relaxation times. Moreover, by decrease of  the relaxation
accuracy $\varepsilon$ the relaxation times can become of any large amount even in the simplest
situations of motion to unique stable fixed point. For every initial state $x$ and given $k$ and
$\varepsilon$ all relaxation times are finite. But the set of relaxation time values for various $x$
and $k$  and given $\varepsilon >0$ can be unbounded. Just in this case we speak about the slow
relaxations.

Let us consider the simplest example. Let us consider the differential equation $\dot x = x^2 -1$ on
the  segment $[-1,1]$ . The point $x = -1$ is stable, the point $x = 1$ is unstable. For any fixed
$\varepsilon > 0,\ \varepsilon < \frac{1}{2}$ the  relaxation times $\tau_{1, 2, 3}, \eta_3$ have the
singularity: $\tau_{1, 2, 3}, \eta_3 (x, k, \varepsilon) \rightarrow \infty$ as $x \rightarrow 1,\ x
< 1$. The times $\eta_1, \eta_2$    remain bounded in this case.

     Let us say that the system has
$\tau_i$- $(\eta_i)$-{\it slow relaxations}, if for some  $\varepsilon > 0$ the function $\tau_i (x,
k, \varepsilon)$ $( \eta_i (x, k, \varepsilon))$ is unbounded from above in $X \times K$, i.e. for
any $t>0$  there  are  such $x \in X,\ k \in K$,    that $\tau_i (x, k, \varepsilon) > t\ (\eta_i(x,
k, \varepsilon) > t)$.

One of the possible reasons of slow relaxations is a sudden jump  in dependence of the $\omega$-limit
set $\omega(x,k)$ of $x,k$ (as well as a jump in dependence  of $\omega(k)$ of $k)$. These
``explosions" (or  bifurcations)  of $\omega$-limit sets are studied in Sec.~\ref{S1}. In the next
Sec.~\ref{S2} we give the theorems,  providing  necessary  and sufficient conditions of slow
relaxations. Let us mention two of them.

\vspace{10pt}

\noindent{\bf Theorem \ref{T2.1}$'$} {\it The system has  $\tau_1$-slow relaxations if and  only if
there is a singularity  of  the  dependence $\omega(x,k)$  of  the following kind: there are such
points $x^* \in X,\ k^* \in K$, sequences $x_i \rightarrow x^*,\ k_i \rightarrow k^*$,  and number
$\delta >0$, that for  any $i, \ y \in \omega (x^*, k^*),\ z \in \omega (x_i, k_i)$   the distance
$\rho(y, z) > \delta$. }

\vspace{10pt}

The singularity of $\omega (x, k)$ described in the statement  of  the theorem indicates that the
$\omega$-limit set $\omega (x, k)$ makes a jump: the distance from any point of $\omega (x_i, k_i)$
to any point  of $\omega (x^*, k^*)$ is greater than $\delta$.

By the next theorem, necessary and sufficient  conditions of  $\tau_3$-slow relaxations are given.
Since $\tau_3 \geq \tau_1$, the  conditions  of $\tau_3$-slow relaxations are weaker than the
conditions of Theorem \ref{T2.1}$'$, and $\tau_3$-slow relaxations are ``more often" than
$\tau_1$-slow relaxation (the relations between different kinds of slow relaxations with
corresponding examples are given below  in  Subsec. \ref{SS3.2}). That is why   the discontinuities
of $\omega$-limit  sets  in the following theorem are weaker.

\vspace{10pt}

 \noindent{\bf Theorem \ref{T2.7}} {\it $\tau_3$-slow relaxations exist if and only if
at least one of the following conditions is satisfied:
\begin{enumerate}
\item{There are points $x^* \in X,\ k^* \in K,\ y^* \in \omega (x^*, k^*)$, sequences $x_i \rightarrow
x^*,\ k_i \rightarrow k^*$
 and number $\delta > 0$ such that for any $i$ and $z \in \omega (x_i,\
k_i)$ the inequality $\rho(y^*,z) > \delta$ is valid\footnote {The existence of one such $y$ is
sufficient, compare it with Theorem \ref{T2.1}$'$.}}
\item{There are $x \in X,\ k \in K$ such that $x
\not \in \omega (x, k)$,  for  any  $t>0$  can be found $y(t) \in X$, for which $f(t, y(t), k) = x\
(y(t)$ is  a  shift  of $x$ over $-t$), and for some $z \in \omega (x, k)$ can be found such a
sequence $t_i \rightarrow \infty$ that $y(t_i) \rightarrow z$. \footnote{That is, the
$(x,k)$-trajectory is a generalized loop: the intersection of its $\omega$-limit set and
$\alpha$-limit set (i.e., the limit set for $t \rightarrow - \infty$) is non-empty, and $x$ is not a
limit point for the $(x,k)$-motion}}
\end{enumerate} }

\vspace{10pt}

An example of the point satisfying the condition 2 is provided by any point lying on the loop, that
is the trajectory starting from  the fixed point and returning to the same point.

Other theorems of Sec.~\ref{S2}  also  establish  connections between slow relaxations and
peculiarities of the limit  behaviour under  different  initial conditions  and  parameter  values.
In general, in topological and differential  dynamics  the main attention is paid to  the  limit
behavior  of  dynamical  systems
\cite{[1],[2],[3],[4],[5],[6],[24],[25],[26],[27],[28],Katok,Katok2}. In applications,  however,  it
is often of importance how rapidly the motion  approaches  the limit regime.  In  chemistry,
long-time delay of reactions far from equilibrium (induction periods) have been studied since
Van't-Hoff \cite{[29]} (the first Nobel Prize laureate in Chemistry). It is necessary to mention the
classical monograph of N.N. Semjonov [30] (also the Nobel Prize laureate in Chemistry), where
induction periods in combustion are studied. From the latest works let us note \cite{[31]}. When
minimizing functions by relaxation methods, the similar delays can cause some problems. The paper
\cite{[32]}, for example, deals with their elimination. In the simplest cases, the slow relaxations
are bound with delays near unstable fixed points. In the general case, there is a complicated system
of interrelations between different types of slow relaxations and other dynamical peculiarities, as
well as of different types of slow relaxations between themselves. These relations are  the subject
of Sects.~\ref{S2},~\ref{S3}. The investigation is performed generally in the way of classic
topological dynamics \cite{[1],[2],[3]}. There are, however, some distinctions:
\begin{itemize}
\item{From the  very beginning not only one system is considered, but also practically more important
case of parameter dependent systems;}
\item {The motion in these systems is defined, generally
speaking, only for positive  times.}
\end{itemize}

The  last circumstance is bound with the fact that for applications (in particular, for chemical
ones) the motion is defined only in a positive invariant  set (in balance polyhedron, for example).
Some results can be accepted for the case of general semidynamical systems
\cite{[33],[34],[35],[36],[37]}, however, for the majority of applications,  the  considered  degree
of generality is more than sufficient.

For a separate semiflow $f$ (without parameter)  $\eta_1$-slow relaxations are impossible, but
$\eta_2$-slow relaxations can appear in a separate system too (Example \ref{E2.1}). Theorem
\ref{T3.1} gives the necessary conditions for $\eta_2$-slow relaxations in systems without parameter.

Let us recall the definition of {\it non-wandering points}. A  point $x^*\in X$   is the
non-wandering point for the semiflow $f$, if for any neighbourhood $U$ of $x^*$ and for any $T>0$
there is such $t>T$ that $f(t,U) \bigcap U \neq \varnothing$. Let us denote by $\omega_f$ the
complete $\omega$-limit set of one semiflow $f$ (instead of $\omega(k)$).

\vspace{10pt}

\noindent{\bf Theorem \ref{T3.1}} {\it Let a semiflow $f$ possess   $\eta_2$-slow  relaxations. Then
there exists a non-wandering point $x^* \in X$ which does not belong to $\overline{\omega_f}$.}

\vspace{10pt}

For smooth systems it is possible to  obtain  results that have no analogy in topological dynamics.
Thus, it is shown in Sec.~\ref{S2}  that ``almost always" $\eta_2$-slow relaxations are absent in one
separately taken  $C^1$-smooth dynamical system (system,  given  by differential equations with
$C^1$-smooth right parts). Let us explain what ``almost always" means in this case. A set $Q$ of
$C^1$-smooth dynamical systems with common phase space is called nowhere-dense in $C^1$-topology, if
for  any  system  from $Q$   an infinitesimal perturbation of right hand parts can be chosen
(perturbation  of  right hand parts and its first derivatives should be smaller than an arbitrary
given $\varepsilon > 0$) such that the perturbed system should not belong  to $Q$ and should exist
$\varepsilon_1 > 0\ (\varepsilon_1 < \varepsilon)$ such that under $\varepsilon_1$-small variations
of right parts (and of first derivatives) the perturbed system could not return in $Q$. The union of
finite number of nowhere-dense sets  is also nowhere-dense. It is not the case for countable  union:
for example, a point on a line  forms nowhere-dense set, but the countable set of rational numbers is
dense on the real line: a rational number is on any segment. However, both on line  and  in many
other cases countable union of nowhere-dense sets can be considered  as  very ``meagre". Its
complement is so-called ``residual set". In particular, for $C^1$-smooth dynamical systems on compact
phase space the union of countable number of nowhere-dense sets has the following property: any
system, belonging to this union, can be eliminated from it  by infinitesimal perturbation. The above
words ``almost always" meant:  except for  union  of countable number of nowhere-dense sets.

In two-dimensional  case  (two  variables), ``almost any" $C^1$-smooth dynamical system is rough,
i.e. its  phase portrait under small  perturbations  is  only  slightly  deformed, qualitatively
remaining the  same.  For rough two-dimensional systems $\omega$-limit sets consist of fixed points
and limit cycles, and the stability of these points and cycles can be verified by linear
approximation. The correlation of  six  different  kinds  of  slow relaxations between themselves for
rough two-dimensional  systems becomes considerably more simple.

\vspace{10pt}

\noindent{\bf Theorem \ref{T3.5}} {\it Let $M$ be $C^{\infty}$-smooth compact manifold, $\dim M =2,\
F$ be a structural  stable smooth dynamical system over $M$, $F|_X$ be  an  associated  with $M$
semiflow   over   connected   compact positive-invariant subset $X \subset M$. Then:
\begin{enumerate}
\item{For $F|_X$  the existence of  $\tau_3$-slow relaxations is equivalent to the existence of
$\tau_{1,2}$- and  $\eta_3$-slow relaxations;}
\item{$F|_X$  does not possess  $\tau_3$-slow
relaxations if  and  only  if $\omega_F \bigcap X$
 consists of one fixed point or of points of one limit cycle;}
\item{$\eta_{1,2}$-slow relaxations are impossible for $F|_X$.}
\end{enumerate}
}

\vspace{10pt}

For smooth rough two-dimensional  systems  it  is  easy  to estimate the measure (area) of  the
region  of  durable  delays $\mu_i(t) = \mbox{mes} \{ x \in X \ | \ \tau_i (x, \varepsilon) >t \}$
under fixed  sufficiently  small $\varepsilon$   and large $t$ (the  parameter  $k$ is  absent
because a separate  system  is studied). Asymptotical behaviour of $\mu_i (t)$ as $t \rightarrow
\infty$ does not  depend on $i$ and $$ \lim_{t \rightarrow \infty} \frac{\ln \mu_i (t)}{t} = - \min
\{\varkappa_1, \ldots, \varkappa_n \}, $$ where $n$ is a number of unstable limit motions (of fixed
points and cycles) in $X$, and the numbers   are determined as follows. We denote by $B_i, \ldots,
B_n$ the unstable limit motions lying in $X$.
\begin{enumerate}
\item{Let $B_i$  be an unstable node or focus. Then $\varkappa_1$ is the  trace of matrix of linear
approximation in the point $b_i$.} \item{Let $b_i$  be a saddle. Then $\varkappa_1$ is positive
eigenvalue  of the matrix of linear approximation in this point.} \item{Let  $b_i$   be  an unstable
limit   cycle.   Then $\varkappa_i$     is characteristic indicator of the cycle (see \cite{[38]}, p.
111).}
\end{enumerate}
Thus, the area of the region  of  initial  conditions,  which result in durable delay of the motion,
in the case of smooth rough two-dimensional systems behaves at large delay  times as $\exp
(-\varkappa t)$, where  $t$ is a  time of  delay, $\varkappa$   is  the  smallest number of
$\varkappa_i, \ldots, \varkappa_n$. If $\varkappa$ is close to zero (the system is close to
bifurcation \cite{[5],[38]}), then this area decreases slowly enough  at large $t$. One can find here
analogy with linear time of  relaxation to a stable fixed point $$ \tau_l = -1/\max \mbox{Re} \lambda
$$ where $\lambda$ runs through all  the  eigenvalues of the matrix of linear approximation  of right
parts in this point, $\max \mbox{Re} \lambda$ is the largest (the  smallest  by  value)  real part of
eigenvalue, $\tau_l \rightarrow \infty$  as $\mbox{Re} \lambda \rightarrow 0$.

However, there are  essential differences. In particular, $\tau_l$   comprises  the  eigenvalues
(with negative real part) of linear approximation matrix in that (stable) point, to  which  the
motion is going, and the asymptotical estimate $\mu_i$  comprises  the  eigenvalues (with positive
real part) of the matrix in that (unstable) point or cycle, near which the motion is retarded.

In typical situations for two-dimensional parameter depending systems the singularity of $\tau_l$
entails existence  of  singularities of relaxation times $\tau_i$ (to this statement can be  given an
exact meaning and it can be proved as a theorem). The inverse  is  not true. As an example should be
noted the   delays  of  motions near  unstable fixed  points.  Besides,  for  systems  of higher
dimensions the situation  becomes  more complicated,  the rough systems cease to be ``typical" (this
was  shown  by  S.~Smale \cite{[39]}, the discussion see in \cite{[5]}), and the limit behaviour even
of rough systems does not come to tending of motion to fixed point or limit cycle. Therefore the area
of reasonable application the linear relaxation time $\tau_l$ to analysis  of transitional processes
becomes in this case even more restricted.

Any real system exists under the permanent perturbing influence of the external world. It is hardly
possible to construct a model taking into account all such perturbations. Besides  that, the model
describes  the  internal properties of the system  only approximately. The discrepancy between the
real  system and the model arising from these two circumstances is  different for different models.
So, for the systems of  celestial  mechanics it can be done  very  small.  Quite the contrary,  for
chemical kinetics, especially for kinetics of heterogeneous catalysis, this discrepancy can be if not
too large but, however, not  such  small to be neglected. Strange as it may seem, the presence of
such an unpredictable divergence of the model and reality can  simplify  the situation: The
perturbations ``conceal" some fine details of dynamics, therefore these details become irrelevant to
analysis of real systems.

Sec.~\ref{S4}  is  devoted  to  the  problems  of  slow relaxations in presence of small
perturbations. As  a  model  of perturbed motion here are taken $\varepsilon$-{\it motions}: the
function  of  time $\varphi (t)$ with values in $X$, defined at $t \geq 0$, is  called
$\varepsilon$-motion $(\varepsilon > 0)$ under  given   value   of $k \in K$,   if   for   any $t
\geq 0,\ \tau \in [0, T]$ the inequality $\rho(\varphi (t+\tau),\ f(\tau, \varphi (t), k)) <
\varepsilon$ holds. In other words, if for an arbitrary point $\varphi (t)$ one considers its motion
on the force  of  dynamical system, this motion will diverge $\varphi(t+\tau)$ from    no  more than
at $\varepsilon$  for $\tau \in [0, T]$. Here  $[0,T]$ is a certain interval of time, its length $T$
is not very important (it is important that  it  is  fixed),  because later we shall consider the
case $\varepsilon \rightarrow 0$.

     There are two traditional approaches to the  consideration of
perturbed motions. One of them is to investigate the motion in the presence of small constantly
acting  perturbations  \cite{[40],[41],[42],[43],[44],[45],[46]},  the other is the study of
fluctuations under the influence of small stochastic  perturbations
\cite{[47],[48],[49],[50],[51],[52]}. The stated results join  the first direction, there to are used
some ideas bound with the second one. The $\varepsilon$-motions were studied earlier in differential
dynamics,  in general  in  connection with the   theory  of Anosov about $\varepsilon$-trajectories
and its applications \cite{[27],[53],[54],[55],[56]}, see also \cite{[57]}.

When studying perturbed motions, we correspond to each point ``a bundle" of  $\varepsilon$-motions,
$\{\varphi (t)\}$, $t \geq 0$ going  out from this point ($\varphi (0) = x$) under given value  of
parameter $k$. The totality of all $\omega$-limit points of these $\varepsilon$-motions (of limit
points of all $\varphi (t)$ as $t \rightarrow \infty$) is denoted by $\omega^{\varepsilon}(x,k)$.
Firstly, it is necessary to notice that $\omega^{\varepsilon}(x,k)$ does not always tend to
$\omega(x,k)$ as $\varepsilon \rightarrow 0$: the set $\omega^0(x,k) = \bigcap_{\varepsilon > 0}
\omega^{\varepsilon}(x,k)$ may not coincide with $\omega(x,k)$. In Sec.~\ref{S4} there are studied
relaxation times of $\varepsilon$-motions and  corresponding slow relaxations. In contrast to the
case of nonperturbed  motion, all natural kinds of slow relaxations are not considered because they
are too numerous (eighteen), and the principal attention  is  paid to two of them, which are analyzed
in more  details  than  in Sec.~\ref{S2}.

The structure of limit sets of one  perturbed  system is studied. The analogy of general perturbed
systems  and Morse-Smale systems as well as smooth rough two-dimensional systems is revealed. Let us
quote in this connection the review by Professor A.M.Molchanow of the thesis \cite{Diss} of
A.N.Gorban\footnote{This paper is the first complete publication of that thesis.} (1981): {\it``After
classic works of Andronov, devoted to the rough systems on the plane, for  a long  time it  seemed
that division of plane into finite number of cells with source and drain is an example of structure
of multidimensional systems too... The most interesting (in the opinion of opponent) is the fourth
chapter ``Slow relaxations of the perturbed systems". Its principal result is approximately as
follows.  If  a complicated dynamical system is made rough (by means of $\varepsilon$-motions), then
some its important properties are similar to the properties of rough systems on the plane. This is
quite positive result, showing in what sense the approach of Andronov can be generalized for
arbitrary systems"}.

To study limit sets of perturbed system, two relations are introduced in \cite{[22]} for general
dynamical systems: of preorder $\succsim$ and of equivalence $\sim$:
\begin{itemize}
\item{$x_1 \succsim x_2$  if  for any $\varepsilon > 0$ there is such a $\varepsilon$-motion
$\varphi(t)$ that $\varphi(0)=x_1$ and $\varphi(\tau)=x_2$  for some $\tau>0$;} \item {$x_1 \sim x_2$
if $x_1 \succsim x_2$ and $x_2 \succsim x_1$.}
\end{itemize}
For smooth dynamical systems with finite number of ``basic attractors" similar relation of
equivalence had been introduced with the help of action functionals in studies on stochastic
perturbations  of dynamical systems (\cite{[52]} p. 222 and further). The concepts of
$\varepsilon$-motions and related topics can be found in \cite{[57]}. For the Morse-Smale systems
this relation is the Smale order \cite{[4]}.

Let $\omega^0 = \bigcup_{x \in X} \omega^0 (x)$ ($k$ is omitted, because only one system is studied).
Let us identify equivalent points in $\omega^0$. The  obtained factor-space  is  totally disconnected
(each  point  possessing  a fundamental   system   of   neighborhoods   open    and closed
simultaneously). Just this space $\omega^0 / \sim$ with the order over it can be considered as a
system of sources and drains analogous  to  the system  of  limit  cycles  and  fixed  points  of
smooth rough two-dimensional dynamical system. The sets $\omega^0 (x)$  can  change  by jump only on
the  boundaries  of  the region  of  attraction  of corresponding ``drains"  (Theorem  \ref{T4.4}).
This totally disconnected factor-space $\omega^0 / \sim$ is the generalization of the Smale diagrams
\cite{[4]} defined for the Morse-Smale systems onto the whole class of general dynamical systems. The
interrelation of six principal kinds of slow relaxations in perturbed system is analogous to their
interrelation in smooth rough two-dimensional system described in Theorem \ref{T3.5}.

Let  us  enumerate  the  most  important   results   of   the investigations being  stated.
\begin{enumerate}
\item{It is not always necessary to search for ``foreign" reasons of slow relaxations, in the first
place one should  investigate  if there are  slow  relaxations  of  dynamical origin in the system.}
\item{One of possible reasons of slow relaxations is the existence  of bifurcations (explosions) of
$\omega$-limit sets. Here, it is necessary to study  the  dependence $\omega(x,k)$ of  limit  set
both on parameters  and initial data.  It is violation  of  the  continuity with respect to $(x,k)\in
X\times K$ that leads to slow relaxations.}
\item{The complicated dynamics can be made ``rough" by perturbations. The useful model of perturbations in
topological dynamics provide the $\varepsilon$-motions. For $\varepsilon \rightarrow 0$ we obtain the
rough structure of sources and drains similar to the Morse-Smale systems (with totally disconnected
compact instead of finite set of attractors).}
\item{The interrelations between the singularities of relaxation
times and other peculiarities of dynamics for general dynamical system under small perturbations are
the same as for the Morse-Smale systems, and, in particular, the same as for rough two-dimensional
systems.}
\item{There is a large quantity of different slow  relaxations, unreducible  to  each other, therefore
for interpretation of experiment it is  important to understand which namely of relaxation times is
large.}
\item{Slow relaxations in real systems often are ``bounded slow",  the relaxation time is large
(essentially greater than could  be expected proceeding from the coefficients of equations and
notions about the characteristic times), but  nevertheless  bounded. When studying such
singularities, appears to be useful  the  following method, ascending to the works  of A.A.~Andronov:
the considered system  is included  in  appropriate  family   for   which slow relaxations are to be
studied in the sense accepted in the present work. This study together with the mention of degree of
proximity of particular systems to the initial one  can  give  an important information.}
\end{enumerate}

\section{Bifurcations (Explosions) of $\omega$-limit Sets}\label{S1}

Let $X$ be a compact metric space with the metrics $\rho$, and $K$ be a compact metric space (the
space of parameters) with the metrics $\rho_K$,
\begin{equation}\label{e1}
f: [0, \infty) \times X \times K \rightarrow X
\end{equation}
be a continuous mapping for any $t \geq 0,\ k \in K$; let mapping
$f (t, \cdot, k): X \rightarrow X$
be homeomorphism of $X$  into subset of $X$ and  under  every $k \in K$
let these homeomorphisms form monoparametric semigroup:
\begin{equation}\label{e2}
f(0, \cdot, k) = {\rm id},\ f(t,f(t',x,k), k) = f(t+t', x,k)
\end{equation}
for any $t, t' \geq 0, x \in X$.

     Below we  call the semigroup of mappings $f(t,\cdot,k)$   under
fixed $k$ a {\it semiflow of homeomorphisms} (or,  for  short,  semiflow), and the mapping (\ref{e1})
a family of semiflows or simply a system  (\ref{e1}). It is obvious that all results, concerning  the
system (\ref{e1}), are valid also in the case when $X$ is  a  phase  space  of  dynamical system,
i.e. when every semiflow can be prolonged along $t$  to  the left onto the whole axis $(- \infty,
\infty)$ up to flow  (to monoparametric group of homeomorphisms of $X$ onto $X$).

\subsection{Extension of Semiflows to the Left} \label{SS1.1}

 It is clear that under fixed $x$ and $k$  the mapping
$f(\cdot, x, k)$: $t \rightarrow f(t, x, k)$ can be, generally speaking, defined  also  for certain
negative $t$, preserving semigroup property  (\ref{e2}).  Really,  consider under fixed $x$ and $k$
the set of all non-negative $t$ for which  there is point $q_i \in X$ such that $f(t, q_i, k) = x$.
Let us denote the  upper  bound  of this set by $T(x,k)$:
\begin{equation}\label{e3}
T(x,k) = \sup \{ t \ | \ \exists q_t \in X,\ f(t,q_t,k) = x\}.
\end{equation}

Under given $t, x,k$ the point $q_t$, if it exists,  has  a  single value,  since  the  mapping $f(t,
\cdot,k): X \rightarrow X$ is   homeomorphism. Introduce the denotation $f(-t,x,k) = q_t$. If
$f(-t,x,k)$ is determined, then    for    any $\tau$     within  $0 \leq \tau \leq t$ the point
$f(-\tau,x,k)$ is  determined: $f(-\tau,x,k) = f(t - \tau, f(-t,x,k), k)$. Let $T(x,k) < \infty,
\quad T(x,k) >t_n >0\quad (n=1,2, \ldots),\quad t_n \rightarrow T$. Let us choose from the  sequence
$f(-t_n, x,k)$ a subsequence converging to some $q^* \in X$ and denote it by $\{ q_j \}$, and the
corresponding times denote by $-t_j \ (q_j = f(-t_j, x,k))$. Owing  to  the continuity of $f$ we
obtain: $f(t_j, q_j, k) \rightarrow f(T(x,k),q^*,k)$,  therefore $f(T(x,k),q^*,k) = x$. Thus,
$f(-T(x,k), x, k) = q^*$.

So, under fixed $x,k$ the mapping $f$ was  determined  by  us  in interval $[-T (x,k), \infty)$, if
$T(x,k)$ is finite, and in  $(-\infty, \infty)$ in  the opposite case. Let us denote by $S$ the set
of all triplets $(t,x,k)$, in which $f$ is now determined. For enlarged mapping $f$ the semigroup
property in following form is valid:

\begin{proposition}\label{P1.1}(Enlarged semigroup property).

A) If $(\tau, x, k)$ and $(t,f(\tau,x,k),k) \in S$, then $(t+ \tau, x,k) \in S$  and  the equality
\begin{equation}\label{e4}
f(t,f(t,x,k), k) = f(t+\tau, x, k)
\end{equation}
is true.

B)   Inversely,   if $(t+\tau,x,k)$    and $(\tau,x,k) \in S$, then $(t,f(\tau,x,k),k) \in S$ and
(\ref{e4}) is true.

Thus, if the left part of the equality (\ref{e4}) makes sense,  then its right part is determined too
and the  equation  is  valid.  If there are determined both the right part and $f(\tau,x,k)$ in the
left part, then the whole left part makes sense and (\ref{e4}) is true.
\end{proposition}
{\bf Proof.} The proof consists  in  consideration  of  several  variants. Since the parameter $k$ is
assumed to be fixed, for the  purpose  of shortening the record it is absent in following formulas.\\
     1. $f(t,f(-\tau,x)) = f(t-\tau,x) \ (t, \tau>0) \ a) \ t> \tau >0$.\\
Let the left  part make sense: $f(-\tau, x)$ is determined. Then, taking into account  that
$t-\tau>0$, we   have $f(t,f(-\tau,x)) = f(t-\tau+\tau, f(-\tau,x)) = f(t-\tau,f (\tau,f (-\tau, x)))
= f(t-\tau,x)$, since $f(\tau,f(-\tau,x)) = x$ by definition.

Therefore the equality 1 is true (the right part makes  sense
since  $t>\tau$)- the part  for the case 1a is proved.  Inversely,  if
$f(-\tau,x)$ is determined, then the whole left part of 1  $(t>0)$  makes
sense, and then according to the proved the equality is true.

The other variants are considered in analogous way.

\begin{proposition}\label{P1.2}
The set $S$ is closed in $(-\infty, \infty) \times X \times K$ and  the mapping  $f : S \rightarrow
X$ is continuous.
\end{proposition}
{\bf Proof}. Denote by  $\langle -T(x,k), \infty)$ the  interval  $[-T(x,k), \infty)$,  if $T(x,k)$
is finite, and the whole axis $(-\infty, \infty)$ in opposite case. Let $t_n \rightarrow t^*,\ x_n
\rightarrow x^*,\ k_n \rightarrow k^*$, and $t_n \in \langle -T (x_n, k_n), \infty)$.   To   prove
the proposition, it should be made certain that $t^* \in \langle  -T (x^*, k^*), \infty)$   and
$f(t_n, x_n, k_n) \rightarrow f(t^*, x^*, k^*)$. If $t^* > 0$,  this  follows  from   the continuity
of $f$ in $[0,\infty) \times X \times K$. Let $t^* \leq 0$. Then it  can  be  supposed that $t_n <0$.
Let us redenote with changing the signs $t_n$  by $-t_n$   and $t^*$ by $-t^*$. Let us choose from
the sequence $f(-t_n, x_n, k_n)$  using  the compactness of $X$ a subsequence converging to some $q^*
\in X$. let us denote it by  $q_j$, and the sequences of corresponding  $t_n, x_n$   and $k_n$ denote
by $t_j, x_j$ and  $k_j$. The sequence  $f(t_j, q_j, k_j)$ converges  to $f(t^*, q^*, k^*)\ (t_j >0,\
t^* >0)$. But $f(t_j, q_j, k_j) = x_j \rightarrow x^*$. That is why $f(t^*, q^*, k^*) = x^*$ and
$f(-t^*, x^*, k^*) = q^*$ is determined. Since $q^*$  is an arbitrary  limit point of $\{q_n\}$, and
the  point  $f(-t^*,x^*,k^*)$, if  it  exists,  is determined by given $t^*, x^*, k^*$   and  has  a
single  value,  the sequence $q_n$  converges to  $q^*$. The proposition is proved.

Later  on  we shall   denominate   the   mapping $f(\cdot, x,k): \ \langle  -T(x,k), \omega)
\rightarrow X$ $k$-{\it motion} of the point $x$ ($(k,x)$-motion), the  image of $(k,x)$-motion --
$k$-{\it trajectory} of the  point $x$ ($(k,x)$-trajectory), the image of the interval $\langle
-T(x,k), 0)$ a negative, and the image of $0,\infty)$  a  positive  $k$-{\it semitrajectory}  of  the
point $x$ ($(k,x)$-semitrajectory). If $T(x,k) = \infty$, then let us call the $k$-motion  of  the
point $x$ the {\it whole} $k$-motion, and the corresponding $k$-trajectory the whole $k$-trajectory.

Let  $(x_n, k_n) \rightarrow (x^*, k^*),\ t_n \rightarrow t^*,\ t_n, t^* >0$ and  for  any  $n$  the
$(k_n, x_n)$-motion  be  determined  in  the  interval  $[-t_n, \infty)$, i.e. $[-t_n, \infty)
\subset \langle  -T (x_n, k_n), \infty)$.
  Then $(k^*, x^*)$-motion is determined   in
$[-t^*, \infty]$. In particular, if all $(k_n,x_n)$-motions are  determined  in $[-\bar t, \infty)\
(\bar t >0)$, then $(k^*, x^*)$-motion is determined in  too.  If $t_n \rightarrow \infty$ and $(k_n,
x_n)$-motion is determined in $[-t_n, \infty)$, then $(k^*, x^*)$-motion is determined in $(-\infty,
\infty)$ and is a whole  motion.  In particular,  if  all   the   $(k_n, x_n)$-motions are   whole,
then $(k^*,x^*)$-motion is whole too. All this is a direct  consequence  of the closure of the set
$S$, i.e.  of the domain  of  definition  of extended mapping $f$. It should be noted that from
$(x_n, k_n) \rightarrow (x^*, k^*)$
 and $[-t^*, \infty) \subset \langle -T(x^*,k^*), \infty)$
does not  follow  that  for any $\varepsilon > 0 \ [-t^* +\varepsilon,
\infty) \subset \langle -T (x_n, k_n), \infty)$
 for $n$ large enough.

Let us note an important property of uniform  convergence  in compact intervals. Let  $(x_n, k_n)
\rightarrow (x^*, k^*)$ and  all $(k_n, x_n)$-motions and  correspondingly  $(k^*, x^*)$-motion  be
determined  in   compact interval $[a,b]$.  Then  $(k_n,  x_n)$-motions  converge  uniformly in
$[a,b]$ to $(k^*,x^*)$-motion:  $f(t,  x_n,  k_n) \rightrightarrows f(t,  x^*, k^*)$.  This  is  a
direct consequence of continuity of the mapping $f:S \rightarrow X$

\subsection {Limit Sets}\label{SS1.2}

\begin{definition}\label{D1.1}Point $p \in X$ is called $\omega$- ($\alpha$-)-{\it limit
point} of the $(k,x)$-motion (correspondingly of the whole $(k,x)$-motion), if there is such sequence
$t_n \rightarrow \infty\  (t_n \rightarrow -\infty)$ that  $f(t_n, x,k) \rightarrow p$  as $n
\rightarrow \infty$.  The totality of all $\omega$- ($\alpha$-)-limit points of $(k,x)$-motion is
called  its $\omega$- ($\alpha$-)-{\it limit set} and is denoted by $\omega(x,k) \ (\alpha(x,k))$.
\end{definition}

\begin{definition}\label{D1.2} A set $W \subset X$ is called $k$-{\it invariant} set, if  for any  $ x \in W$
the $(k,x)$-motion is whole and the whole  $(k,x)$-trajectory belongs $W$.  In  similar way,  let us
call a set $V \subset X$ $(k,+)$-{\it invariant} (($k$,{\it positive})-{\it invariant}), if for any
$x \in V,\ t>0\ f(t,x,k)\ \in \ V$.
\end{definition}

\begin{proposition}\label{P1.3} The sets $\omega (x,k)$ and
$\alpha (x,k)$ are $k$-invariant.
\end{proposition}
{\bf Proof}. Let  $p \in \omega (x,k),\  t_n \rightarrow \infty,\ x_n = f(t_n, x,k) \rightarrow p$.
Note   that $(k,x_n)$-motion is determined at least in $[-t_n, \infty)$. Therefore, as  it was noted
above, $(k,p)$-motion is determined in $(-\infty, \infty)$, i.e. it  is whole. Let us show that the
whole  $(k,p)$-trajectory  consists  of $\omega$-limit points of $(k,x)$-motion. Let $f(\bar t, p,k)$
be an arbitrary point of $(k,p)$-trajectory. Since $t \rightarrow \infty$,  from  some $n$is
determined  a sequence   $f(\bar t + t_n, x,k)$).   It   converges   to $f(\bar t,p,k)$,    since
$f(\bar t + t_n, x,k) = f(\bar t,f (t_n, x,k),k)$
 (according to Proposition  \ref{P1.1}),
$f(t_n, x, k) \rightarrow p$
 and $f:S \rightarrow X$ is continuous (Proposition \ref{P1.2}).

     Now, let $q \in \alpha (x,k),\  t_n \rightarrow - \infty$
and $x_n = f(t_n, x,k) \rightarrow q$. Since (according to the definition of $\alpha$-limit points)
$(k,x)$-motion is  whole,  then all $(k,x_n)$-motions are whole too.  Therefore,  as  it  was  noted,
$(k,q)$-motion is whole. Let us show that every  point  $f(\bar t, q,k)$ of $(k,q)$-trajectory is
$\alpha$-limit for $(k,x)$-motion.  Since  $(k,x)$-motion is whole, then the semigroup property and
continuity  of $f$   in $S$ give $$ f(\bar t + t_n, x,k) = f(\bar t, f (t_n, x, k), k) \rightarrow
f(\bar t, q,k), $$ and since $\bar t + t_n \rightarrow - \infty$, then $f(\bar t, q,k)$ is
$\alpha$-limit point of $(k,x)$-motion. Proposition \ref{P1.3} is proved.

Further we need  also   the   complete   $\omega$-limit   set $\omega (k) : \omega (k) = \bigcup_{x
\in X} \omega (x,k)$. The set $\omega (k)$ is $k$-invariant, since it is  the union of $k$-invariant
sets.

\begin{proposition}\label{P1.4}
The sets $\omega (x,k),\ \alpha (x,k)$  (the  last  in  the
case  when
$(k,x)$-motion  is  whole)  are  nonempty,  closed   and
connected.
\end{proposition}
The proof practically literally coincides with the  proof  of similar statements for usual dynamical
systems (\cite{[6]}, p.356-362). The set $\omega (k)$ might be unclosed.

\begin{example}\label{E1.1} (The set $\omega (k)$ might be unclosed).   Let  us  consider  the system given by the
equations $\dot x = y(x-1),\ \dot y = -x (x-1)$ in  the  circle $x^2 + y^2 \leq 1$
 on the plane.

The complete $\omega$-limit set is $\omega =  \{  (1,0)  \}  \bigcup  \{ (x,y) \ | \ x^2 + y^2 <1
\}$. It is unclosed. The closure of $\omega$ coincides with the  whole  circle ($x^2 + y^2 \leq 1$),
the boundary of $\omega$ consists of two trajectories:  of  the  fixed  point
 $(1,0) \in \omega$ and of the loop
$\{ (x,y) \ | \ x^2 + y^2 = 1,\ x \neq 1 \} \nsubseteq \omega$
\end{example}

\begin{proposition}\label{P1.5} The sets $\partial \omega (k),\ \partial  \omega  (k)  \setminus  \omega (k)$
and $\partial \omega (k) \bigcap \omega (k)$ are $(k,+)$-invariant. Furthermore, if $\partial \omega
(k) \setminus \omega (k) \neq \varnothing$,  then $\partial \omega (k) \bigcap \omega (k) \neq
\varnothing \ (\partial \omega (k) = \overline{\omega (k)} \setminus {\rm int}  \omega (k)$ is the
boundary of the set $\omega (k))$.
\end{proposition}
Let us note that for the propositions \ref{P1.4} and \ref{P1.5} to be true, the compactness  of $X$
is important, because  for  non-compact  spaces analogous propositions are incorrect, generally
speaking.

To study slow relaxations, we need also  sets  that consist of $\omega$-limit  sets $\omega (x,k)$ as
of elements (the sets of $\omega$-limit  sets):
\begin{eqnarray}\label{e5}
& \Omega (x,k) = \{\omega(x',k) \ | \
\omega (x', k) \subset \omega (x,k), \ x' \in X \}; & \nonumber\\
& \Omega (k) = \{ \omega (x,k) \ | \ x \in X \}, &
\end{eqnarray}
$\Omega (x,k)$ is the set of all $\omega$-limit sets, lying in $\omega (x,k)$, $\Omega (k)$ is  the
set of $\omega$-limit sets of all $k$-motions.

\subsection{Convergences in the Spaces of Sets}\label{SS1.3}

Further  we consider  the  connection  between slow relaxations  and  violations  of  continuity  of
the  dependencies $\omega (x,k),\  \omega(k),\ \Omega (x,k),\ \Omega(k)$. Let  us  introduce
convergences  in spaces of sets and investigate the mappings continuous with respect to them. One
notion of continuity, used by us,  is  well  known (see \cite{[58]} Sec.~18 and \cite{[59]}  Sec.~43,
lower semicontinuity).  Two other ones  are  some  more ``exotic".  In  order  to  reveal  the
resemblance and distinctions between these  convergences,  let us consider  them  simultaneously (all
the statements,  concerning lower semicontinuity, are variations of known ones, see
\cite{[58],[59]}).

Let us denote the set of all nonempty subsets of $X$ by $B(X)$, and the set of all nonempty subsets
of $B(X)$ by $B(B(X))$.

     Let us introduce in $B(X)$ the following  proximity  measures:
let $p, q \in B (X)$, then
\begin{equation}\label{e6}
d(p,q) = \sup_{x \in p} \ \inf_{y \in q} \rho (x,y);
\end{equation}
\begin{equation}\label{e7}
r(p,q) = \inf_{x \in p, y \in q} \rho (x,y).
\end{equation}

The ``distance" $d(p,q)$ represents ``a half" of known Hausdorff metrics (\cite{[59]}, p.223):
\begin{equation}\label{e8}
\mbox{dist} (p,q) = \max \{ d (p,q),\ d(q,p) \}.
\end{equation}
It should be noted that, in general, $d(p,q) \neq d(q,p)$. Let us determine in $B(X)$ converges using
the introduced proximity measures. Let $q_n$ be a sequence of points of $B(X)$. We say that $q_n\
d$-converges to $p \in B(X)$, if $d(p,q_n) \rightarrow 0$. Analogously, $q_n$ $r$-converges to $p \in
B(X)$, if $r(p,q_n) \rightarrow 0$. Let us notice that $d$-convergence defines topology in $B(X)$
with a countable base in every point and the continuity with respect to this topology is equivalent
to $d$-continuity ($\lambda$-topology \cite{[58]}, p.183). As a basis of neighborhoods of the point
$p \in B(X)$ in this topology can be taken, for example, the family of sets $\{ q \in B(X) \ | \
d(p,q) < 1/n\ (n = 1,2, \ldots) \}$. The topology conditions can be easily verified, since the
triangle inequality
\begin{equation}\label{e9}
d(p,s) \leq d(p,q) + d(q,s)
\end{equation}
is true (in regard to these conditions see, for example, \cite{[60]}, p.19-20), $r$-convergence does
not determine topology in $B(X)$. To prove this, let us use the following obvious property of
convergence in topological spaces: if $p_i \equiv p,\ q_i \equiv q$ and $s_i \equiv s$ are constant
sequences of the points of topological space and $p_i \rightarrow q,\ q_i \rightarrow s$, then $p_i
\rightarrow s$. This property is not valid for $r$-convergence. To construct an example, it is enough
to take two points $x, y \in X$ ($x \neq y$) and to make $p = \{x \}, \ q = \{x,y \}, \ s = \{y \}$.
Then $r(p,q) = r(q,s) = 0, \ r(p,s) = \rho (x,y) > 0$. Therefore $p_i \rightarrow q, \ q_i
\rightarrow s, \ p_i \not \rightarrow s$, and $r$-convergence does not determine topology for any
metric space $X \neq \{x\}$.

Introduce also a proximity measure in $B(B(X))$ (that is the set of nonempty subsets of $B(X)$): let
$P, Q \in B(B(X))$, then
\begin{equation}\label{e10}
D(P,Q) = \sup_{p \in P}\ \inf_{q \in Q} r(p,q).
\end{equation}
Note that the formula (\ref{e10}) is similar to the formula (\ref{e6}), but in (\ref{e10}) appears
$r(p,q)$ instead of $\rho (x,y)$. The expression (\ref{e10}) can be somewhat simplified by
introducing the following denotations. Let $Q \in B(B(X))$. Let us define $SQ = \bigcup_{q \in Q} q,
SQ \in B(X)$; then
\begin{equation}\label{e11}
D(P,Q) = \sup_{p \in P} r (p, SQ).
\end{equation}
Let us introduce convergence in $B(B(X))$ ($D$-convergence): $$Q_n \rightarrow P, \:\mbox{if} \:
D(P,Q_n) \rightarrow 0.$$ $D$-convergence, as well as $r$-convergence, does not determine topology.
This can be illustrated in the way similar to that used for $r$-convergence. Let $x, y \in x, x \neq
y, P = \{ \{ x \} \}, Q = \{ \{ x,y \} \}, R = \{ \{y \} \}, P_i = P, Q_i = Q$. Then $D(Q, P) = D(R,
Q) = 0,\ P_i \rightarrow Q,\ Q_i \rightarrow R,\ D(R,P) = \rho (x,y) > 0, P_i \not \rightarrow R$.

Later we need the following criteria of convergence of sequences in $B(X)$ and in $B(B(X))$.

\begin{proposition}\label{P1.6} (see \cite{[58]}). The sequence of sets $q_n \in B(X)$ $d$-converges to $p \in B(X)$ if and
only if $\inf_{y \in q_n} \rho(x,y) \rightarrow 0$ as $n \rightarrow \infty$ for any $x \in p$.
\end{proposition}
\begin{proposition}\label{P1.7} The sequence of sets $q_n \in B(X)$ $r$-converges to $p \in B(X)$ if and only
if there are such $x_n \in p$ and $y_n \in q_n$ that $\rho (x_n, y_n) \rightarrow 0$ as $n
\rightarrow \infty$.
\end{proposition}
This  follows immediately from the definition of $r$-proximity.

Before treating the criterion of $D$-convergence,
let us prove the following
topological lemma.

\begin{lemma}\label{L1.1}Let $p_n, q_n \ (n = 1,2, \ldots)$
be subsets of compact metric space $X$ and $r(p_n, q_n) > \varepsilon > 0$ for any $n$. Then there
are such $\gamma > 0$ and an infinite set of indices $J$ that $r(p_N, q_n) > \gamma$ for $n \in J$
and for some number $N$.
\end{lemma}
{\bf Proof.} Choose in $X\ \varepsilon / 5$-network $M$; let to each $q \subset X$ correspond $q^M
\subset M$:
\begin{equation}\label{e12}
q^M = \left\{ m \in M \bigm| \inf_{x \in q} \rho (x,m) \leq \varepsilon /5
\right\}.
\end{equation}
For any two sets $p, q \subset X$ $r(p^M, q^M) + \frac{2}{5} \varepsilon \geq r(p,q)$. Therefore
$r(p^M_n, q^M_n) > 3 \varepsilon / 5$. Since the number of different pairs $p^M,q^M$ is finite ($M$
is finite), there exists an infinite set $J$ of indices $n$, for which the pairs $p^M_n, q^M_n$
coincide: $p^M_n = p^M,\ q^M_n = q^M$ as $n \in J$. For any two indices $n, l \in J\ r(p^M_n, q^M_l)
= r(p^M, q^M) > 3 \varepsilon / 5$, therefore $r(p_n, q_l) > \varepsilon / 5$, and this fact
completes the proof of the lemma. It was proved more important statement really: there exists such
infinite set $J$ of indices that for any $n, l \in J\ r(p_n, q_l) > \gamma$ (and not only for one
$N$).

\begin{proposition}\label{P1.8} The sequence of sets $Q_n \in B(B(X))$ $D$-converges to $p \in B(X))$ if and
only if $\inf_{q \in Q} r(p,q) \rightarrow 0$ for any $p \in P$.
\end{proposition}
{\bf Proof.} In one direction this is obvious: if $Q_n \rightarrow P$, then according to definition
$D(P, Q_n) \rightarrow 0$, i.e. the upper bound by $p \in P$ of the value $\inf_{q \in Q_n} r(p,q)$
tends to zero and all the more for any $p \in P\ \inf_{q \in Q} r(p,q) \rightarrow 0$. Now, suppose
that for any $p \in P\ \inf_{q \in Q_n} r(p,q) \rightarrow 0$. If $D(P, Q_n) \not \rightarrow 0$,
then one can consider that $D(P, Q_n) > \varepsilon > 0$. Therefore (because of (\ref{e11})) there
are such $p_n \in P$ for which $r(p_n, SQ_n) > \varepsilon\  \left( SQ_n = \bigcup_{q \in Q_n} q
\right)$. Using Lemma \ref{L1.1}, we conclude that for some $N\ r(p_N, SQ_n) > \gamma > 0$, i.e.
$\inf_{q \in Q_n}\ r(p_N, q) \not \rightarrow 0$. The obtained contradiction proves the second part
of Proposition \ref{P1.8}.

Everywhere further, if there are no another mentions, the convergence in $B(X)$ implies
$d$-convergence, and the convergence in $B(B(X))$ implies $D$-con\-ver\-gen\-ce, and as continuous
are considered the functions with respect to these convergences.

\subsection {Bifurcations of $\omega$-limit Sets}\label{SS1.4}

\begin{definition}\label{D1.3}. Let us say that the system (\ref{e1}) possesses:\\ A) $\omega (x,k)$-{\it
bifurcations}, if $\omega(x,k)$ is not continuous function in $X \times K$;\\ B) $\omega (k)$-{\it
bifurcations}, if $\omega (k)$ is not continuous function in $K$;\\ C) $\Omega (x,k)$-{\it
bifurcations}, if $\Omega(x,k)$ is not continuous function in $X \times K$;\\ D) $\Omega (k)$-{\it
bifurcations}, if $\Omega (k)$ is not continuous function in $K$.
\end{definition}

The points of $X \times K$ or $K$, in which the functions $\omega (x,k)$, $\omega (k)$, $\Omega
(x,k)$, $\Omega (k)$ are not $d$- or not $D$-continuous, we call {\it the points of bifurcation}. The
considered discontinuities in the dependencies $\omega (x,k)$,\ $\omega (k)$,\ $\Omega (x,k)$,\
$\Omega (k)$ could be also called {\it ``explosions"} of $\omega$-limit sets (compare with the
explosion of the set of non-wandering points in differential dynamics (\cite{[26]}, Sec.~6.3.,
p.185-192, which, however, is a violation of semidiscontinuity from above).

\begin{proposition}\label{P1.9} A) If the system (\ref{e1}) possesses $\Omega (k)$-bifurcations, then it
possesses $\Omega (x,k)$-, $\omega (x,k)$- and $\omega (x,k)$-bifurcations.\\ B) If the system
(\ref{e1}) possesses $\Omega (x,k)$-bifurcations, then it possesses $\omega (x,k)$-bifurcations too.
\\ C) If the system (\ref{e1}) possesses $\omega (k)$-bifurcations, then it possesses $\omega
(x,k)$-bi\-fur\-ca\-tions.
\end{proposition}
It is convenient to illustrate Proposition \ref{P1.9} by the scheme (the word ``bifurcation" is
omitted on the scheme):
\begin{equation}\label{e13}
\begin{array}{ccc}
\begin{picture}(10,5)
\put(2,3){\line(10,0){10}} \put(2,3){\vector(0,-3){6}}
\end{picture}
& \Omega (k)\!\! &
\begin{picture}(10,5)
\put(7,3){\line(-8,0){10}} \put(7,3){\vector(0,-3){6}}
\end{picture}              \\
\Omega (x,k) & &  \omega (k)\\
\begin{picture}(10,5)
\put(2,0){\line(0,3){3}}
\put(2,0){\vector(3,0){10}}
\end{picture}
& \omega (x,k) &
\begin{picture}(10,5)
\put(7,0){\line(0,3){3}}
\put(7,0){\vector(-1,0){10}}
\end{picture}
\end{array}
\end{equation}

{\bf Proof.} Let us begin from the item $C$. Let the system (\ref{e1}) (family of semiflows) possess
$\omega (k)$-bifurcations. This means that there are such $k^* \in K$ (point of bifurcation),
$\varepsilon >0,\ x^* \in \omega (k^*)$ and sequence $k_n \in K,\ k_n \rightarrow k^*$, for which
$\inf_{y \in \omega (x_0, k_n)} \rho (x^*, y) > \varepsilon$ for any $n$ (according to Proposition
\ref{P1.6}). The point $x^*$ belongs to some $\omega(x_0, k^*) \ (x_0 \in X)$. Note that $\omega
(x_0, k_n) \subset \omega (k_n)$, consequently, $\inf_{y \in \omega (k_n)} \rho(x^*, y) >
\varepsilon$, therefore the sequence $\omega (x_0, k_n)$ does not converge to $\omega (x_0, k^*)$:
there exist $\omega (x,k)$-bifurcations, and the point of bifurcation is $(x_0, k^*)$.

Prove the statement of the item B. Let the system (\ref{e1}) possess $\Omega (x,k)$-bifurcations.
Then, (according to Proposition \ref{P1.8}) there are such $(x^*, k^*) \in X \times K$ (the point of
bifurcation), $\omega (x_0, k^*) \subset \omega (x^*, k^*)$ and sequence $(x_n, k_n) \rightarrow
(x^*, k^*)$ that $$ r(\omega (x_0, k^*), S\,\Omega (x_n, k_n)) > \varepsilon > 0 \ \mbox{for any} \
n. $$
 But the last means that
$r (\omega (s_0, k^*), \omega(x_n, k_n)) > \varepsilon > 0$ and, consequently, $$ \inf_{y \in \omega
(x_n, k_n)} \rho(\xi,y)> \varepsilon \ \mbox{for any} \ \xi \in \omega (x^0, k^*). $$ Since $\xi \in
\omega (x^*, k^*)$, the existence of $\omega (x, k)$-bifurcations follows and $(x^*,k^*)$ is the
point of bifurcation.

Prove the statement of the item A. Let the system (\ref{e1}) possess $\Omega (k)$-bifurcations. Then
there are $k^* \in K$ (the point of bifurcation), $\varepsilon > 0$ and sequence of points $k_n, k_n
\rightarrow k^*$, for which $D(\Omega (k^*), \Omega (k_n)) > \varepsilon$ for any $n$, that is for
any $n$ there is such  $x_n \in X$ that $r(\omega (x_n, k^*), \omega (k_n)) > \varepsilon$ (according
to (\ref{e11})). But by Lemma \ref{L1.1} there are such $\gamma >0$ and natural $N$ that for infinite
set $J$ of indices $r(\omega (x_N, k^*), \omega (k_n)) > \gamma$ for $n \in J$. All the more
$r(\omega(x_N, k^*), \omega (x_N, k_n)) > \gamma\ (n \in J)$, consequently, there are $\Omega (x,
k)$-bifurcations: $$
\begin{array}{c}
(x_N, k_n) \rightarrow (x_N, k^*)$ as $n \rightarrow \infty, \ n \in J;\\
\vspace{1mm} D (\Omega (x_N, k^*), \ \Omega (x_N, k_n)) =
 \sup_{\omega (x, k^*) \subset \Omega (x_N, k^*)}
r(\omega (x, k^*), \omega (x_N, k_n)) \geq\\
\vspace{1mm} \geq r( \omega (x_N, k^*),\ \omega (x_N, k_n)) > \gamma.\\
\end{array}
$$ The point of bifurcation is $(x_N, k^*)$.

We are only to show that if there are $\Omega (k)$-bifurcations, then $\omega (k)$-bifurcations
exist. Let us prove this. Let the system (\ref{e1}) possess $\Omega (k)$-bifurcations. Then, as it
was shown just above, there are such $k^* \in K, x^* \in X,\ \gamma > 0\ (x^* = x_N)$ and a sequence
of points $k_n \in K$ that $k_n \rightarrow k^*$ and $r(\omega (x^*, k^*),\ \omega (k_n)) > \gamma$.
All the more, for any $\xi \in \omega (x^*, k^*)$ $\inf_{y \in \omega (k_n)} \rho (\xi, y) > \gamma$,
therefore $d(\omega (k^*), \omega(k_n)) > \gamma$ and there are $\omega (k)$-bifurcations ($k^*$ is
the point of bifurcation). Proposition \ref{P1.9} is proved.

\begin{proposition}\label{P1.10} The system (\ref{e1}) possesses $\Omega (x,k)$-bifurcations if and only if
$\omega (x,k)$ is not $r$-continuous function in $X \times K$.
\end{proposition}
{\bf Proof.} Let the system (\ref{e1}) possess $\Omega (x,k)$-bifurcations, then there are $(x^*,
k^*) \in X \times K$, the sequence $(x_n, k_n) \in X \times K, (x_n, k_n) \rightarrow (x^*, k^*)$ for
which for any $n$ $$ D(\Omega(x^*, k^*), \Omega(x_n, k_n)) > \varepsilon >0. $$ The last means that
for any $n$ there is $x^*_n \in X$ for which $\omega(x^*_n, k^*) \subset \omega (x^*, k^*)$, and
$r(\omega (x^*_n, k^*), \omega (x_n, k_n)) > \varepsilon$. From Lemma \ref{L1.1} follows the
existence of such $\gamma >0$ and natural $N$ that for infinite set $J$ of indices $r(\omega ( x^*_N,
k^*), \omega(x_n, k_n)) > \gamma$ as $n \in J$. Let $x^*_0$ be an arbitrary point of $\omega(x^*_N,
k^*)$. As it was noted already, $(k^*, x^*_0)$-trajectory lies in $\omega(x^*_N<k^*)$ and because of
the closure of the last $\omega(x^*_0, k^*) \subset \omega (x^*_N, k^*)$. Therefore $r(\omega(x_n,
k_n))
> \gamma$ as $n \in J$. As $x^*_0 \in \omega (x^*, k^*)$, there is such sequence $t_i \rightarrow
\infty, t_i >0$,  that $f(t_i, x^*, k^*) \rightarrow x^*_0$ as $i \rightarrow \infty$. Using the
continuity of $f$, choose for every $i$ such $n(i) \in J$ that $\rho (f(t_i, x^*, k^*), f(t_i,
x_{n(i)}, k_{n(i)})) < 1/i$. Denote $f(t_i, x_{n(i)}, k_{n(i)}) = x'_i,\ k_{n(i)} = k'_i$. Note that
$\omega (x'_i, k'_i) = \omega (x_{n(i)}, k_{n(i)})$. Therefore for any $i\ r(\omega (x^*_0, k^*),
\omega (x'_i, k'_i)) > \gamma$. Since $(x'_i, k'_i) \rightarrow (x^*_0, k^*)$, we conclude that
$\omega(x,k)$ is not $r$-continuous function in $X \times K$.

Let us emphasize that the point of $\Omega (x,k)$-bifurcations
can be not the point of $r$-discontinuity.

Now, suppose that $\omega (x,k)$ is not $r$-continuous in $X \times K$. Then there exist $(x^*, k^*)
\in X \times K$, sequence of points $(x_n, k_n) \in X \times K$, $(x_n, k_n) \rightarrow (x^*, k^*)$,
and $\varepsilon > 0$, for which $r(\omega(x^*, k^*), \omega(x_n, k_n)) > \varepsilon$ for any $n$.
But, according to (\ref{e11}), from this it follows that $D(\Omega (x^*, k^*), \Omega (x_n, k_n)) >
\varepsilon$ for any $n$. Therefore $(x^*, k^*)$ is the point of $\Omega(x,k)$-bifurcation.
Proposition \ref{P1.10} is proved.

The $\omega (k)$- and $\omega(x,k)$-bifurcations can be called bifurcations with appearance of new
$\omega$-limit points, and $\Omega (k)$- and $\Omega (x,k)$-bifurcations with appearance of
$\omega$-limit sets. In the first case there is such sequence of points $k_n$ (or ($x_n, k_n$)),
converging to the point of bifurcation $k^*$ (or $(x^*, k^*$)) that there is such point $x_0 \in
\omega (k^*)$ (or $x_0 \in \omega (x^*, k^*$)) which is removed away from all $\omega(k_n)\
(\omega(x_n, k_n))$ more than at some $\varepsilon > 0$. It could be called the ``new" $\omega$-limit
point. In the second case, as it was shown, the existence of bifurcations is equivalent to existence
of a sequence of the points $k_n$ (or ($x_n, k_n) \in X \times K$), converging to the point of
bifurcation $k^*$ (or ($x^*, k^*$)), together with existence of some set $\omega (x_0, k^*) \subset
\omega (k^*)\ (\omega( x_0, k^*) \subset \omega(x^*, k^*))$, being $r$-removed from all $\omega
(k_n)\  (\omega (x_n, k_n))$ more than at $\gamma >0$:  $\rho(x,y) > \gamma$ for any $x \in \omega
(x_0, k^*)$ and $y \in \omega (k_n)$. It is natural to call the set $\omega (x_0, k^*)$ the ``new"
$\omega$-limit set. A question arises: are there bifurcations with appearance of new $\omega$-limit
points, but without appearance of new $\omega$-limit sets? The following example gives positive
answer to this question.

\begin{figure}[t]
\centering{
\includegraphics[width=110mm,height=56mm]{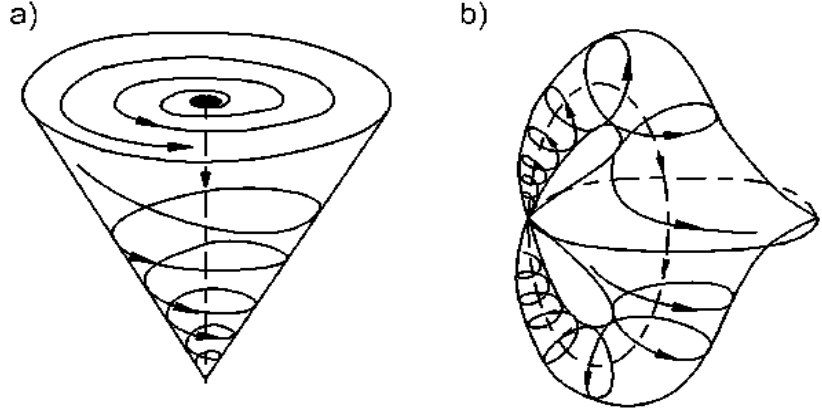}
\caption{\label{Fig.1} $\omega (x,k)$-, but not $\Omega (x,k)$-bifurcations:  {\it a} - phase
portrait of the system (\ref{e14}); {\it b} - the same portrait after gluing all fixed points.}}
\end{figure}

\begin{example}\label{E1.2} $(\omega (x,k)$-, but not $\Omega (x,k)$-bifurcations).
Consider at first the system, given
in the cone $x^2 + y^2 \leq z^2,\ 0 \leq z \leq 1$
by differential equations (in cylindrical coordinates)
\begin{eqnarray}\label{e14}
& & \dot{r} = r(2z - r - 1)^2 - 2r (1-r)(1-z);\nonumber\\
& &\dot{\varphi} = r \cos \varphi + 1;\\
& &\dot{z} = -z(1-z)^2. \nonumber
\end{eqnarray}
The solutions of (\ref{e14}) under initial conditions $0 \leq z(0) \leq 1,\ 0 \leq r(0) \leq z(0)$
and arbitrary $\varphi$ tend as $t \rightarrow \infty$ to their unique $\omega$-limit point (this
point is the equilibrium $z = r = 0$). If $0 < r(0) < 1$, then as $t \rightarrow \infty$ the solution
tends to the circumference $z = r = 1$. If $z (0) = 1,\ r(0) = 0$, then the $\omega$-limit point is
unique: $z = 1,\ r =0$. If $z(0) = r(0) =1$, then the $\omega$-limit point is also unique: $z = r =
1,\ \varphi = \pi$ (Fig. \ref{Fig.1}). Thus, $$ \omega (r_0, \varphi_0, z_0) = \left\{
\begin{array}{ll}
(z = r = 0), &                 \mbox{if}\ z_0 < 1;\\
\{(r, \varphi, z) \ |\ r = z = 1 \}, & \mbox{if}\ z_0 = 1,\ r_0 \neq 0,1;\\
(z = r = 1),\ \varphi = \pi,  & \mbox{if}\ z_0 = r_0 = 1;\\
(r = 0, z = 1), &              \mbox{if}\ z_0 = 1, \ r_0 =0.
\end{array}
\right. $$ Consider the sequence of points of the cone $(r_n, \varphi_n, z_n) \rightarrow (r^*,
\varphi^*, 1),\ r^* \neq 0,1$ and $z_n < 1$ for all $n$. For any point of the sequence the
$\omega$-limit set includes one point, and for $(r^*, \varphi, 1)$ the set includes the
circumference. If all the positions of equilibrium were identified, then there would be $\omega
(x,k)$-, but not $\omega (x,k)$-bifurcations.
\end{example}

The correctness of the identification procedure should be grounded. Let the studied semiflow $f$ have
fixed points $x_i, \ldots, x_n$. Define a new semiflow $\tilde{f}$ as follows: $$ \tilde{X} = X
\setminus \{ x_i, \ldots, x_n \} \bigcup \{ x^* \} $$ is a space obtained from $X$ when the points
$x_i, \ldots, x_n$ are deleted and a new point $x^*$ is added. Let us give metrics over $\tilde{X}$
as follows: let $x, y \in \tilde{X}, \ x \neq x^*$, $$ \tilde{\rho} (x,y) = \left\{
\begin{array}{ll}
\min \left\{ \rho (x,y), \min_{1 \leq j \leq n} \rho(x, x_j) +
\min_{1 \leq j \leq n} \rho (y, x_j) \right\}, & \mbox{if}\ y \neq x^*;\\
\min_{1 \leq j \leq n} \rho (x, x_j),          & \mbox{if}\ y = x^*.
\end{array}
\right. $$ Let $\tilde{f} (t,x) = f(t,x)$ if $x \in X \bigcap \tilde{X},\ \tilde{f} (t, x^*) = x^*$.

\begin{lemma}\label{L1.2}
The mapping $\tilde{f}$ determines semiflow in $\tilde{X}$.
\end{lemma}
{\bf Proof.} Injectivity and semigroup property are obvious from the corresponding properties of $f$.
If $x \in X \bigcap \tilde{X},\ t \geq 0$ then the continuity of $\tilde{f}$ in the point $(t,x)$
follows from the fact that $\tilde{f}$ coincides with $f$ in some neighbourhood of this point. The
continuity of $\tilde{f}$ in the point $(t,x^*)$ follows from the continuity of $f$ and the fact that
any sequence converging in $\tilde{X}$ to $x^*$ can be divided into finite number of sequences, each
of them being either (a) a sequence of points $X \bigcap \tilde{X}$, converging to one of $x_j$ or
(b) a constant sequence, all elements of which are $x^*$ and some more, maybe, a finite set. Mapping
$\tilde{f}$ is a homeomorphism, since it is continuous and injective, and $\tilde{X}$ is compact.

\begin{proposition}\label{P1.11} Let each trajectory lying in $\omega (k)$ be recurrent for any $k$. Then the
existence of $\omega (x,k)$- ($\omega (k)$-)-bifurcations is equivalent to the existence of $\Omega
(x,k)$- ($\Omega (k)$-)-bifurcations. More exact,\\ A) if $(x_n, k_n) \rightarrow (x^*, k^*)$  and
$\omega (x_n, k_n) \not \rightarrow \omega(x^*,k^*)$,   then $\Omega(x_n, k_n) \not \rightarrow
\Omega (x^*, k^*)$\footnote{Let us recall that below the convergence in $B(X)$ implies
$d$-convergence, and the convergence in $B(B(X))$ implies $D$-convergence, and continuity is
considered as continuity with respect to these convergences, if there are no other mentions.},\\ B)
if $k_n \rightarrow k^*$ and $\omega(k_n) \not \rightarrow \omega(k^*)$, then $\Omega (k_n) \not
\rightarrow \Omega(k^*)$.
\end{proposition}
{\bf Proof.} A) Let $(x_n, k_n) \rightarrow (x^*, k^*)$, $\omega(x_n, k_n) \not \rightarrow \omega
(x^*, k^*)$. Then, according to Proposition \ref{P1.6}, there is such $\tilde x \in (x^*, k^*)$ that
$\inf_{y \in \omega (x_n, k_n)} \rho (\tilde x, y) \not \rightarrow 0$. Therefore from $\{(x_n, k_n)
\}$  we  can  choose  a subsequence (denote it as $\{(x_m, k_m) \}$) for which there exists such
$\varepsilon >0$ that $\inf_{y \in \omega(x_m, k_m)} \rho (\tilde x, y)
> \varepsilon$   for any $m = 1, 2< \ldots$. Denote by $L$ the set
of all limit points of sequences of the  kind $\{y_m\},\ y_m \in \omega (x_m, k_m)$. The set $L$   is
closed  and   $k^*$-invariant.  Note  that $\rho^* (\tilde x, L) \geq \varepsilon$. Therefore
$\omega(\tilde x, k^*) \bigcap L = \varnothing$ as $\omega(\tilde x, k^*)$ is  a  minimal  set
(Birkhoff's theorem, see \cite{[6]}, p.404). From this follows the existence of such $\delta>0$ that
$r(\omega(\tilde x, k^*), L) > \delta$ and from  some $M\ r(\omega(\tilde x, k^*),\ (x_m, k_m)) >
\delta/2$ (when $m>M$). Therefore (Proposition \ref{P1.8}) $\Omega(x_m, k_m) \not \rightarrow
\Omega(x^*, k^*)$.

     B) The proof practically literally coincides with that for the part $A$ (it should be  substituted
$\omega(k)$ for $\omega(x,k)$).

\begin{corollary}\label{C1.1}Let for every pair $(x,k) \in X \times K$ the $\omega$-limit set be
minimal: $\Omega(x,k) = \{ \omega(x,k) \}$. Then  the  statements A, B of Proposition \ref{P1.11} are
true.
\end{corollary}
{\bf Proof}. According to one of  Birkhoff's  theorems  (see  \cite{[6]}, p.402),  each  trajectory
lying in  minimal  set  is  recurrent. Therefore, Proposition \ref{P1.11} is applicable.

\section{Slow Relaxations} \label{S2}

\subsection {Relaxation Times}\label{SS2.1}

The principal object of our consideration is {\it the relaxation time}.

\begin{proposition}\label{P2.1} For any $x \in X,\ k \in K$ and $\varepsilon > 0$
 the numbers
$\tau_i(x,k,\varepsilon)$ and $\eta_i(x,k,\varepsilon)$ $(i = 1, 2, 3)$ are  defined.   The
inequalities $\tau_i \geq \eta_i$, $\tau_1 \leq \tau_2 \leq \tau_3$, $\eta_1 \leq \eta_2 \leq \eta_3$
are true.
\end{proposition}
{\bf Proof.} If $\tau_i, \eta_i$   are  defined,  then  the   validity   of inequalities is obvious
$(\omega(x,k) \subset \omega (k)$, the time of the first  entry in the $\varepsilon$-neighbourhood of
the set of limit points is included into the time of being outside of this neighbourhood, and the
last is not larger than the time of final entry  in  it).  The  numbers  $\tau_i,\ \eta_i$ are
definite (bounded): there  are  $t_n  \in  [0,   \infty),\   t_n \rightarrow \infty$ and $y \in
\omega (x,k)$, for which $f(t_n, x, k) \rightarrow y$ and from some $n$ $\rho (f(t_n, x, k), y) <
\varepsilon$, therefore the sets $\{t>0 \ | \ \rho^* (f(t,x,k), \omega (x,k)) < \varepsilon \}$ and
$\{t>0 \ | \ \rho^* (f(t,x,k), \omega (k)) < \varepsilon \}$ are nonempty. Since $X$ is compact,
there is such $t(\varepsilon) >0$ that  for $t >t (\varepsilon)\  \rho^* (f(t,x,k), \omega (x,k)) <
\varepsilon$. Really, let us suppose the contrary: there are such $t_n >0$ that $t_n \rightarrow
\infty$ and $ \rho^* (f(t_n, x, k), \omega(x,k)) > \varepsilon$.  Let  us choose from the sequence
$f(t_n, x,k)$ a convergent  subsequence  and denote its limit $x^*;\ x^*$ satisfies the definition of
$\omega$-limit point of $(k,x)$-motion, but it  lies  outside  of  $\omega(x,k)$. The  obtained
contradiction proves the required, consequently, $\tau_3$ and $\eta_3$   are defined. According to
the proved, the sets
\begin{eqnarray*}
& &\{ t>0 \ | \ \rho^* (f (t,x,k), \omega (x,k)) \geq \varepsilon \},\\
& &\{ t>0 \ | \ \rho^* (f(t,x,k), \omega (k)) \geq \varepsilon \}
\end{eqnarray*}
are bounded. They are measurable because of  the  continuity  with respect to $t$ of the functions
$\rho^* (f(t,x,k), \omega (x,k))$ and $\rho^* (f(t,x,k), \omega(k))$. The  proposition  is  proved.
Note  that  the   existence (finiteness) of $\tau_{2,3}$    and $\eta_{2,3}$ is associated with  the
compactness of $X$.

\begin{definition}\label{D2.1}  We  say  that  the  system  (\ref{e1})   possesses $\tau_i$- ($\eta_i$-)-{\it
slow relaxations}, if for some $\varepsilon >0$ the  function $\tau_i(x,k,\varepsilon)$
(correspondingly $\eta_i (x,k, \varepsilon)$) is not bounded above in $X \times K$.
\end{definition}

\begin{proposition}\label{P2.2} For any semiflow ($k$ is fixed) the function $\eta_1 (x, \varepsilon)$ is
bounded in $X$ for every $\varepsilon >0$.
\end{proposition}
{\bf Proof.} Suppose the contrary. Then there is such  sequence  of points $x_n  \in  X$  that  for
some  $\varepsilon  > 0\ \eta_1 (x_n, \varepsilon) \rightarrow \infty$. Using the compactness of $X$
 and, if  it  is  needed,  choosing  a  subsequence,  assume that $x_n
\rightarrow x^*$. Let us  show  that  for  any  $t>0\  \rho^*(f(t,x^*),  \omega  (k))  > \varepsilon
/ 2$. Because of  the property of uniform continuity on limited segments there  is  such $\delta =
\delta (\tau) >0$ that $\rho(f(t,x^*),f(t,x)) <  \varepsilon / 2$ if $0 \leq t \leq \tau$ and
$\rho(x,x^*) < \delta$.  Since $\eta_1 (x_n, \varepsilon) \rightarrow \infty$  and $x_n \rightarrow
x^*$,  there  is  such $N$   that $\rho(x_N, x^*) < \delta$ and $\eta_1(x_N, \varepsilon) > \tau$,
 i.e. $\rho^*(f(t,x_N), \omega (k)) \geq \varepsilon$  under
$0 \leq t \leq \tau$.  From  this  we obtain   the   required:   for $0 \leq t \leq \tau$
$\rho^*(f(t,x^*), \omega (k)) > \varepsilon / 2$ or $\rho^*(f(t,x^*), \omega(k)) > \varepsilon / 2$
for any $t>0$, since $\tau$ was chosen  arbitrarily. This contradicts to the finiteness of $\eta_1
(x^*, \varepsilon / 2)$ (Proposition \ref{P2.1}). Proposition \ref{P2.2} is proved.  For $\eta_{2,3}$
and $\tau_{1,2,3}$   does not exist proposition  analogous  to Proposition \ref{P2.2}, and slow
relaxations are  possible  for  one semiflow too.

\begin{figure}[t]
\centering {
\includegraphics[width=120mm,height=131mm]{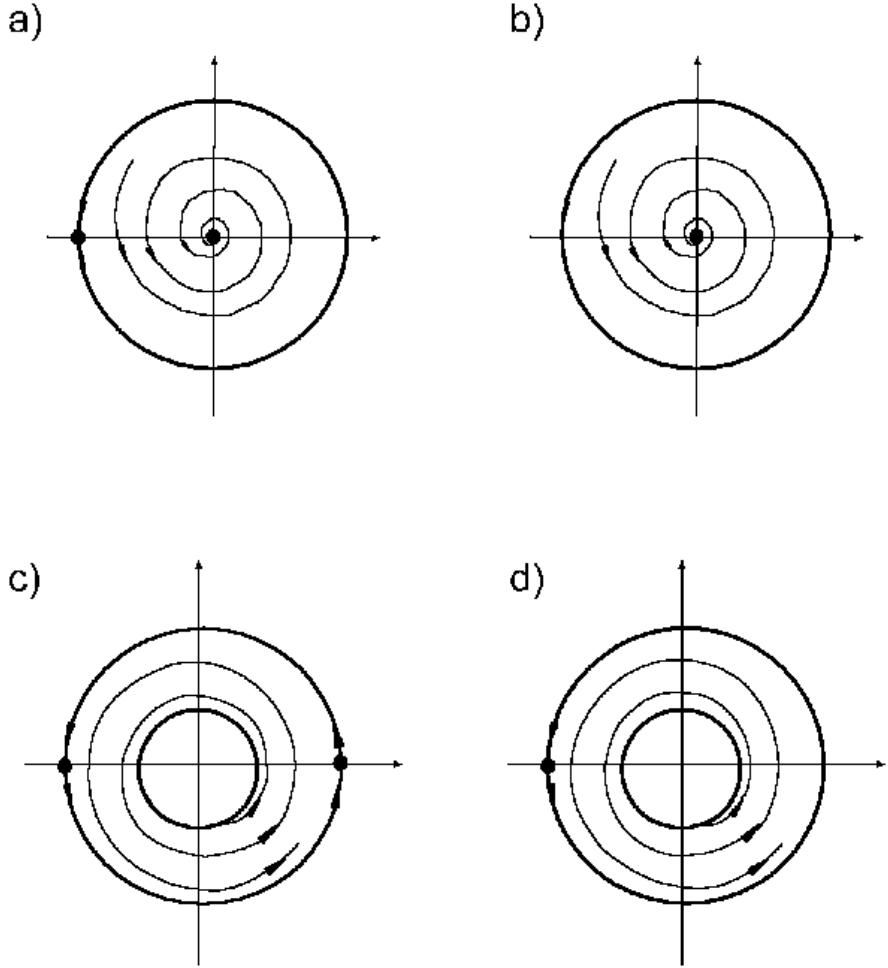}
\caption{\label{Fig.2} Phase portraits of the systems:  {\it a} - (\ref{e15}); {\it b} - (\ref{e16});
{\it c} - (\ref{e17}); {\it d} - (\ref{e18})}}
\end{figure}

\begin{example}\label{E2.1} ( $\eta_2$-slow relaxations for one  semiflow).  Let  us
consider a system on the plane in the circle $x^2 + y^2 \leq 1$ given by the equations  in  the polar
coordinates
\begin{eqnarray}\label{e15}
& &\dot r = -r(1-r)(r\ \cos \varphi + 1);\nonumber\\
& &\dot \varphi = r\ \cos \varphi +1.
\end{eqnarray}
The complete $\omega$-limit set consists of two fixed points $r=0$ and $r=1,\ \varphi = \pi$ (Fig.
\ref{Fig.2},a), $\eta_2((r, \varphi), 1/2) \rightarrow \infty$ as $r \rightarrow 1,\  r < 1$.
\end{example}

The  following  series  of  simple  examples  is   given   to
demonstrate the  existence  of  slow  relaxations  of  some  kinds
without some other kinds.

\begin{example}\label{E2.2} ($\eta_3$- but not  $\eta_2$-slow
relaxations). Let us  rather
modify the previous example, substituting unstable limit cycle for
the boundary loop:
\begin{eqnarray}\label{e16}
& & \dot r = -r (1-r); \nonumber\\
& & \dot \varphi = 1.
\end{eqnarray}
Now the  complete   $\omega$-limit   set   includes   the   whole    boundary circumference and the
point $r=0$ (Fig. \ref{Fig.2},b), the time of the system being outside of  its
$\varepsilon$-neighborhood is limited for  any $\varepsilon >0$. Nevertheless, $\eta_3 ((r, \varphi),
1/2) \rightarrow \infty$ as $r \rightarrow 1,\  r \neq 1$
\end{example}

\begin{example}\label{E2.3} ($\tau_1$,   but  not     $\eta_{2,3}$-slow
relaxations).  Let  us
analyze in the ring \linebreak
\mbox{$\frac{1}{2} \leq x^2 + y^2 \leq 1$}
a system  given  by  differential
equations in polar coordinates
\begin{eqnarray}\label{e17}
& &\dot r = (1-r)(r \cos \varphi +1) (1 - r\cos \varphi);\nonumber\\
& &\dot \varphi = ( r \cos \varphi +1) (1 - r \cos \varphi).
\end{eqnarray}
In  this  case the  complete   $\omega$-limit  set  is  the  whole  boundary circumference  $r = 1$
(Fig. \ref{Fig.2},c). Under $r =1,\ \varphi \to \pi,\ \varphi > \pi\ \tau_1 (r,  \varphi, 1/2) \to
\infty$ since for these points $\omega(r,\varphi) = \{(r=1,\ \varphi = 0)\}$.
\end{example}

\begin{example}\label{E2.4} ($\tau_3$, but not $\tau_{1,2}$   and not
$\eta_3$-slow  relaxations).
Let us modify the preceding example of the  system  in  the  ring,
leaving  only   one   equilibrium   point   on   the   boundary
circumference $r=1$:
\begin{eqnarray}\label{e18}
& &\dot r = (1-r) (r \cos \varphi + 1);\nonumber\\
& &\dot \varphi = r \cos \varphi +1.
\end{eqnarray}
In this case under $r=1,\ \varphi \to \pi,\ \varphi \to \pi\ \tau_3 ((r, \varphi), 1/2) \to \infty$
and  $\tau_{1,2}$    remain limited  for   any   fixed   $\varepsilon >0$, because   for   these
points $\omega(r, \varphi) = \{(r=1,\ \varphi = \pi)\}$ (Fig. \ref{Fig.2},d).  $\eta_{2,3}$   are
limited, since the   complete $\omega$-limit set is the circumference $r=1$.
\end{example}

\begin{figure}[t]
\centering {
\includegraphics[width=128mm,height=66mm]{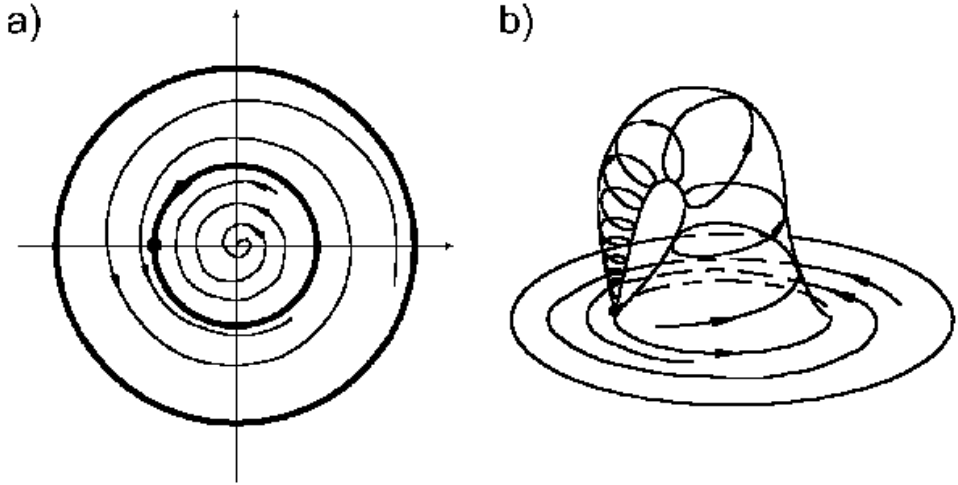}
\caption{\label{Fig.3} Phase portrait of the system (\ref{e19}):\protect\\ {\it a} - without gluing
the fixed points; {\it b} - after gluing.}}
\end{figure}

\begin{example}\label{E2.5} ($\tau_2$, but not $\tau_1$ and not $\eta_2$-slow relaxations). We could not
find a simple example on the  plane  without  using  Lemma \ref{L1.2}. Consider at first a semiflow
in the  circle $x^2 + y^2 \leq 2$ given by the equations
\begin{eqnarray}\label{e19}
& &\dot r = -r(1-r)^2 [(r \cos \varphi +1)^2 + r^2 \sin \varphi]; \nonumber\\
& &\dot \varphi = (r \cos \varphi + 1)^2 + r^2 \sin^2 \varphi.
\end{eqnarray}
the $\omega$-limit sets of this system are as follows (Fig. \ref{Fig.3},a): $$ \omega(r_0, \varphi_0)
= \left\{
\begin{array}{ll} \mbox{circumference}\ r=1, & \mbox{if}\ r_0 >1;\\ \mbox{point}\ (r=1,\ \varphi =
\pi), & \mbox{if} \ r_0 =1;\\ \mbox{point}\ (r=0), & \mbox{if}\ r_0 <1.
\end{array}
\right. $$ Let us identify the fixed points $(r=1, \varphi = \pi)$  and $(r=0)$  (Fig.
\ref{Fig.3},b). We obtain that under $r \rightarrow 1,\ r < 1\ \tau_2(r,\varphi,1/2) \rightarrow
\infty$, although $\tau_1$   remains bounded as well as $\eta_2$. However, $\eta_3$  is unbounded.
\end{example}

     The  majority  of  the  above  examples  is  represented   by
nonrough systems, and there are serious reasons for this nonroughness. In rough systems on a plane
$\tau_{1,2,3}$- and $\eta_3$-slow relaxations can occur only simultaneously (see Subsection
\ref{SS3.3}).

\subsection {Slow Relaxations and Bifurcations of $\omega$-limit Sets}\label{SS2.2}

In  the  simplest situations  the  connection  between  slow relaxations and bifurcations of
$\omega$-limit sets is obvious. We should mention the case when the motion tending to  its
$\omega$-limit set  is retarded near unstable equilibrium position. In general  case  the situation
becomes more complicated  at least  because  there  are several relaxation times (and consequently
several corresponding kinds of slow relaxations). Except  that,  as  it  will  be  shown below,
bifurcations are not  a  single possible  reason  of  slow relaxation appearance. Nevertheless, for
the time of the first entering (both for the proper time $\tau_1$  and for the non-proper one
$\eta_1$) the connection between bifurcations and slow relaxations is manifest.

\begin{theorem}\label{T2.1}
The system (\ref{e1}) possesses  $\tau_1$-slow relaxations  if and only if it possesses
$\Omega(x,k)$-bifurcations.
\end{theorem}
{\bf Proof.} Let the system possess $\Omega(x,k)$-bifurcations, $(x^*, k^*)$ be the point of
bifurcation. This means that there are such $x' \in X, \varepsilon >0$ and sequence of points $(x_n,
k_n) \in X \times K$, for  which $\omega (x', k^*) \subset \omega (x^*, k^*),\ (x_n, k_n) \rightarrow
(x^*, k^*)$, and $r(\omega (x', k^*), \omega(x_n, k_n)) > \varepsilon$ for   any  $n$.    Let $x_0
\in \omega (x',k^*)$. Then $\omega(x_0, k^*) \subset \omega (x', k^*)$ and $r(\omega(x_0, k^*),
\omega(x_n, k_n)) > \varepsilon$ for any $n$. Since $x_0 \in \omega(x^*,k^*)$, there is such sequence
$t_i >0,\ t \rightarrow \infty$,  for which $f(t_i, x^*,\ k^*) \rightarrow x_0$. As for every $i\
f(t_i, x_n, k_n) \rightarrow f(t_i, x^*, k^*)$, then there is such sequence $n(i)$   that $f(t_i,
x_{n(i)}, k_{n(i)}) \rightarrow x_0$ as $i \rightarrow \infty$. Denote $k_{n(i)}$     as $k'_i$ and
$f(t_i, x_{n(i)}, k_{n(i)})$ as $y_i$. It is obvious that $\omega(y,k'_i) = \omega(x_{n(i)},
k_{n(i)})$. Therefore $r(\omega(x',k^*), \omega(y_i, k'_i)) > \varepsilon$.

Let us show that for any $\tau>0$ there is such $i$ that $\tau_1(y_i, k'_i, \varepsilon / 2) > \tau$.
To  do that, let us use the  property  of  uniform  continuity  of  $f$  on compact  segments  and
choose   such  $\delta > 0$   that $\rho(f(t,x_0,k^*),\ f(t,y_i, k'_i)) < \varepsilon / 2$ if $0 \leq
t \leq \tau,\ \rho (x_0, y_i) + \rho_K (k^*,k'_i) < \delta$. The   last inequality is true from some
$i_0$ (when $i>i_0$), since  $y_i \rightarrow x_0,\ k'_i \rightarrow k^*$. For any $t \in (-\infty,
\infty)$ $f(t, x_0, k^*) \in \omega (x', k^*)$, consequently, $\rho^* (f(t, y_i, k'_i),\omega(y_i,
k'_i)) > \varepsilon / 2$ for $i>i_0,\ 0 \leq t \leq \tau$, therefore   for   these $i\ \tau_1 (y_i,
k'_i, \varepsilon / 2) > \tau$. The existence of $\tau_1$-slow relaxations is proved.

     Now, let us suppose that there are  $\tau_1$-slow relaxations: there
can be found  such  a  sequence  $(x_n, k_n) \in X \times K$  that  for  some $\varepsilon >0$
$\tau_1 (x_n, k_n, \varepsilon) \rightarrow \infty$. Using the compactness of $X \times K$, let us
choose  from this  sequence  a  convergent  one,  preserving  the  denotations: $(x_n, k_n)
\rightarrow (x^*, k^*)$.
 For any $ y \in \omega(x^*, k^*)$ there is such $n=n(y)$ that when
$n> n(y)\ \rho^* (y, \omega(x_n, k_n)) > \varepsilon / 2$. Really, as $y \in \omega(x^*,k^*)$,  there
is  such $t>0$ that $\rho(f(t,x^*, k^*), y) < \varepsilon / 4$.     Since $(x_n, k_n) \rightarrow
(x^*, k^*),\ \tau_1 (x_n, k_n, \varepsilon) \rightarrow \infty$, there is such $n$ (we denote it by
$n(y)$)   that for $n>n(y) \ \rho^*(\bar t, x_n, k_n)) < \varepsilon / 4,\ \tau_1(x_n, k_n,
\varepsilon) > t$.  Thereby, since $\rho^* (f(\bar{t}, x_n, k_n), \omega(x_n, k_n)) > \varepsilon$,
then $\rho^* (f(\bar{t}, x^*,k^*), \omega(x_n, k_n)) > 3 \varepsilon / 4$, and, consequently,
$\rho^*(y, \omega(x_n, k_n)) > \varepsilon / 2$. Let $y_i, \ldots, y_m$  be $\varepsilon / 4$-network
in $\omega(x^*, k^*)$. Let $N = \max\ n(y_i)$. Then  for $n>N$
 and for any $i\ (1 \leq i \leq m)\ \rho^*(y_i, \omega(x_n, k_n)) > \varepsilon / 2$.
Consequently  for any  $y \in \omega(x^*, k^*)$  for $n>N\ \rho^*(y, \omega(x_n, k_n) > \varepsilon /
4$,   i.e.   for $n>N$ $r(\omega(x^*, k^*), \omega(x_n, k_n)) > \varepsilon / 4$. The existence of
$\Omega(x,k)$-bifurcations  is proved (according to Proposition \ref{P1.8}). Using Theorem \ref{T2.1}
and Proposition \ref{P1.10} we obtain the following theorem.

\vspace{10pt}

\noindent{\bf Theorem \ref{T2.1}$'$} {\it The system (\ref{e1}) possesses  $\tau_1$-slow relaxations
if and only if $\omega(x,k)$  is not $r$-continuous function in $X \times K$.}

\begin{theorem}\label{T2.2}The system (\ref{e1}) possesses $\eta_1$-slow relaxations if and only if it
possesses $\Omega(k)$-bifurcations.
\end{theorem}
{\bf Proof.} Let  the  system  possess $\Omega(k)$-bifurcations.   Then (according to Proposition
\ref{P1.8})  there  is  such  sequence  of parameters $k_n \rightarrow k^*$  that for some
$\omega(x^*, k^*) \in \Omega (k^*)$ and $\varepsilon
>0$  for  any $n$ $r(\omega(x^*, k^*), \omega(k_n)) > \varepsilon$. Let $x_0 \in \omega(x^*, k^*)$.
Then for any $n$ and $t \in (-\infty, \infty)\ \rho^*(f(t, x_0, k^*), \omega(k_n)) > \varepsilon$
because $f(t, x_0, k^*) \in \omega(x^*, k^*)$. Let  us  prove that $\eta_1(x_0, k_n, \varepsilon / 2)
\rightarrow \infty$   as $n \rightarrow \infty$.  To  do  this,  use  the   uniform continuity of $f$
on compact segments and  for  any $\tau > 0$   find  such $\delta = \delta (\tau) > 0$
   that  $\rho(f(t,x_0, k^*), f(t,x_0, k_n)) < \varepsilon / 2$ if
$0 \leq t \leq \tau$ and $\rho_K(k^*, k_n) < \delta$.
 Since  $k_n \rightarrow k^*$,  there  is  such
$N = N(\tau)$   that  for $n>N\ \rho_K (k_n, k) < \delta$. Therefore for  $n>N,\ 0 \leq t \leq \tau\
\rho^*(f(t,x_0, k_n), \omega(k_n)) > \varepsilon / 2$. The existence of  $\eta_1$-slow relaxations is
proved.

     Now, suppose that there exist  $\eta_1$-slow
relaxations: there  are such $\varepsilon>0$ and sequence $(x_n, k_n) \in X \times K$   that
$\eta_1(x_n, k_n, \varepsilon) \rightarrow \infty$.  Use  the compactness of $X \times K$ and turn to
converging  subsequence  (retaining the same denotations): $(x_n, k_n) \rightarrow (x^*, k^*)$. Using
the way  similar  to the proof of Theorem \ref{T2.1}, let us show that for any $y \in \omega(x^*,
k^*)$ there is such $n=n(y)$ that if $n>n(y)$, then $\rho^*(y, \omega(k_n)) > \varepsilon / 2$.
Really, there is  such $\tilde t > 0$  that $\rho(f(\tilde t, x^*, k^*),y) < \varepsilon/4$. As
$\eta_1(x_n, k_n, \varepsilon) \rightarrow \infty$   and $(x_n, k_n) \rightarrow (x^*, k^*)$,
 there is such $n=n(y)$ that  for $n>n(y) \ \rho(f(\tilde t, x^*, k^*),
f(\tilde t, x_n, k_n)) < \varepsilon / 4$ and $\eta_1(x_n, k_n, \varepsilon) > \tilde t$.  Thereafter
we   obtain
\begin{eqnarray*}
\lefteqn{\rho^*(y, \omega(k_n)) \geq} \\
& & \geq \rho^*(f(t, x_n, k_n), \omega(k_n)) -
\rho(y,f(\tilde t, x^*, k^*)) - \rho(f(\tilde t, x^*, k^*),
f(\tilde t, x_n, k_n)) > \varepsilon / 2.
\end{eqnarray*}
 Further   the   reasonings   about  $\varepsilon / 4$-network
of the set $\omega(x^*, k^*)$ (as in the proof  of  Theorem \ref{T2.1}) lead to the  inequality
$r(\omega(x^*, k^*), \omega(k_n)) > \varepsilon / 4$  for  $n$  large enough. On  account  of
Proposition  \ref{P1.8}  the  existence  of $\Omega(k)$-bifurcations is proved, therefore is proved
Theorem \ref{T2.2}.

\begin{theorem}\label{T2.3}If the system (\ref{e1}) possesses  $\omega(x,k)$-bifurcations then it possesses
$\tau_2$-slow relaxations.
\end{theorem}
{\bf Proof.} Let the system (\ref{e1}) possess $\omega(x,k)$-bifurcations: there is such sequence
$(x_n, k_n) \in X \times K$ and such $\varepsilon>0$ that $(x_n, k_n) \rightarrow (x^*, k^*)$ and $$
\rho^*(x', \omega(x_n, k_n)) > \varepsilon \ \mbox{for any} \ n \ \mbox{and some} \  x' \in
\omega(x^*,k^*). $$ Let $t>0$.  Define the following auxiliary function:
\begin{equation}\label{e20}
\Theta (x^*, x', t, \varepsilon) = \mbox{mes} \{t' \geq 0\ | \ t' \leq t, \ \rho
(f(t', x^*,k^*), x') < \varepsilon / 4 \},
\end{equation}
$\Theta (x^*, x', t, \varepsilon)$ is ``the  time  of  residence"  of $(k^*, x^*)$-motion   in
$\varepsilon / 4$-neighbourhood of $x$ over the time segment $[0,t]$. Let  us  prove that $\Theta
(x^*, x', t, \varepsilon) \rightarrow \infty$ as $t \rightarrow \infty$. We need the  following
corollary of continuity of $f$ and compactness of $X$

\begin{lemma}\label{L2.1}Let $x_0 \in X,\ k \in K,\ \delta > \varepsilon >0$. Then there is such $t_0 >
0$ that for any $x \in X$   the  inequalities $\rho(x, x_0) < \varepsilon$ and  $0 \leq t' < t_0$
lead to $\rho(x_0, f(t', x,k)) < \delta$.
\end{lemma}
{\bf Proof.} Let us suppose the contrary: there are such  sequences $x_n$  and $t_n$  that $\rho(x_0,
x_n) < \varepsilon,\ t'_n \rightarrow 0$, and $\rho(x_0, f(t'_n,x_n, k)) \geq \delta$. Due to the
compactness of $X$ one can choose from the sequence $x_n$  a  convergent one.  Let  it  converge  to
$\bar x$.  The  function $\rho(x_0,f(t,x,k))$   is continuous.   Therefore $\rho(x_0,f(t'_n,x_n,k))
\rightarrow \rho(x_0, f (0,x,k)) = \rho(x_0, \bar x)$. Since $\rho(x_0, x_n) < \varepsilon$, then
$\rho(x_0, \bar x) \leq \varepsilon$. This contradicts to the  initial supposition
$(\rho(x_0,f(t'_n,x_n,k)) \geq \delta \geq \varepsilon)$.

Let us  return  to  the  proof  of  Theorem  \ref{T2.3}.  Since $x' \in \omega (x^*, k^*)$, then
there is such monotonic sequence $t_{j} \rightarrow \infty$ that  for any $j$ $\rho(f(t_{j},
x^*,k^*),x') < \varepsilon / 8$. According to  Lemma  \ref{L2.1} there is $t_0 > 0$ for which
$\rho(f(t_{j} + \tau, x^*,k^*),x') < \varepsilon / 4$ as  $0 \leq \tau \leq t_0$. Suppose (turning to
subsequence, if it  is  necessary)  that $t_{j+1} - t_{j} > t_0$. $\Theta(x^*, x',t, \varepsilon) >
jt_0$ if $t>t_{j} + t_0$. For any $j=1,2, \ldots$  there is  such $N(j)$ that $\rho(f(t,x_n,k_n),
f(t,x^*,k^*)) < \varepsilon / 4$ under the conditions  $n>N(j),\ 0 \leq t \leq t_{j} + t_0$. If
$n>N(j)$, then $\rho(f(t,x_n,k_n), x') < \varepsilon / 2$ for $t_j \leq t \leq t_j + t_0\ (i \leq
j)$. Consequently, $\tau_2 (x_n, k_n, \varepsilon / 2) > jt_0$ if $n>N(j)$. The existence of $\tau_2$
slow relaxations is proved.

\begin{theorem}\label{T2.4} If the system  (\ref{e1})  possesses  $\omega(k)$-bifurcations, then it possesses
$\eta_2$-slow relaxations too.
\end{theorem}
{\bf Proof.} Let the system (\ref{e1})  possess  $\omega(k)$-bifurcations:  there are such sequence
$k_n \in K$ and  such $\varepsilon > 0$   that $k_n > k^*$    and $\rho^*(x', \omega(k_n)) >
\varepsilon$ for  some $x' \in \omega(k^*)$  and any $n$. The point $x'$   belongs to the
$\omega$-limit set of some motion: $x' \in \omega (x^*, k^*)$. Let $\tau>0$  and $t^*$ be  such  that
$\Theta (x^*,x',t^*, \varepsilon) > \tau$ (the existence of such $t^*$  is  shown  when  proving
Theorem \ref{T2.3}). Due to the uniform continuity of $f$  on  compact intervals there is such $N$
that $\rho(f(x^*,k^*),f(t,x^*,k_n)) < \varepsilon / 4$  for $0 \leq t \leq t^*,\ n>N$.
 But    from    this    fact  it  follows     that
$\eta_2(x^*,k_n, \varepsilon / 2) \geq \Theta (x^*, x',t^*, \varepsilon)
> \tau\ (n>N)$.
Because  of  the  arbitrary choice of $\tau$ Theorem \ref{T2.4} is proved.

The two following theorems provide  supplementary  sufficient
conditions of $\tau_2$ - and $\eta_2$ -slow relaxations.

\begin{theorem}\label{T2.5}If for the system (\ref{e1}) there are  such $x \in X, k \in K$ that $(k,x)$-motion
is whole and  $\alpha(x,k) \not \subset \omega (x,k)$, then  the  system (1.1) possesses
$\tau_2$-slow relaxations.
\end{theorem}
{\bf Proof.} Let be such $x$ and $k$ that  $(k,x)$-motion  is  whole  and $\alpha (x,k) \not \subset
\omega (x,k)$. Let us denote by $x^*$  an  arbitrary  $\alpha$-,  but  not  $\omega$-limit point of
$(k,x)$-motion. Since $\omega(x,k)$ is closed, $\rho^*(x^*, \omega(x,k)) > \varepsilon >0$. Define an
auxiliary    function $$ \varphi(x,x^*, t, \varepsilon) = \mbox{mes}\{t' \ | \ -t \leq t' \leq 0,\
\rho(f(t',x,k),x^*) < \varepsilon / 2 \}. $$ Let us  prove  that $\varphi (x,x^*, \varepsilon)
\rightarrow \infty$ as $t \rightarrow \infty$. According  to  Lemma   \ref{L2.1} there is such $t_0
>0$  that $\rho(f(t,y,k), x^*) < \varepsilon / 2$ if $0 \leq t \leq t_0$ and $\rho(x^*,y) <
\varepsilon / 4$. Since $x^*$  is  $\alpha$-limit point of $(k,x)$-motion, there is such sequence
$t_j < 0,\ t_{j+1} - t_j < -t_0$, for which $\rho(f(t_j,x,k), x^*) < \varepsilon / 4$. Therefore, by
the way used  in  the proof of Theorem \ref{T2.3} we obtain: $\varphi(x,x^*,t_j, \varepsilon) >jt_0$.
This  proves Theorem \ref{T2.5}, because $\tau_2(f(-t,x,k),k, \varepsilon / 2) \geq \varphi (x,x^*,t,
\varepsilon)$.

\begin{theorem}\label{T2.6}If for the system (\ref{e1}) exist such $x \in X, k \in K$ that $(k,x)$-motion is
whole and  $\alpha(x,k) \not \subset \overline{\omega (k)}$,  then  the  system  (1.1) possesses
$\eta_2$-slow relaxations.
\end{theorem}
{\bf Proof.} Let   $(k,x)$-motion   be   whole   and $$ \alpha(x,k) \not \subset \overline{\omega
(k)}, \ x^* \in \alpha(x,k) \setminus \overline{\omega (k)}, \ \rho^*(x^*, \overline{\omega (k)}) =
\varepsilon >0. $$ As in the proof of  the  previous theorem,  let  us   define   the   function
$\varphi (x,x^*,t, \varepsilon)$.   Since $\varphi (x,x^*,t, \varepsilon) \rightarrow \infty$ as $t
\rightarrow \infty$ (proved   above)   and $\eta_2(f(-t,x,k),k, \varepsilon / 2) \geq \varphi
(x,x^*,t, \varepsilon)$, the theorem is proved.

Note that the conditions of the  theorems  \ref{T2.5},  \ref{T2.6}  do  not imply bifurcations.

\begin{figure}[t]
\centering {
\includegraphics[width=40mm,height=45mm]{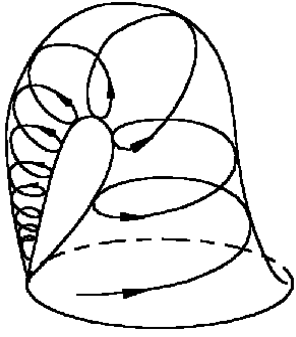}
\caption{\label{Fig.4} Phase portrait of the system (\ref{e15}) after gluing the fixed points.} }
\end{figure}

\begin{example}\label{E2.6}   ($\tau_2$-, $\eta_2$-slow relaxations without bifurcations). Examine the
system given by the set of equations (\ref{e15}) in  the circle $x^2 + y^2 \leq 1$ (see Fig.
\ref{Fig.2},a, Example \ref{E2.1}). Identify the fixed  points $r=0$  and $r=1,\ \varphi = \pi$ (Fig.
\ref{Fig.4}). The  complete $\omega$-limit set of the system obtained consists of one fixed point.
For initial data $r_0 \rightarrow 1, r_0 < 1$ ($\varphi_0$ is  arbitrary)  the  relaxation  time
$\eta_2 (r_0, \varphi_0, 1/2) \rightarrow \infty$ (hence, $\tau_2(r_0, \varphi_0, 1/2) \rightarrow
\infty)$.
\end{example}

Before analyzing $\tau_3, \eta_3$-slow relaxations, let us define {\it Poisson's stability} according
to \cite{[6]}, p.363: $(k,x)$-motion is {it Poisson's   positively stable} ($P^+$-{\it stable}), if
$x \in \omega(x,k)$.

Note that any $P^+$-stable motion is whole.

\begin{lemma}\label{L2.2}If for the system (\ref{e1}) exist  such  $x \in X,\ k \in K$ that
$(k,x)$-motion is whole but  not   $P^+$-stable, then  the  system  (\ref{e1}) possesses
$\tau_3$-slow relaxations.
\end{lemma}
{\bf Proof.} Let $\rho^*(x,\omega(x,k)) = \varepsilon >0$ and $(k,x)$-motion be  whole. Then
\noindent
\begin{eqnarray*}
& \tau_3(f(-t,x,k), k,\varepsilon) \geq t, &\\
& \mbox{since} \
f(t,f(-t,x,k),k) =x \ \mbox{and} \
\rho^*(x,\omega(f(-t,x,k),k)) = \varepsilon &
\end{eqnarray*}
(because $\omega(f(-t,x,k),k) = \omega(x,k))$. Therefore $\tau_3$-slow relaxations exist.

\begin{lemma}\label{L2.3}If for the system (\ref{e1}) exist  such $x \in X,\ k \in K$ that $(k,x)$-motion is
whole  and $x \not \in \omega(k)$, then  this  system  possesses $\eta_3$ -slow relaxations.
\end{lemma}
{\bf Proof.}
Let $\rho^*(x,\omega(k)) = \varepsilon >0$ and
$(k,x)$-motion  be  whole.  Then
\begin{eqnarray*}
& \eta_3(f(-t,x,k),k, \varepsilon)) \geq t, & \\
& \mbox{since} \ f(t,f(-t,x,k),k) = x \ \mbox{and} \ \rho^*(x, \omega(k)) =
\rho^*(x,\overline{\omega(k)}) = \varepsilon. &
\end{eqnarray*}
 Consequently,  $\eta_3$-slow relaxations exist.

\begin{lemma}\label{L2.4}Let for the system (\ref{e1}) be  such $x_0 \in X,\ k \in K$   that $(k_0,
x_0)$-motion is whole. If $\omega(x,k)$ is $d$-continuous function in $X \times K$ (there are no
$\omega(x,k)$-bifurcations), then:
\begin{enumerate}
\item{$\omega(x^*, k_0) \subset \omega (x_0, k_0)$ for any $x^* \in \alpha(x_0, k_0)$, that is
$\omega(\alpha(x_0, k_0), k_0) \subset \omega (x_0, k_0)$;}
\item{In particular, $\omega(x_0, k_0) \bigcap \alpha (x_0, k_0) \neq \varnothing$.}
\end{enumerate}
\end{lemma}
{\bf Proof.} Let $x^* \in \alpha (x_0, k_0)$. Then there are such $t_n >0$ that $t_n \rightarrow
\infty$ and $x_n = f(-t_n, x_0, k_0) \rightarrow x^*$. Note  that  $\omega(x_n, k_0) = \omega(x_0,
k_0)$.  If $\omega(x^*, k_0) \not \subset \omega (x_0, k_0)$, then,  taking  into  account  closure
of $\omega(x_0, k_0)$, we would obtain  inequality  $d(\omega(x^*,k_0), \omega(x_0, k_0)) >0$. In
this  case   $x_n \rightarrow x^*$,  but   $\omega(x_n, k_0) - / \rightarrow \omega (x^*, k_0)$, i.e.
there   is $\omega(x,k)$-bifurcation. But according to the assumption there  are  no
$\omega(x,k)$-bifurcations. The obtained contradiction proves  the  first statement of the lemma. The
second  statement  follows  from  the facts that $\alpha(x_0, k_0)$ is closed,  $k_0$-invariant and
nonempty.  Really, let $x^* \in \alpha (x_0, k_0)$. Then $\overline{f((-\infty, \infty), x^*, k_o)}
\subset \alpha(x_0, k_0)$ and, in particular, $\omega(x^*, k_0) \subset \alpha(x_0, k_0)$.
 But it has been proved that $\omega(x^*, k_0) \subset \omega(x_0, k_0)$.
Therefore $\omega(x_0, k_0) \bigcap \alpha (x_0, k_0) \supset \omega (x^*, k_0) \neq \varnothing$.

\begin{theorem}\label{T2.7}The system (\ref{e1}) possesses  $\tau_3$-slow relaxations
if and only if at least one of the following conditions is satisfied:
\begin{enumerate}
\item{There are $\omega(x,k)$-bifurcations;}
\item{There are such $x \in X,\ k \in K$ that $(k,x)$-motion is whole but not $P^+$-stable.}
\end{enumerate}
\end{theorem}
{\bf Proof.} If there exist $\omega(x,k)$-bifurcations, then the existence of  $\tau_3$-slow
relaxations  follows  from  Theorem  \ref{T2.3}  and  the inequality $\tau_2 (x,k, \varepsilon) \leq
\tau_3 (x,k, \varepsilon)$. If the condition 2  is  satisfied, then the existence of $\tau_3$-slow
relaxations follows from  Lemma \ref{L2.2}. To finish the proof, it  must  be ascertained that if the
system (\ref{e1}) possesses  $\tau_3$-slow  relaxations  and  does  not possess
$\omega(x,k)$-bifurcations,  then  there  exist  such $x \in X,\ k \in K$ that $(k,x)$-motion is
whole and not $P^+$-stable.  Let  there  be  $\tau_3$-slow relaxations and $\omega(x,k)$-bifurcations
be absent. There can be chosen such convergent (because  of  the compactness  of $X \times K$)
sequence $(x_n, k_n) \rightarrow (x^*, k^*)$ that $\tau_3(x_n, k_n, \varepsilon) \rightarrow \infty$
for some $\varepsilon >0$.  Consider  the sequence $y_n = f(\tau_3(x_n, k_n, \varepsilon), x_n,
k_n)$.  Note  that $\rho^*(y_n, \omega(x_n, k_n)) = \varepsilon$. This follows from the definition of
relaxation time and continuity of the function $\rho^*(f(t,x,k),s)$ of $t$  at any $(x,k) \in X
\times K,\ s \subset X$. Let  us choose from the sequence $y_n$   a  convergent  one  (preserving the
denotations $y_n, x_n, k_n$). Let us denote its limit: $y_n \rightarrow x_0$. It is clear that $(k^*,
x_0)$-motion is whole. This follows from the results of Subection 1.1 and the fact that $(k_n,
y_n)$-motion  is   defined  in  the time interval $[-\tau_3(x_n, k_n, \varepsilon), \infty)$, and
$\tau_3 (x_n, k_n, \varepsilon) \rightarrow \infty$ as $n \rightarrow \infty$. Let  us prove that
$(k^*, x_0)$-motion  is  not  $P^+$-stable,  i.e. $x_0 \not \in \omega(x_0, k^*)$. Suppose the
contrary: $x_0 \in \omega (x_0, k^*)$. Since $y_n \rightarrow x_0$, then there is such $N$  that
$\rho(x_0, y_n) < \varepsilon / 2$  for   any   $n \geq N$. For the   same $n \geq N\ \rho^* (x_0,
\omega(y_n, k_n)) > \varepsilon / 2$, since $\rho^*(y_n, \omega(y_n, k_n)) = \varepsilon$. But from
this  fact and  from  the  assumption  $x_0 \in \omega (x_0, k^*)$ it follows  that   for $n \geq N \
d(\omega (x_0, k^*), \omega (y_n, k_n)) > \varepsilon / 2$, and   that   means   that   there   are
$\omega(x,k)$-bifurcations. So far as it  was  supposed $d$-continuity  of $\omega(x,k)$, it was
proved that $(k^*, x_0)$-motion is  not $P^+$-stable,  and this completes the proof of the theorem.

Using Lemma \ref{L2.4}, Theorem \ref{T2.7}  can  be  formulated  as follows.

\vspace{10pt}

\noindent{\bf Theorem \ref{T2.7}$'$} {\it The system (\ref{e1}) possesses  $\tau_3$-slow relaxations
if and only if at least one of the following conditions is satisfied:
\begin{enumerate}
\item{There are $\omega(x,k)$-bifurcations;}
\item{There are such $x \in X,\ k \in K$  that $(k,x)$-motion is whole but not $P^+$-stable    and
possesses     the     following     property: $\omega(\alpha(x,k),k) \subset \omega (x,k)$.}
\end{enumerate}
}

\vspace{10pt}

     As an example of motion satisfying the  condition  2  can  be
considered a trajectory going from a fixed point to the same point
(for example, the  loop  of  a  separatrice),  or  a  homoclinical
trajectory of a periodical motion.

\begin{theorem}\label{T2.8}The system (\ref{e1}) possesses  $\eta_3$-slow relaxations  if and only if at least
one of the following conditions is satisfied:
\begin{enumerate}
\item{There are $\omega(k)$-bifurcations;}
\item{There are such $x \in X,\ k \in K$ that  $(k,x)$-motion is  whole  and $x \not \in
\overline{\omega (k)}$.}
\end{enumerate}
\end{theorem}
{\bf Proof.} If there are $\omega(k)$-bifurcations, then, according to Theorem \ref{T2.4}, there are
$\eta_2$- and all the more  $\eta_3$-slow relaxations. If the condition 2 holds, then the  existence
of $\eta_3$-slow relaxations follows from Lemma \ref{L2.3}.  To  complete the proof, it must be
established that if  the  system  (\ref{e1})  possesses $\eta_3$-slow relaxations and does not
possess  $\omega(k)$-bifurcations then the condition 2 of the theorem holds: there  are such $x \in
X,\ k \in K$ $(k,x)$-motion is  whole  and  $x \not \in \overline{\omega(k)}$. Let  there be $\eta_3$
-slow relaxation and $\omega(k)$-bifurcations be absent.  Then  we  can choose such convergent
(because of  the  compactness  of $X \times K$) sequence $(x_n, k_n) \rightarrow (x^*, k^*)$ that
$\eta_3(x_n, k_n, \varepsilon) \rightarrow \infty$ for some $\varepsilon >0$. Consider the sequence
$y_n = f(\eta_3(x_n, k_n, \varepsilon), x_n, k_n)$. Note  that $\rho^*(y_n, \omega(k_n)) =
\varepsilon$. Choose from the sequence  $y_n$   a  convergent one  (preserving  the denotations $y_n,
x_n, k_n)$. Let us denote its limit by $x_0: y_n \rightarrow x_0$.  From the results of Subection 1.1
and the fact that $(k_n, y_n)$-motion is defined at least on the segment $[-\eta_3(x_n, k_n,
\varepsilon), \infty)$ and $\eta_3(x_n, k_n, \varepsilon) \to \infty$ we  obtain that  $(k^*,
x_0)$-motion  is whole.  Let  us  prove  that $x_0 \not \in \overline{\omega(k)}$.
 Really, $y_n \rightarrow x_0$, hence there is such $N$ that  for  any
$n \geq N$
the  inequality $\rho(x_0, y_n) < \varepsilon / 2$   is   true.   But
$\rho^*(y_n, \overline{\omega(k_n)}) = \varepsilon$,
consequently for $n>N\ \rho^*(x_0, \overline{\omega(k_n)})
> \varepsilon / 2$. If $x_0$  belonged  to $\overline{\omega(k^*)}$,
then for $n>N$ the inequality $d \overline{(\omega(k^*)}, \omega(k)n)) > \varepsilon / 2$ would be
true  and there  would  exist $\omega(k)$-bifurcations.  But  according   to   the assumption they do
not exist. Therefore is proved that $x_0 \not \in \overline{\omega(k^*)}$.

     Formulate now some corollaries from $\omega(k)$ the proved theorems.

\begin{corollary}\label{C2.1} Let any trajectory from $\omega(k)$ be recurrent  for any $k \in K$ and there be
not  such $(x,k) \in X \times K$ that $(k,x)$-motion  is whole,  not    $P^+$-stable   and
$\omega(\alpha(x,k), k) \subset \omega(x,k)$    (or   weaker, $\omega (x,k) \bigcap \alpha (x,k) \neq
\varnothing$). Then the existence  of   $\tau_3$-slow  relaxations  is equivalent to the existence of
$\tau_{1,2}$-slow relaxations.
\end{corollary}
Obviously,  this  follows  from  Theorem \ref{T2.7}  and   Proposition \ref{P1.11}.

\begin{corollary}\label{C2.2}   Let   the   set $\omega(x,k)$  be    minimal
$(\Omega(x,k) = \{ \omega (x,k) \} )$
 for  any $(x,k) \in X \times K$   and  there  be  not  such
$(x,k) \in X \times K$  that  $(k,x)$-motion  is   whole,   not  $P^+$-stable and $\omega (\alpha
(x,k), k) \subset \omega (x,k)$  (or   weaker, $\alpha(x,k) \bigcap \omega(x,k) \neq \varnothing$).
Then   the existence of $\tau_{3}$-slow relaxations is equivalent to the existence of $\tau_{1,2}$
-slow relaxations.
\end{corollary}
     This follows from Theorem \ref{T2.7} and Corollary  \ref{C1.1}  of
Proposition \ref{P1.11}.

\section {Slow Relaxations of One Semiflow}\label{S3}

In this section we study one semiflow $f$. Here and further we denote by $\omega_f$ and $\Omega_f$
the  complete $\omega$-limit sets of one semiflow $f$   (instead of $\omega(k)$  and $\Omega(k)$).

\subsection {$\eta_2$-slow Relaxations}\label{SS3.1}

As it was shown (Proposition \ref{P2.2}), $\eta_1$-slow relaxations of one semiflow are impossible.
Also was given  an  example  of $\eta_2$-slow relaxations in one system (Example \ref{E2.1}). It is
be proved below that a set of smooth systems possessing $\eta_2$-slow relaxations  on  a compact
variety is a set of first category in $C^1$-topology. As for general dynamical systems, for them is
true the following theorem.

Let us recall the definition of {\it non-wandering points}.

\begin{definition}\label{Dnonwan}A  point $x^*\in X$   is the non-wandering point for the semiflow $f$, if for any
neighbourhood $U$ of $x^*$ and for any $T>0$ there is such $t>T$ that $f(t,U) \bigcap U \neq
\varnothing$.
\end{definition}

\begin{theorem}\label{T3.1}
Let a semiflow $f$ possess   $\eta_2$-slow  relaxations. Then there exists a non-wandering point $x^*
\in X$ which does not belong to $\overline{\omega_f}$.
\end{theorem}
{\bf Proof.} Let for some $\varepsilon >0$ the function $\eta_2 (x, \varepsilon)$ be unlimited  in
$X$.
 Consider a sequence $x_n \in X$ for which $\eta_2 (x_n, \varepsilon)
\rightarrow \infty$. Let $V$ be a closed subset of the set $\{x \in X \ | \ \rho^*\ (x, \omega_f)
\geq \varepsilon \}$. Define an auxiliary  function: the  residence  time   of   $x$-motion   in the
intersection of the closed $\delta$-neighbourhood of the point $y \in V$  with $V$:
\begin{equation}\label{e21}
\psi (x,y, \delta, V) = \mbox{mes} \{ t>0 \ | \ \rho(f(t,x), y) \leq \delta,\
f(t,x) \in V \}.
\end{equation}
From the inequality $\psi (x,y, \delta, V) \leq \eta_2 (x, \varepsilon)$ and the  fact  that  finite
$\eta_2 (x, \varepsilon)$ exists for each $x \in X$ (see Proposition \ref{P2.1}) it follows that the
function $\psi$ is  defined  for  any $x, y, \delta >0$  and $V$   with  indicated properties ($V$ is
closed, $r(V, f) \geq \varepsilon$).

Let us fix some $\delta >0$.  Suppose that $V_0 = \{ x \in X \ | \ \rho^* (x, \omega_f) \geq
\varepsilon \}$.

Let $ \bar{U}_{\delta} (y_j)$  be  a closed sphere of radius $\delta$ centered in $y_j \in V_0$).
Consider a finite covering of $V_0$ with closed spheres centered in some points from  $V_0$:  $V_0
\subset \bigcup^k_{j=1} \bar{U}_{\delta} (y_i)$. The inequality
\begin{equation}\label{e22}
\sum^k_{j=1} \psi (x,y_i, \delta, V_0) \geq \eta_2 (x, \varepsilon)
\end{equation}
is true (it is obvious: being in  $V_0$, $x$-motion is  always  in  some of $\bar U_{\delta}(y_i)$).
 From  (\ref{e22}) it follows  that $\sum^k_{j=1} \psi (x,y_i, \delta, V_0)
\rightarrow \infty$ as $n \rightarrow \infty$. Therefore there is $j_0\ (1 \leq j_0 \leq k)$ for
which there is such subsequence $\{ x_{m(i)} \} \subset \{ x_n \}$ that $\psi (x_{m(i)}, y_{j_0},
\delta, V_0) \rightarrow \infty$. Let $y^*_0 = y_{j_0}$.

Note that if $\rho(x,y^*_0) < \delta$  then for any $I>0$ there  is
$t>T$ for which
$$
f(t, \bar U_{2\delta} (x)) \bigcap \bar U_{2 \delta} (x) \neq
\varnothing.
$$

Let us denote  $V_1 = \bar U_{\delta} (y^*_0) \bigcap V_0$. Consider the finite covering of $V_1$
with closed spheres of radius $\delta / 2$ centered in some points from  $V_1$: $V_1 =
\bigcup^{k_1}_{j=1} \bar U_{\delta / 2} (y^1_j)$; $y^1_j \in V_1$. The following inequality is true:
\begin{equation}\label{e23}
\sum^{k_1}_{j=1} \psi (x,y^1_j, \delta / 2, V_1) \geq
\psi (x, y^*_0, \delta, V_0).
\end{equation}
Therefore exists $j'_0\ (1 \geq j'_0 \geq k_1)$
 for which  there  is  such  sequence
$\{x_{l(i)} \} \subset \{ x_{m(i)} \} \subset \{ x_n \}$ that $\psi(x_{l(i)}, y^1_{j'_0}, \delta / 2,
V_1) \rightarrow \infty$ as $i \rightarrow \infty$. We denote $y^*_1 = y^1_{j'_0}$.

\sloppy
Note that if $\rho(x,y^*_1) \leq \delta / 2$ then for any $T>0$ there is  such $t>T$ that $$
f(t, \bar U_{\delta} (x)) \bigcap \bar U_{\delta} (x) \neq \varnothing. $$

Let us denote  $V_2 = \bar U_{\delta / 2} (y^*_1) \bigcap V_1$ and   repeat   the   construction,
substituting $\delta / 2$ for $\delta, \delta / 4$ for $\delta / 2,\ V_{1,2}$    for $V_{0,1}$.

Repeating this constructing  further,  we obtain  the fundamental sequence $y^*_0, y^*_1, \ldots$. We
denote its limit  $x^*$, $\rho^*(x^*, \omega_f) \geq \varepsilon$ because $x^*\in V_0$ .  The point
$x^*$ is non-wandering: for any its neighbourhood $U$  and for  any $T>0$ there is such $t>T$ that
$f(t,U) \bigcap U \neq \varnothing$. Theorem \ref{T3.1} is proved.

The inverse is not true in general case.

\begin{figure}[t]
\centering {
\includegraphics[width=40mm,height=70mm]{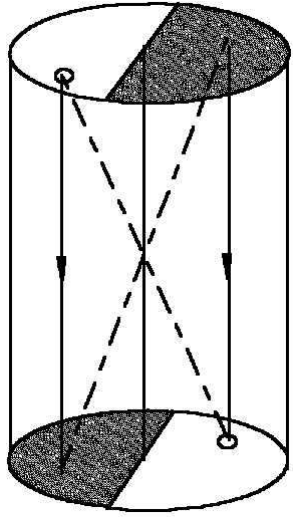}
\caption{\label{Fig.5} Phase space of the system (Example \ref{E3.1}). All the points of the axis are
non-wandering; $ \bigcirc $ is the place of delay near fixed points.} }
\end{figure}

\begin{example}\label{E3.1}  (The existence of  non-wandering  point $x^* \not \in \overline{\omega_f}$
without  $\eta_2$-slow relaxations). Consider a cylinder in $R^3:\ x^2 + y^2 \leq 1,\ -1 \leq z \leq
1$. Define in it a motion by the equations $\dot x = \dot y = 0,\ \dot z = " (x,y,z)$, where $"$ is a
smooth function, $" \geq 0$, and it is equal to zero only at (all) points of the sets $(z=-1, x \leq
0)$ and $(z=1, x \geq 0)$. Since the  sets are  closed,  such function exists (even infinitely
smooth). Identify the opposite bases of the cylinder, preliminary turning them at angle $\pi$.  In
the  obtained dynamical system  the  closures  of trajectories,  consisting of more than one  point,
form  up Zeifert foliation  (Fig. \ref{Fig.5})  (see, for example, \cite{[5]}, p.158).

 Trajectory  of the  point  $(0,0,0)$
is a loop, tending at $t \rightarrow \pm \infty$ to one point which is  the   identified centers   of
cylinder bases. The trajectories of all other nonfixed points are  also loops, but before to close
they make two turns near the trajectory $(0,0,0)$ The nearer is the initial point of motion to
$(0,0,0)$, the larger is  the time interval between it and the point of following entering of this
motion in small neighborhood of $(0,0,0)$ (see Fig. \ref{Fig.5}).
\end{example}

\subsection {Slow Relaxations and Stability}\label{SS3.2}

Let us recall the definition of Lyapunov stability of closed invariant set given by Lyapunov (see
\cite{[25]}, p.31-32), more general approach  is given in \cite{[61]}.

\begin{definition}\label{D3.1} A closed  invariant  set $W \subset X$ is {\it Lyapunov
stable} if and only if for any $\varepsilon >0$ there is such
$\delta = \delta(\varepsilon) >0$ that  if $\rho^*(x,W) < \delta$
 then the inequality $ \rho^*(f(t,x),W) < \varepsilon$ is true for all
$t \geq 0$.
\end{definition}

     The following lemma follows directly from the definition.

\begin{lemma}\label{L3.1}A closed invariant set $W$ is Lyapunov stable if and
only  if  it  has  a  fundamental  system  of positive-invariant  closed
neighborhoods:  for  any  $\varepsilon$  there  are  such
$\delta > 0$  and  closed positive-invariant set $V \subset X$  that
\begin{equation}\label{e24}
\{x \in X \ | \ \rho^*(x,W) < \delta \} \subset V \subset \{ x \in X \ | \ \rho^* (x,W)
< \varepsilon \}.
\end{equation}
\end{lemma}
To get the set $V$, one can take for  example  the  closure  of following (it is obviously
positive-invariant)  set: $\{f(t,x) \ | \ \rho^* (x,W) \leq \delta = \delta (\varepsilon / 2),\ t \in
[0, \infty) \}$, i.e.  of  the  complete  image   (for   all $t \leq 0$)   of $\delta$-neighbourhood
of $W$, where $\delta (\varepsilon)$ is  that  spoken  about  in  Definition \ref{D3.1}.

The following lemma can be deduced from  the  description  of Lyapunov stable sets (\cite{[25]},
Sec.~11, p.40-49).

\begin{lemma}\label{L3.2}Let a closed invariant set
$W \subset X$ be  not  Lyapunov stable. Then for any $\lambda >0$ there is  such $y_0 \in X$  that
the $y_0$-motion  is whole, $\rho^*(y_0, W) < \lambda,\ d(\alpha(y_0),W) < \lambda$ (i.e. the
$\alpha$-limit set  of the $y_0$-motion lies in $\lambda$-neighbourhood of $W$), and $y_0 \not \in
W$.
\end{lemma}

\begin{definition}\label{izol}An $f$-invariant set $W \subset X$ for the semiflow $f$ is called
isolated, if there is such $\lambda > 0$ that for any $ y \in \omega_f$ from the condition $\rho^*(y,
W) < \lambda$ it follows that $y \in W$, that is, any $\omega$-limit point $ y \in \omega_f$ from the
$\lambda$-neighbourhood of $W$ belongs to $W$.
\end{definition}

\begin{theorem}\label{T3.2}
If for semiflow $f$ there exists closed isolated and not Lyapunov stable invariant set $W \subset X$
then this semiflow possesses $\eta_3$-slow relaxations.
\end{theorem}
{\bf Proof.} Let  $W$  be  a  closed  invariant  isolated  Lyapunov unstable set.  Let $\lambda > 0$
be  the  value  from  the  definition  of isolation. Then Lemma \ref{L3.2} guarantees  the existence
of such $y_0 \in X$  that $y_0$-motion is whole, $\rho^*(y_0, W) < \lambda$ and $y_0 \not \in W$. It
gives  (due to closure of $W$) $\rho^*(y_0, W) = d > 0$.  Let $\delta = \min \{ d/2, (\lambda - d) /
2 \}$.  Then $\delta$-neighbourhood of the point $y_0$  lies outside of the set $W$, but  in its
$\lambda$-neighbourhood, and the last is free from the points  of  the set $\omega_f \setminus W$
(isolation of $W$). Thus, $\delta$-neighbourhood of the  point $y_0$ is free from the points  of the
set  $\omega_f \subseteq W \bigcup (\omega_f \setminus W)$,  consequently $y_0 \not \in
\overline{\omega_f}$. Since  $y_0$-motion is whole, Theorem  \ref{T2.8}  guarantees  the presence of
$\eta_3$-slow relaxations. Theorem \ref{T3.2} is proved.

\begin{lemma}\label{L3.3}Let $X$ be connected and $\overline{\omega_f}$ be disconnected, then
$\overline{\omega_f}$ is not Lyapunov stable.
\end{lemma}
{\bf Proof.} Since $\overline{\omega_f}$ is  disconnected,  there  are  such  nonempty closed $W_1,
W_2$  that $\overline{\omega_f} = W_1 \bigcup W_2$ and $W_1 \bigcap W_2 = \varnothing$. Since any
$x$-trajectory is connected and $\overline{\omega_f}$ is invariant, then and $W_1$ and $W_2$   are
invariant  too. The sets $\omega(x)$ are connected (see Proposition \ref{P1.4}),  therefore  for any
$x \in X \ \omega(x) \subset W_1$ or $\omega(x) \subset W_2$. Let us prove that at least one of  the
sets $W_i \ (i = 1,2)$ is not stable. Suppose the contrary: $W_1$ and $W_2$  are stable. Define for
each of them {\it attraction domain}:
\begin{equation}\label{e25}
At (W_i) = \{x \in X \ | \ \omega(x) \subset W_i \}.
\end{equation}
It is obvious that $W_i \subset At (W_i)$ owing to closure and invariance of $W_i$. The sets $At
(W_i)$ are open due to the stability of $W_i$.  Really, there are non-intersecting closed
positive-invariant neighborhoods $V_i$  of the sets $W_i$, since the last do not intersect and  are
closed  and stable (see Lemma \ref{L3.1}). Let $x \in At(W_i)$. Then there is such $t \geq$ that
$f(t,x) \in \mbox{int} V_i$. But because of the continuity of $f$  there  is  such neighbourhood of
$x$ in $X$ that for each its point $x'\ f(t,x) \in \mbox{int} V_i$. Now positive-invariance  and
closure  of $V_i$  ensure $\omega(x') \subset W_i$,  i.e. $x' \in At (W_i)$. Consequently, $x$ lies
in $At(W_i)$ together  with  its neighbourhood  and  the  sets  $At(W_i)$  are  open   in  $X$. Since
$At(W_i) \bigcup At (W_2) = X,\ At (W_1) \bigcap At(W_2) = \varnothing$, the obtained result
contradicts to the connectivity of $X$. Therefore at least one of the sets $W$ is not Lyapunov
stable. Prove that from this follows unstability of $\overline{\omega_f}$. Note that if a closed
positive-invariant  set $V$  is  union  of  two non-intersecting closed sets, $V = V_1 \bigcup V_2,\
V_1 \bigcap V_2 = \varnothing$, then $V_1$ and $V_2$ are also positive-invariant  because  of  the
connectivity   of   positive semitrajectories.  If $\overline{\omega_f}$ is stable,  then  it possess
fundamental system of closed positive-invariant neighborhoods $V_1 \supset V_2 \supset \ldots V_n
\supset \ldots $. Since $\overline{\omega_f} = W_1 \bigcup W_2,\ W_1 \bigcap W_2 = \varnothing$ and
$W_i$  are  nonempty  and closed, then from some $N$ $V_n = V'_n \bigcup V''_n = \varnothing$ for $n
\geq N$, and the families  of the sets $V'_n \supset V'_{n+1} \supset \ldots,\ V''_n \supset
V''_{n+1} \ldots$  form  fundamental  systems  of neighborhoods of $W_1$  and $W_2$  correspondingly.
So long  as $V'_n,V''_n$ are closed positive-invariant neighborhoods, from this follows stability of
both  $W_1$ and  $W_2$,  but  it  was  already  proved  that  it  is impossible. This contradiction
shows that $\overline{\omega_f}$ is  not  Lyapunov stable and completes the proof of the lemma.

\begin{theorem}\label{T3.3}
Let $X$ be connected and  $\overline{\omega_f}$  be disconnected.  Then the semiflow $f$ possesses
$\eta_3$- and     $\tau_{1,2,3}$-slow relaxations.
\end{theorem}
{\bf Proof.} The first part (the existence of  $\eta_3$-slow  relaxations) follows from Lemma
\ref{L3.3} and Theorem \ref{T3.2} (in the last as  a closed invariant set  one  should  take
$\overline{\omega_f}$). Let  us  prove  the existence  of  $\tau_1$-slow relaxations.  Let $\omega_f$
be   disconnected: $\overline{\omega_f} = W_1 \bigcup W_2,\ W_1 \bigcap W_2 = \varnothing, \ W_i \ (i
= 1,2)$ are  closed   and,   consequently, invariant due to the connectivity of  trajectories.
Consider  the sets $At(W_i)$ (\ref{e25}). Note that at least one  of  these  sets $At(W_i)$ does not
include any neighbourhood  of  $W_i$. Really,  suppose  the contrary: $At(W_i)\ (i = 1,2)$ includes a
$\varepsilon$-neighbourhood  of  $W_i$.  Let $x \in At(W_i),\ \tau = \tau_1 (x, \varepsilon / 3)$
 be the time of the first entering of the $x$-motion
into $\varepsilon / 3$-neighbourhood of the set $\omega(x) \subset W_i$. The point $x$  possesses
such neighbourhood $U \subset X$ that for any $y \in U\ \rho(f(t,x), f(t,y)) < \varepsilon / 3$ as $y
\in U,\ 0 \leq t \leq \tau$. Therefore $d(f(\tau, U), W_i) \leq 2 \varepsilon / 3,\ U \subset
At(W_i)$. Thus, $x$ lies in $At(W_i)$
 together with its neighbourhood: the sets $At(W_i)$ are  open.
This contradicts to the connectivity of $X$, since $X = At(W_1) \bigcup At (W_2)$ and $At (W_1)
\bigcap At (W_2) = \varnothing$. To  be  certain,  let $At(W_1)$  contain  none neighbourhood of
$W_1$. Then (owing to the compactness of $X$ and  the closure  of $W_1$)  there  is  a  sequence $x_i
\in At (W_2),\ x_i \rightarrow y \in W_1,\ \omega(x_i) \subset W_2,\ \omega(y) \subset W_1$. Note
that $r((x_i), \omega(y)) \geq r(W_1, W_2) >0$, therefore there are $\Omega(x)$-bifurcations ($y$  is
the  bifurcation  point)  and, consequently, (Theorem \ref{T2.1}) there  are  $\tau_1$-slow
relaxations. This yields the existence of $\tau_{2,3}$-slow relaxations $(\tau_1 \leq \tau_2 \leq
\tau_3)$.

\subsection {Slow Relaxations in Smooth Systems}\label{SS3.3}

In this subsection we  consider  the application of the approach  to  the  semiflows associated with
smooth   dynamical systems developed above. Let $M$ be  a  smooth  (of  class $C^{\infty}$)
finite-dimensional manifold, $F:\ (-\infty, \infty) \times M \rightarrow M$ be  a  smooth  dynamical
system  over  $M$, generated by vector  field  of class $C^1, X$  be  a  compact  set
positive-invariant with respect to the system $F$ (in particular, $X = M$ if $M$ is compact). The
restriction of $F$ to the set  we call the semiflow over $X$, associated with $F$, and denote it by
$F|_X$.

We shall often use the following condition: the  semiflow $F|_X$ has not non-wandering points at the
boundary of $X$ $(\partial X)$; if $X$ is positive-invariant submanifold of $M$ with smooth
boundary, $\mbox{int} X \neq \varnothing$,   then this follows, for example, from the requirement of
transversality of the vector field corresponding to the system $F$ and the boundary of $X$. All the
below results are valid, in particular, in the  case when $X$ is the whole manifold $M$  and $M$ is
compact (the  boundary  is empty).

\begin{theorem}\label{T3.4}
The complement of the set of smooth dynamical systems on compact manifold $M$ possessing the
following attribute  1, is the set of first category  (in $C^1$-topology  in  the  space  of smooth
vector fields).

\noindent{\bf Attribute  1.} Every semiflow $F|_X$  associated  with  a system $F$ on any compact
positive-invariant set $X \subset M$ without non-wandering points on $\partial X$ has not
$\eta_2$-slow relaxations.
\end{theorem}

This theorem is a direct consequence of the Pugh closing lemma, the density theorem \cite{[9],[26]},
and Theorem \ref{T3.1} of the present work.

Note  that  if $X$ is positive-invariant  submanifold  in $M$ with  smooth boundary, $\mbox{int} X
\neq \varnothing$, then  by  infinitesimal  (in  $C^1$-topology) perturbation of $F$ preserving
positive-invariance of $X$ one can obtain that semiflow over $X$, associated with the perturbed
system,  would  not have non-wandering points on $\partial X$. This  can  be  easily  proved  by
standard in differential topology reasonings about transversality. In  the  present  case  the
transversality  of  vector  field  of ``velocities" $F$ to the boundary of $X$ is meant.

The structural stable systems  over  compact  two-dimensional manifolds are studied much better than
in general case  \cite{[62],[63]}. They possess a number of characteristics which do  not  remain  in
higher  dimensions.  In  particular,   for   them   the   set   of non-wandering points consists of a
finite number of  limit  cycles and fixed points, and  the  ``loops"  (trajectories  whose $\alpha$-
and $\omega$-limit  sets  intersect,  but  do  not  contain  points  of   the trajectory itself) are
absent. Slow relaxations in  these  systems also are different from the relaxations  in the  case  of
higher dimensions.

\begin{theorem}\label{T3.5}
Let $M$ be $C^{\infty}$-smooth compact manifold, $\dim M =2,\ F$ be a structural  stable smooth
dynamical system over $M$, $F|_X$ be  an  associated  with $M$  semiflow   over   connected   compact
positive-invariant subset $X \subset M$. Then:
\begin{enumerate}
\item{For $F|_X$  the existence of  $\tau_3$-slow relaxations is equivalent to the existence of
$\tau_{1,2}$- and  $\eta_3$-slow relaxations;}
\item{$F|_X$  does not possess  $\tau_3$-slow
relaxations if  and  only  if $\omega_F \bigcap X$
 consists of one fixed point or of points of one limit cycle;}
\item{$\eta_{1,2}$-slow relaxations are impossible for $F|_X$.}
\end{enumerate}
\end{theorem}
{\bf Proof.} To prove the part 3, it is sufficient to refer to  Theorem \ref{T3.1} and Proposition
\ref{P2.2}. Let us prove  the  first  and the second parts. Note that $\omega_{F|_X} = \omega_F
\bigcap X$. Let $\omega_F \bigcap X$  consist  of  one fixed point or of points of one limit cycle.
Then $\omega(x) = X \bigcap \omega_F$  for any $x \in X$. Also there are not such $x \in X$ that
$x$-motion would be  whole but  not   $P^+$-stable  and $\alpha(x) \bigcap \omega (x) \neq
\varnothing$ (owing  to  the  structural stability).  Therefore  (Theorem  \ref{T2.7}) $\tau_3$-slow
relaxations   are impossible. Suppose now that $\omega_F \bigcap X$ includes  at  least  two  limit
cycles or a cycle and a fixed point or two fixed points. Then $\omega_{F|_X}$ is disconnected, and
using Theorem \ref{T3.3}  we  obtain  that $F|_X$ possesses $\eta_3$-slow  relaxations. Consequently,
exist  $\tau_3$-slow relaxations. From Corollary \ref{C2.1},  i.e.  the  fact  that every trajectory
from $\omega_F$ is a fixed point or a limit cycle and also from the fact that rough two-dimensional
systems  have  no  loops  we conclude that  $\tau_1$-slow relaxations  do  exist. Thus,  if $\omega_F
\bigcap X$  is connected, then $F|_X$ has not even  $\tau_3$-slow relaxations, and if $\omega_F
\bigcap X$ is disconnected, then there are $\eta_3$ and $\tau_{1,2,3}$-slow relaxations. Theorem
\ref{T3.5} is proved.

In general case (for structural stable systems with $\dim\ M>2$) the statement 1 of Theorem
\ref{T3.5} is not always true.  Really,  let  us consider topologically transitive $U$-flow $F$  over
the manifold $M$ \cite{[64]}; $\omega_F = M$, therefore $\eta_3 (x, \varepsilon) = 0$ for any $X \in
M,\ \varepsilon >0$. The set of limit cycles is dense in $M$. Let us choose two different  cycles
$P_1$ and $P_2$, whose stable $(P_1)$ and unstable $(P_2)$ manifolds intersect (such cycles exist,
see for example \cite{[4],[28]}). For the point $x$ of  their intersection $\omega (x) = P_1,\ \alpha
(x) = P_2$,  therefore $x$-motion  is  whole and not Poisson's positive stable, and (Lemma
\ref{L2.2}) $\tau_3$-slow relaxations exist.  And what is more, there exist $\tau_1$-slow relaxations
too.  These appears because the motion beginning at point near $P_2$ of $x$-trajectory delays near
$P_2$ before to enter small neighbourhood of  $P_1$. It  is easy to prove the  existence  of  $\Omega
(x)$-bifurcations  too.  Really, consider a sequence $t_1 \rightarrow \infty$, from the corresponding
sequence $F(t_i,x)$ choose    convergent subsequence: $F(t_j, x) \rightarrow y \in P_2, \ \omega (y)
= P_2,\ \omega (F(t_j, x)) = P_1$, i.e. there  are both $\tau_1$-slow  relaxations   and $\Omega
(x)$-bifurcations. For $A$-flows a weaker version of the statement 1 of Theorem \ref{T3.5} is valid
($A$-flow is called  a  flow satisfying S.Smale $A$-axiom \cite{[4]}, in regard to  $A$-flows  see
also  \cite{[28]}, p.106-143).

\begin{theorem}\label{T3.6}
Let $F$ be an $A$-flow over compact  manifold  $M$.  Then for any compact connected
positive-invariant $X \subset M$ which does  not  possess non-wandering points of $F|_X$   on  the
boundary  the existence  of $\tau_3$-slow  relaxations   involves   the   existence   of
$\tau_{1,2}$-slow relaxations for $F|_X$.
\end{theorem}
{\bf Proof.} Note that $\omega_{F|_X} = \omega_f \bigcap \mbox{int} X$. If $\omega_F \bigcap
\mbox{int} X$  is  disconnected, then,  according  to  Theorem \ref{T3.3}, $F|_X$  possesses
$\eta_3$- and $\tau_{1,2,3}$-slow relaxations. Let $\omega_F \bigcap \mbox{int} X$ be connected. The
case when it consists of one fixed point or of points of one limit cycle  is trivial: there are no
any slow relaxations. Let $\omega_F \bigcap \mbox{int} X$ consist of one non-trivial (being neither
point nor cycle)  basic  set  (in regard to these basic sets see \cite{[4],[28]}): $\omega_F \bigcap
\mbox{int} X = \Omega_0$.
 Since  there are no non-wandering points over
$\partial X$, then every  cycle  which  has point in $X$ lies entirely in $\mbox{int} X$. And due to
positive-invariance  of $X$, unstable manifold of such cycle lies in $X$. Let $P_1$  be  some  limit
cycle  from $X$.  Its  unstable  manifold  intersects  with  stable manifold of some other cycle $P_2
\subset X$  \cite{[4]}. This  follows  from  the existence of hyperbolic structure on  $\Omega_0$
(see also \cite{[28]},  p.110). Therefore there is such $x \in X$  that   $\omega (x) = P_2,\ \alpha
(x) = P_1$. From this follows the existence of $\tau_1$- (and   $\tau_{2,3}$-)-slow relaxations.  The
theorem is proved.

{\bf Remark}. We have  used  only  very  weak  consequence  of  the hyperbolicity of the set of
non-wandering points: the existence in any non-trivial (being neither point  nor  limit  cycle)
isolated connected  invariant  set  of  two  closed  trajectories,   stable manifold of one of which
intersects  with  unstable  manifold  of another one. It seems very likely that the systems for which
the statement of Theorem \ref{T3.6}  is  true  are  typical,  i.e.  the complement of their set in
the space of flows is a  set  of  first category (in $C^1$- topology).

\section {Slow Relaxation of Perturbed Systems}\label{S4}

\subsection {Limit Sets of $\varepsilon$-motions}\label{SS4.1}

As models of perturbed  motions  let  us take {\it $\varepsilon$-motions}. These motions are
mappings $f^{\varepsilon}:\ [0, \infty) \rightarrow X$, which during some fixed time $T$  depart from
the real motions at most at $\varepsilon$.

\begin{definition}\label{D4.1}  Let  $x \in X,\ k \in K,\ \varepsilon >0,\ T>0$.   The   mapping
$f^{\varepsilon} :\ [0, \infty) \rightarrow X$ is called $(k, \varepsilon, T)$-{\it motion} of the
point $x$  for the  system (\ref{e1}) if $f^{\varepsilon} (0) = x$ and for any $t \geq 0,\ \tau \in
[0,T]$
\begin{equation}\label{e26}
\rho(f^{\varepsilon} (t+ \tau),\ f (\tau, f^{\varepsilon} (t), k)) <
\varepsilon.
\end{equation}
\end{definition}

We call $(k, \varepsilon, T)$-motion of the point $x$ $(k, x, \varepsilon, T)$-motion and  use  the
denotation $f^{\varepsilon} (t|x, k, T)$. It  is obvious  that  if $y = f^{\varepsilon} (\tau| x, k,
T)$ then   the   function  $f^* (t) = f^{\varepsilon} (t + \tau| x, k, T)$ is $(k, y,
\varepsilon)$-motion.

The condition (\ref{e26}) is fundamental in  study  of  motion  with constantly functioning
perturbations.  Different restrictions  on the value of perturbations of  the right  parts  of
differential equations (uniform restriction, restriction at the average  etc.,  see \cite{[65]},
p.184 and further) are  used  as  a  rule in order to obtain estimations analogous to (\ref{e26}). On
the base of these estimations the further study is performed.

Let us introduce two auxiliary functions:
\begin{equation}\label{e27}
\varepsilon (\delta, t_0) = \sup \{ \rho (f(t,x,k), f(t,x',k')) \ | \ 0 \leq t \leq t_0, \ \rho(x,
x') < \delta,\ \rho_K (k, k') < \delta \};
\end{equation}
\begin{equation}\label{e28}
\delta (\varepsilon, t_0) = \sup \{\delta \geq 0 \ | \ \varepsilon (\delta, t_0) \leq \varepsilon \}.
\end{equation}

     Due to the compactness of $X$  and $K$ the following statement  is
true.

\begin{proposition}\label{P4.1} A) For any $\delta >0$   and $t_0 >0$ the function $\varepsilon
(\delta, t_0)$ is  defined;  $\varepsilon (\delta, t_0) \rightarrow 0$ as $\delta \rightarrow 0$
uniformly  over any  compact segment $t_0 \in [t_1, t_2]$.\\ B) For any $\varepsilon
>0$ and $t_0 >0$ the function $\delta (\varepsilon, t_0) > 0$ is defined (is finite).
\end{proposition}
{\bf Proof.} A)  Let  $\delta >0,\ t_0 >0$.  Finiteness  of $\varepsilon (\delta, t_0)$ follows
immediately from the compactness of $x$. Let $\delta_i > 0,\ \delta_i \rightarrow 0$.  Let  us prove
that $\varepsilon (\delta_i, t_0) \rightarrow 0$. Suppose the contrary. In this case one  can choose
in $\{ \delta_i \}$ such subsequence that  corresponding $\varepsilon (\delta_i, t_0)$  are separated
from zero by common constant: $\varepsilon (\delta_i, t_0) > \alpha > 0$. Let us  turn to this
subsequence, preserving the same denotations. For every $i$ there are such $t_i, x_i, x'_i, k_i,
k'_i$ that $0 \leq t_i \leq t_0,\ \rho (x_i, x'_i) < \delta_i,\ \rho_K(k_i, k'_i) < \delta_i$ and
$\rho(f(t_i, x_i, k_i), f(t_i,x'_i, k'_i)) > \alpha >0$.   The   product $[0,t_0] \times X \times X
\times K \times K$ is compact. Therefore from the sequence $(t_i, x_i, x'_i, k_i, k'_i)$
 one can choose a convergent subsequence.  Let  us
turn to it preserving the denotations: $(t_i, x_i, x'_i, k_i, k'_i) \rightarrow (\tilde t, x_0, x'_0,
k_0, k'_0)$. It is obvious that $\rho(x_0, x'_0) = \rho_K(k_0, k'_0) = \rho_K (k_0, k'_0) = 0$,
therefore $x_0 = x'_0, \ k_0 = k'_0$. Consequently, $f(\tilde t, x_0, k_0) = f(t,x'_0, k'_0)$. On the
other hand, $\rho(f(t_i, x_i, k_i), f(t_i,x'_i, k'_i)) > \alpha >0$, therefore $\rho(f(\tilde t, x_0,
k_0), f(\tilde t, x'_0, k'_0)) \geq \alpha >0$ and $f(\tilde t, x_0, k_0) \neq f(\tilde t, x'_0,
k'_0)$. The  obtained  contradiction  proves  that $\varepsilon (\delta_i, t_0) \rightarrow 0$.

     The uniformity of tending to $0$ follows from the fact that for
any $t_1, t_2 > 0,\ t_1 < t_2$ the inequality $\varepsilon (\delta, t_1)
\leq \varepsilon (\delta, t_2)$ is true, $\varepsilon (\delta, t)$
is a monotone function.

The statement of the point B) follows from the point A).

The following estimations of divergence of  the  trajectories are true. Let $f^{\varepsilon}
(t|x,k,T)$ be $(k,x, \varepsilon, T)$-motion. Then\footnote{Here and further trivial verifications
that follow directly from the triangle inequality are omitted.}
\begin{equation}\label{e29}
\rho(f^{\varepsilon} (t|x, k, T), f(t,x,k)) \leq \chi (\varepsilon, t, T),
\end{equation}
where $\chi (\varepsilon, t, T) = \sum^{[t/T]}_{i=0} \varkappa_i,\ \varkappa_0 = \varepsilon,
\varkappa_i = \varepsilon(\varkappa_{i-1}, T) + \varepsilon$.

     Let $f^{\varepsilon_1} (t|x_1, k_1, T),\ f^{\varepsilon_2} (t| x_2,
k_2, T)$ be   correspondingly $(k_1, x_1, \varepsilon_1, T)$- and
$(k_2, x_2, \varepsilon_2, T)$-motion.
Then
\begin{eqnarray}\label{e30}
\lefteqn{\rho(f^{\varepsilon_1} (t|x_1,k_1,T),
f^{\varepsilon_2}(t|x_2,k_2,T)) \leq} \nonumber \\
& & \leq \varepsilon (\max \{ \rho(x_1, x_2, , \rho_K (k_1, k_2) \}, T) +
\chi (\varepsilon_1, t, T) + \chi (\varepsilon_2, t, T).
\end{eqnarray}
From Proposition  \ref{P4.1} it follows  that $\chi (\varepsilon, t, T) \rightarrow 0$ as
$\varepsilon \rightarrow 0$ uniformly over any compact segment $t \in [t_1, t_2]$.

     Let $T_2 > T_1 > 0,\ \varepsilon > 0$. Then   any
$(k,x, \varepsilon, T_2)$-motion    is $(k,x, \varepsilon, T_1)$-motion,   and   any $(k,x,
\varepsilon, T_1)$-motion      is $(k,x, \chi (\varepsilon, T_2, T_1), T_2)$-motion. Since  we are
interested   in perturbed motions behavior at $\varepsilon \rightarrow 0$, and $\chi (\varepsilon,
T_2, T_1) \rightarrow 0$ as $\varepsilon \rightarrow 0$, then the choice of $T$ is unimportant.
Therefore let us  fix some $T > 0$ and omit references to  it  in  formulas $((k,x,
\varepsilon)$-motion instead of $(k, x, \varepsilon, T)$-motion and $f^{\varepsilon} (t| x,k)$
instead of $f^{\varepsilon} (t|x,k,T))$.

The  following  propositions   allows us   ``to   glue   together" $\varepsilon$-motions.
\begin{proposition}\label{P4.2}
Let $\varepsilon_1, \varepsilon_2 > 0,\ f^{\varepsilon_1} (t| x,k)$ be
$\varepsilon_1$-motion, $\tau > 0$, $f^{\varepsilon_2} (t| f^{\varepsilon_1}
(\tau | x, k), k)$
be  $\varepsilon_2$-motion.
Then the mapping
$$
f^* (t) =
\left\{ \begin{array}{ll}
f^{\varepsilon_1} (t| x,k), & \mbox{if}\ 0 \leq t \leq \tau;\\
f^{\varepsilon_2} (t-\tau | f^{\varepsilon_1} (\tau | x, k), k), &
\mbox{if}\ t \geq \tau,
\end{array}
\right. $$ is $(k,x, 2 \varepsilon_1 + \varepsilon_2)$-motion.
\end{proposition}
\begin{proposition}\label{P4.3}
Let $\delta, \varepsilon_1, \varepsilon_2 > 0,\ f^{\varepsilon_1} (t|x,k)$
be  $\varepsilon_1$-motion, $\tau > 0$, $f^{\varepsilon_2} (t | y, k')$
 be  $\varepsilon_2$-motion, $\rho_K(k,k')< \delta$,
$\rho(y, f^{\varepsilon_1} (\tau | x,k)) < \delta$. Then  the mapping $$ f^* (t) = \left\{
\begin{array}{ll} f^{\varepsilon_1} (t|x,k), & \mbox{if}\ 0 \leq t < \tau;\\ f^{\varepsilon_2} (t-
\tau|y, k), & \mbox{if}\ t \geq \tau,
\end{array}
\right.
$$
is $(k,x, 2 \varepsilon_1 + \varepsilon_2 + \varepsilon (\delta, T))$-motion.
\end{proposition}
\begin{proposition}\label{P4.4} Let $\delta_j, \varepsilon_j > 0$, the numbers
$\varepsilon_j, \delta_j$  are bounded above, $x_j \in X$, $k_j \in K$, $k^* \in K$, $\tau_0 > T$, $j
= 0, 1, 2, \ldots,\ i = 1, 2, \ldots$, $f^{\varepsilon_j} (t|x_j, k_j)$   be the
$\varepsilon_j$-motions, $\rho (f^{\varepsilon_j} (\tau_j | x_j,k_j), x_{j+1}) < \delta_j,\ \rho_K
(k_j, k^*) < \delta_j / 2$.
 Then the mapping
$$
f^* (t) = \left\{ \begin{array}{ll}
f^{\varepsilon_0} (t | x_0, k_0), & \mbox{if}\ 0 \leq t < \tau_0;\\
f^{\varepsilon_j} \left( t- \sum^{i-1}_{j=0} \tau_j | x_j, k_j \right), &
\mbox{if}\ \sum^{i-1}_{j=0} \tau_j \leq t < \sum^i_{j=0} \tau_j,
\end{array}
\right. $$ is $(k^*, x_0, \beta)$-motion, where $$ \beta = \sup_{0 \leq j < \infty} \{
\varepsilon_{j+1} + \varepsilon (\varepsilon_j + \delta_j + \delta_{j+1} + \varepsilon (\delta_j,
T),T) \}. $$
\end{proposition}

The proof of the propositions \ref{P4.2}-\ref{P4.4} follows  directly  from the definitions.

\begin{proposition}\label{P4.5} Let $x_i \in X,\ k_i \in K,\ k_i \rightarrow k^*,\ \varepsilon_i > 0,\
\varepsilon_i \rightarrow 0,\ f^{\varepsilon_i} (t | x_i, k_i)$ be $(k_i, x_i,
\varepsilon_i)$-motions, $t_i >0, \ t_i > t_0,\ f^{\varepsilon_i} (t_i| x_i, k_i) \rightarrow x^*$.
Then $(k^*,x^*)$-motion  is  defined  over  the segment $[-t_0, \infty)$  and $f^{\varepsilon_i} (t_0
+ t| x_i, k_i)$
 tends to $f(t, x^*, k^*)$  uniformly  over  any  compact
segment from $[-t_0, \infty)$.
\end{proposition}
{\bf Proof.} Let us choose from  the  sequence $\{x_i \}$   a  convergent subsequence  (preserving
the  denotations): $x_i \rightarrow x_0$. Note   that $f^{\varepsilon_i} (t|x_i, k_i) \rightarrow
f(t, x_0, k^*)$
  uniformly  over   any   compact   segment $t \in [t_1,\ t_2] \subset
[0, \infty)$; this follows from the estimations  (\ref{e30})  and  Proposition   \ref{P4.1}.
Particularly, $f(t_0, x_0, k^*) = x^*$. Using   the injectivity of $f$, we obtain that $x_0$ is a
unique limit point of the sequence    $\{ x_i \}$,    therefore $f^{\varepsilon_i} (t_0 + t| y_i,
k_i)$   tends to $f (t, x^*, k^*) = f(t_0 + t,x_0, k^*)$
  uniformly  over  any   compact   segment
$t \in [t_1, t_2] \subset [ -t_0, \infty)$.

\begin{proposition}\label{P4.6} Let  $x_i \in X,\ k_i \in K, k_i \rightarrow K^*,\ \varepsilon_i > 0,\
f^{\varepsilon_i} (t | x_i, k_i)$ be $(k_i, x_i, \varepsilon_i)$-motions, $t_i > 0,\ t_i \rightarrow
\infty,\ f^{\varepsilon_i} (t_i | x_i, k_i) \rightarrow x^*$.    Then $(k^*, x^*)$-motion is whole
and the sequence $f^{\varepsilon_i} (t+ t_i | x_i, k_i)$ defined for $t>t_0$  for any  $t_0$, from
some $i(t_0)$ (for $i \geq i(t_0)$) tends to $f(t, x^*, k^*)$ uniformly over any compact segment.
\end{proposition}
{\bf Proof.} Let $t_0 \in (-\infty, \infty)$. From some $i_0$ $t_i > -t_0$.  Let  us  consider the
sequence  of  $(k_i, f^{\varepsilon_i} (t_i + t_0 |x_i, k_i), \varepsilon_i)$-motions:
$f^{\varepsilon_i} (t | f^{\varepsilon_i} (t_i + t_0 | x_i, k_i), k_i) \stackrel{\mbox{def}}{=}
f^{\varepsilon_i} (t+t_i + t_0 | x_i, k_i)$.

Applying to the sequence the precedent proposition, we obtain
the required statement (due to the arbitrariness of $t_0$).

\begin{definition}\label{D4.2}   Let $x \in X,\ k \in K,\ \varepsilon > 0,\ f^{\varepsilon} (t | x, k)$  be
$(k,x, \varepsilon)$-motion. Let us call $y \in X\ \omega$-{\it limit point} of this
$\varepsilon$-motion, if there is such a sequence $t_i \rightarrow \infty$ that $f^{\varepsilon} (t_i
| x, k) \rightarrow y$. We denote  the  set of all $\omega$-limit points of $f^{\varepsilon} (t | x,
k)$ by $\omega( f^{\varepsilon} (t| x, k))$, the set of all $\omega$-limit  points  of  all  $(k,x,
\varepsilon)$-motions under  fixed $k,x, \varepsilon$  by $\omega^{\varepsilon} (x,k)$, and $$
\omega^0 (x,k) \stackrel{\mbox{def}}{=} \bigcap_{\varepsilon>0} \omega^{\varepsilon} (x, k). $$
\end{definition}

\begin{proposition}\label{P4.7}
For any $\varepsilon > 0$, $\gamma > 0$, $x \in X$, $k \in K$
$$
\overline{\omega^{\varepsilon} (x, k)} \subset
\omega^{\varepsilon + \gamma} (x,k).
$$
\end{proposition}
{\bf Proof.} Let $y \in \overline{\omega^{\varepsilon} (x,k)}$. For any $\delta > 0$  there  are
$(k,x,\varepsilon)$-motion $f^{\varepsilon} (t | x, k)$
 and subsequence $t_i \rightarrow \infty$, for which
$\rho( f^{\varepsilon} (t_i | x, k), y) < \delta$.  Let $\delta = \frac{1}{2} \delta (\gamma, T)$. As
$(k,x, \varepsilon + \gamma)$-motion let us choose $$ f^* (t) = \left\{ \begin{array}{ll} f^* (t | x,
k), & \mbox{if}\ t \neq t_i;\\ y, & \mbox{if}\ t = t_i (i = 1,2, \ldots, t_{i+1} - t_i >T).
\end{array}
\right. $$ $y$ is the $\omega$-limit point of $ f^* (t)$, therefore, $y \in \omega^{\varepsilon +
\gamma} (x,k)$.

\begin{proposition}\label{P4.8}
For any $x \in X,\ k \in K$ the set $\omega_0 (x,k)$  is  closed
and $k$-invariant.
\end{proposition}
{\bf Proof.} From  Proposition   \ref{P4.7} it  follows
\begin{equation}\label{e31}
\omega^0 (x,k) = \bigcap_{\varepsilon > 0}
\overline{\omega^{\varepsilon} (x,k)}.
\end{equation}
Therefore $\omega^0 (x,k)$ is closed. Let us prove that it is $k$-invariant. Let $y \in \omega^0
(x,k)$. Then there are such sequences $\varepsilon_j > 0,\ \varepsilon_j \rightarrow 0, \ t_i^j
\rightarrow \infty \ \mbox{as} \   i \rightarrow \infty \ (j=1,2, \ldots)$ and   such   family   of
$(k,x, \varepsilon_j)$-motions $f^{\varepsilon_j} (t | x, k)$  that $f^{\varepsilon_j} (t^j_i | x,
k)\rightarrow y \ \mbox{as} \ i \rightarrow \infty$ for any $j = 1, 2, \ldots$.  From  Proposition
\ref{P4.6} it follows that $(k,y)$-motion is whole. Let $z = f(t_0, y, k)$. Let  us  show that $z \in
\omega^0 (x,k)$. Let $\gamma > 0$. Construct $(k,x, \gamma)$-motion which has $z$ as its
$\omega$-limit  point.

Let  $t_0 > 0$.  Find  such $\delta_0 > 0, \ \varepsilon_0 > 0$    that $\chi (\varepsilon_0, t_0 +
T, T) + \varepsilon (\delta_0, t_0 + T) < \gamma / 2$ (this is  possible  according  to Proposition
\ref{P4.1}). Let us take $\varepsilon_j < \varepsilon_0$ and choose from  the  sequence $t^j_i\ (i =
1, 2, \ldots)$
 such monotone subsequence $t_l\ (l = 1, 2, \ldots)$ for  which
$t_{l+1} - t_l > t_0 +T$ and $\rho (f^{\varepsilon_j} (t_l | x, k), y) < \delta_0$. Let $$ f^* (t) =
\left\{ \begin{array}{ll} f^{\varepsilon_j} (t| x, k), & \mbox{if}\ t \not \in [t_l, t_l+t_0] \
\mbox{for any}\ l = 1, 2, \ldots; \\ f(t-t_l, y, k), & \mbox{if} \ t \in [t_l, t_l + t_0] \ (l=1,2,
\ldots).
\end{array}
\right. $$  $z$ is the $\omega$-limit point of this $(k,x,\gamma)$-motion.

If $t_0 < 0$, then at  first  it  is  necessary  to  estimate  the divergence of the  trajectories
for ``backward  motion". Let $\delta > 0$. Let us denote
\begin{equation}\label{e32}
\tilde{\varepsilon} (\delta, t_0, k) = \sup \left\{ \rho(x, x') \left|
\inf_{0 \leq t \leq -t_0} \right. \{ \rho(f (t, x, k), f(t, x', k) \}
\leq \delta \right\}.
\end{equation}

\begin{lemma}\label{L4.1}For any  $\delta > 0,\ t_0 < 0$ and
$k \in K\ \tilde{\varepsilon} (\delta, t_0, k)$   is  defined (finite). $\tilde{\varepsilon} (\delta,
t_0, k) \rightarrow 0$ as $\delta \rightarrow 0$ uniformly by $k \in K$ and by $t_0$   from  any
compact segment $[t_1, t_2] \subset (-\infty, 0]$.
\end{lemma}
The proof can be easily  obtained  from  the  injectivity  of
$f(t, \cdot, k)$
 and compactness of $X, K$  (similarly  to  Proposition
\ref{P4.1}).

Let us return to the proof of Proposition \ref{P4.8}. Let $t_0 < 0$. Find        such $\varepsilon_0
> 0$ and $\delta_0 > 0$   that $\tilde{\varepsilon} (\chi (\varepsilon_0, T - t_0, T), t_0, K) +
\tilde{\varepsilon} (\delta_0, t_0 - T, k) < \gamma / 2$. According to Proposition \ref{P4.1} and
Lemma \ref{L4.1} this is possible. Let  us  take $\varepsilon_j < \varepsilon_0$
 and choose from the sequence  $t^j_i \ (i=1,2, \ldots)$
such  monotone subsequence $t_l \ (l = 1, 2, \ldots)$ that $t_l > -t_0,\ \rho(f^{\varepsilon_i} (t_l
| x, k), y) < \delta_0$  and $t_{l+1} - t_l > T -t_0$. Suppose $$ f^* (t) = \left\{ \begin{array}{ll}
f^{\varepsilon_j} (t| x, k), & \mbox{if} \ t \not \in [t_l + t_0,t_l] \ \mbox{for\ any}\ l = 1, 2,
\ldots; \\ f(t-t_l, y, k), & \mbox{if} \ t \in [t_l + t_0, t_l] \ (l = 1, 2, \ldots).
\end{array}
\right. $$ where $f^* (t)$ is $(k,x,\gamma)$-motion, and $z$ is the $\omega$-limit point of this
motion.

Thus, $z \in \omega^{\gamma} (x, k)$ for any $\gamma > 0$. The proposition is proved.

\begin{proposition}\label{P4.9} Let $x \in \omega^0 (x,k)$. Then for any $\varepsilon > 0$ there exists
periodical $(k,x, \varepsilon)$-motion.\footnote{It is a version of $C^0$-closing lemma.}
\end{proposition}
{\bf Proof.} Let $x \in \omega^0 (x,k),\ \varepsilon > 0, \ \delta = \frac{1}{2} \delta
(\frac{\varepsilon}{2}, T)$. There is $(x,k,\delta)$-motion, and $x$  is its $\omega$-limit point: $x
\in (f^{\delta} (t), x,k)$. There  is  such $t_0 > T$ that $\rho(f^{\delta} (t_0 | x, k), x) <
\delta$. Suppose $$ f^* (t) = \left\{ \begin{array}{ll} x, & \mbox{if}\ t = nt_0,\ n = 0,1,2,
\ldots;\\ f^{\delta} (t-nt_0|x,k), & \mbox{if}\ nt < t < (n+1) t_0.
\end{array}
\right. $$ Here $f^* (t)$ is a periodical $(k,x, \varepsilon)$-motion with the period $t_0$.

Thus, if $x \in \omega^0(x,k)$, then $(k,x)$-motion possesses the property of chain recurrence
\cite{[57]}. The inverse statement is also true:  if for any $\varepsilon > 0$ there is a periodical
$(k,x,\varepsilon)$-motion,  then $x \in \omega^0 (x,k)$ (this is obvious).

\begin{proposition}\label{P4.10}  Let $x_i \in X,\ k_i \in K,\
k_i \rightarrow k^*,\ \varepsilon_i > 0,\ \varepsilon_i \rightarrow 0,\
f^{\varepsilon_i} (t | x_i, k_i)$
 be $(k_i, x_i, \varepsilon_i)$-motion, $y_i \in \omega
(f^{\varepsilon_i} (t | x_i, k_i)),\ y_i \rightarrow y^*$. Then $y^* \in \omega^0 (y,k)$.
 If in addition $x_i \rightarrow x^*$ then $y^* \in \omega^0 (x,k)$.
\end{proposition}
{\bf Proof.} Let $\varepsilon > 0$ and $\delta = \frac{1}{2} \delta ( \frac{\varepsilon}{2}, T)$. It
is possible to find such $i$ that $\varepsilon_i < \delta / 2,\ \rho_K (k_i, k^*) < \delta$, and
$\rho(f^{\varepsilon_i} (t_j | x_i, k_i), y^*) < \delta$ for  some  monotone  sequence $t_j
\rightarrow \infty,\ t_{j+1} - t_j > T$. Suppose $$ f^* (t) = \left\{ \begin{array}{ll} y^*, &
\mbox{if}\ t = t_j - t_1 \ (j = 1, 2, \ldots);\\ f^{\varepsilon_i} (t + t_1 | x_i, k_i), &
\mbox{otherwise},
\end{array}
\right. $$ where $f^*(t)$ is $(k^*, y^*, \varepsilon)$-motion, $y^* \in \omega (f^*)$. Since
$\varepsilon >0$   was  chosen arbitrarily, $y^* \in \omega^0 (y^*, k^*)$. Suppose now that $x_i
\rightarrow x^*$  and let us  show that $y^* \in \omega^0 (x^*, k^*)$. Let $\varepsilon >0, \ \delta
= \frac{1}{2} \delta (\frac{\varepsilon}{2}, T)$. Find  such $i$  for  which $\varepsilon_{\tau} <
\delta/2, \ \rho(x_i, x^*) < \delta, \ \rho_K (k_i, k^*) < \delta$
 and  there  is  such   monotone
subsequence $t_j \rightarrow \infty$ that $t_1 >T, \ t_{j+1} - t_j > T;\ \rho (f^{\varepsilon_i} (t_j
| x_i, k_i), y^*) < \delta$. Suppose $$ f^* (t) = \left\{
\begin{array}{ll}
x^*, & \mbox{if}\ t = 0;\\
y^*, & \mbox{if}\ t = t_j(j = 1, 2, \ldots);\\
f^{\varepsilon_i} (t|x_i, k_i), & \mbox{otherwise,}
\end{array}
\right. $$ where  $f^* (t)$  is $(k^*, x^*, \varepsilon)$-motion  and $y^* \in \omega (f^*)$.
Consequently, $y \in \omega^0 (x^*, k^*)$.

\begin{corollary}\label{C4.1} If $x \in X,\ k \in K,\ y^* \in \omega^0 (x,k)$
then $y^* \in \omega^0 (y^*, k)$.
\end{corollary}
\begin{corollary}\label{C4.2} Function $\omega^0 (x,k)$ is upper semicontinuous in $X \times K$.
\end{corollary}
\begin{corollary}\label{C4.3} For any $k \in K$
\begin{equation}\label{e33}
\omega^0 (k) \stackrel{\mbox{def}}{=} \bigcup_{x \in X} \omega^0 (x,k) =
\bigcup_{x \in X} \bigcap_{\varepsilon>0} \omega^{\varepsilon} (x,k) =
\bigcap_{\varepsilon>0} \bigcup_{x \in X} \omega^{\varepsilon} (x,k).
\end{equation}
\end{corollary}
{\bf Proof}. Inclusion $\bigcup_{x \in X}\ \bigcap_{\varepsilon >0} \omega^{\varepsilon} (x, k)
\subset \bigcap_{\varepsilon >0} \bigcup_{x \in X} \omega^{\varepsilon} (x,k)$  is obvious. To prove
the equality, let us take arbitrary element $y$ of  the  right part of this inclusion. For any
natural $n$ there is such $x_n \in X$ that $y \in \omega^{1/n} (x_n, k)$. Using   Proposition
\ref{P4.10},   we    obtain $y \in \omega^0 (y, k) \subset \bigcup_{x \in X} \ \bigcap_{\varepsilon
>0} \omega^{\varepsilon} (x,k)$, and this proves the corollary.

\begin{corollary}\label{C4.4} For any $k \in K$ the  set $\omega^0 (k)$  is  closed  and $k$-invariant, and
the function $\omega^0 (k)$ is upper semicontinuous in $K$.
\end{corollary}
{\bf Proof.} $k$-invariance of $\omega^0 (k)$ follows from the $k$-invariance of $\omega^0 (x,k)$ for
any $x \in X,\ k \in K$ (Proposition  \ref{P4.8}),  closure  and semicontinuity follow from
Proposition \ref{P4.10}.

     Note that the statements analogous to  the  corollaries  4.2.
and 4.4 are incorrect for the true limit sets $\omega (x,k)$ and $\omega (k)$.

\begin{proposition}\label{P4.11}
Let $k \in K, \ Q \subset \omega^0 (k)$ and $Q$ be  connected.  Then
$Q \subset \omega^0 (y,k)$ for any $y \in Q$.
\end{proposition}
{\bf Proof.} Let $y_1, y_2 \in Q,\ \varepsilon >0$. Construct a periodical $\varepsilon$-motion which
passes through the points $y_1, y_2$. Suppose $\delta = \frac{1}{2} \delta (\frac{\varepsilon}{2},
T)$. With $Q$ being connected, there  is  such  finite  set $\{x_1, \ldots, x_n \} \subset Q$  that
$x_1 = y_1,\ x_n = y_2$ and  $\rho (x_i, x_{i+1}) < \frac{1}{2} \delta\ (i=1, \ldots, n-1)$ and for
every $i = 1, \ldots,n$ there is  a  periodical  $(k, x_i, \delta/2)$-motion $f^{\delta/2} (t |x_i,
k)$ (see Proposition \ref{P4.9} and Corollary \ref{C4.1}). Let us choose for every $i = 1, \ldots, n$
such $T_i > T$ that $f^{\delta/2} (T_i | x_i, k) = x_i$.
   Construct  a periodical $(k,y_1, \varepsilon)$-motion
passing through the  points $x_1, \ldots, x_n$
with the period $T_0 = 2 \sum^n_{t=1} T_i - T_1 - T_n$: let
$0 \leq t \leq T_0$, suppose
$$
f^* (t) = \left\{ \begin{array}{ll}
f^{\delta/2} (t x_1, k), & \mbox{if}\ 0 \leq t < T;\\
f^{\delta/2} \left(t- \sum^{j-1}_{i=1} T_i| x_i, k \right), & \mbox{if}\
\sum^{j-1}_{i=1} T_i \leq t < \sum^j_{i=1} T_i \ (j=2, \ldots, n);\\
f^{\delta/2} \left( t- \sum^{n-1}_{i=1} T_i + \sum^n_{i = j+1}
T_i | x_j, k \right), & \mbox{if} \ \sum^{n-1}_{i=1} T_i +
\sum^n_{i=j+1} T_i \leq t < \\
 \ & \ \ \ < \sum^{n-1}_{i=1} T_i + \sum^n_{i=j} T_i.
\end{array}
\right. $$ If   $mT_0 \leq t < (m+1) T_0$,  then $f^* (t) = f^* (t - mT_0)$. $f^* (t)$   is
periodical $(k,y_1, \varepsilon)$-motion passing  through $y_2$.  Consequently  (due  to  the
arbitrary choice of $\varepsilon > 0$), $y_2 \in \omega^0 (y,k)$ and  (due  to  the  arbitrary choice
of $y_2 \in Q)\ Q \subset \omega^0 (y_1, k)$. The proposition is proved.

\begin{definition}\label{D4.3} Let us say  that  the  system  (\ref{e1})  possesses $\omega^0 (x,k)$-
($\omega^0 (k)$-)-{\it bifurcations}, if the function $\omega^0 (x,k)$ ($\omega^0 (k)$)  is not lower
semicontinuous (i.e. $d$-continuous) in $X \times K$.  The point in which the lower semicontinuity
gets broken is called the point of (corresponding) bifurcation.
\end{definition}

\begin{proposition}\label{P4.12} If the system (\ref{e1}) possesses $\omega^0$-bifurcations, then it
possesses $\omega^0 (x,k)$-bifurcations.
\end{proposition}
{\bf Proof.} Assume that $\omega^0 (k)$-bifurcations exist. Then  there  are such $k^* \in K$ (the
point of bifurcation), $x^* \in \omega^0 (k^*)$, $\varepsilon > 0$, and  sequence $k_i \rightarrow
k$,  that $\rho^* (x^*, \omega^0 (k_i)) > \varepsilon$ for   any $i = 1, 2, \ldots$.   Note   that
$\omega^0 (x^*, k_i) \subset \omega^0 (k_i)$,   consequently, $\rho^* (x^*, \omega^0(x^*, k_i)) >
\varepsilon$  for  any $i$. However $x^* \in \omega^0 (x^*, k^*)$  (Corollary  \ref{C4.1}). Therefore
$d(\omega^0 (x^*, k^*),\ \omega^0 (x^*, k_i)) > \varepsilon,\ (x^*, k^*)$
 is the point of $\omega^0 (x,k)$-bifurcation.

\begin{proposition}\label{P4.13}
The sets of all points of discontinuity  of
the functions $\omega^0 (x,k)$ and $\omega^0 (k)$
are subsets of first  category  in $X \times K$
 and $K$ correspondingly. For each $k \in K$ the set of such
$x \in X$ that $(x,k)$
 is the point of $\omega^0 (x,k)$-bifurcation is $(k,+)$-invariant.
\end{proposition}
{\bf Proof.} The statement that the  sets  of  points  of $\omega^0 (x,k)$- and $\omega^0
(k)$-bifurcations are the sets of first category follows from the upper semicontinuity of the
functions $\omega^0(x,k)$ and $\omega^0 (k)$ and from known  theorems  about  semicontinuous
functions  (\cite{[59]}, p.78-81). Let us prove $(k,+)$-invariance. Note  that  for  any $ t > 0\ \
\omega^0 (f (t, x,k),k) = \omega^0 (x,k)$. If $(x_i, k_i) \rightarrow (x,k)$,
 then $(f (t,x_i, k_i), k_i) \rightarrow (f (t, x, k), k)$.
 Therefore,   if  $(x,k)$is   the    point    of
$\omega^0 (x,k)$-bifurcation,  then $(f (t, x,k), k)$  is  also  the  point  of $\omega^0
(x,k)$-bifurcation for any $t > 0$.

     Let $(x_0, k_0)$ be the point of $\omega^0 (x,k)$-bifurcation,
$\Gamma$ be  a  set of such $\gamma > 0$ for which there exist $x^* \in \omega^0 (x_0, k_0)$ and such
sequence $(x_i, k_i) \rightarrow (x_0, k_0)$ that $\rho^* (x^*, \omega^0 (x_i, k_i)) \geq \gamma$ for
all $i = 1, 2, \ldots$. Let  us call the number $\tilde{\gamma} = \sup \Gamma$ {\it the value of
discontinuity}  of $\omega^0 (x,k)$ in the point $(x_0, k_0)$.

\begin{proposition}\label{P4.14} Let $\gamma > 0$. The set  of  those $(x, k) \in X \times K$,  in which the
function $\omega^0 (x,k)$ is not  continuous  and  the  value  of discontinuity $\tilde{\gamma} \geq
\gamma$, is nowhere dense in $X \times K$.
\end{proposition}
The proof can be easily obtained from the upper semicontinuity of the  functions $\omega^0 (x,k)$ and
from  known  results  about semicontinuous functions (\cite{[59]}, p.78-81).

\begin{proposition}\label{P4.15} If there is such $\gamma > 0$  that for any $\varepsilon > 0$ there are
$(x,k) \in X \times K$ for which $d(\omega^{\varepsilon} (x,k), \omega^0 (x,k)) > \gamma$ then  the
system  (\ref{e1}) possesses $\omega^0 (x,k)$-bifurcations with the discontinuity $\tilde{\gamma}
\geq \gamma$.
\end{proposition}
{\bf Proof.} Let the statement of the proposition be true for  some $\gamma > 0$.
 Then there are sequences $\varepsilon_i > 0,\ \varepsilon_i \rightarrow 0$
and $(x_i, k_i) \in X \times K$,  for which $d(\omega^{\varepsilon_i} (x_i, k_i), \omega^0 (x_i,
k_i)) > \gamma$. For every $i = 1,2,\ldots$  choose  such point $y_i \in \omega^{\varepsilon_i} (x_i,
k_i)$ that $\rho^*(y_i, \omega^0 (x_i, k_i)) > \gamma$. Using the compactness of $X$  and $K$, choose
subsequence  (preserving  the  denotations)  in such a way that the new  subsequences $y_i$  and
$(x_i, k_i)$  would  be convergent: $y_i \rightarrow y_0,\ (x_i, k_i) \rightarrow (x_0, k_0)$.
According to Proposition \ref{P4.10} $\ y_0 \in \omega^0 (x_0, k_0)$. For  any  $\varkappa > 0\
\rho^* (y_0, \omega^0 (x_i, k_i)) > \gamma - \varkappa$ from  some $i = i (\varkappa)$. Therefore
$(x_0, k_0)$ is the point of $\omega^0 (x,k)$-bifurcation with the discontinuity $\tilde{\gamma} \geq
\gamma$.

\subsection {Slow Relaxations of $\varepsilon$-motions}\label{SS4.2}

Let $\varepsilon >0,\ f^{\varepsilon} (t|x, k)$ be $(k,x, \varepsilon)$-motion, $\gamma > 0$. Let us
define  the following relaxation times:
\begin{eqnarray}\label{e34}
& & \mbox{(a)} \ \ \tau^{\varepsilon}_1 (t | x, k), \gamma) =
\inf \{ t \geq 0 \  | \ \rho^*
(f^{\varepsilon} (t | x, k), \omega^{\varepsilon} (x,k)) < \gamma \};
\nonumber \\
& & \mbox{(b)} \ \ \tau^{\varepsilon}_2 (f^{\varepsilon} (t | x, k), \gamma ) =
\overline{\mbox{mes}} \{ t \geq 0 \ | \ \rho^* (f^{\varepsilon}
(t | x, k), \omega^{\varepsilon} (x,k)) \geq \gamma \};
\nonumber \\
& & \mbox{(c)} \ \ \tau^{\varepsilon}_3 (f^{\varepsilon} (t | x, k), \gamma) = \inf \{ t \geq 0
\ | \ \rho^* (f^{\varepsilon} (t'|x,k),
\omega^{\varepsilon} (x,k)) < \gamma \  \mbox{for}\ t' > t \};\\
& & \mbox{(d)} \ \ \eta^{\varepsilon}_1 (f^{\varepsilon} (t|x,k), \gamma ) = \inf \{t \geq 0 \ |\
\rho^* (t|x, k), \omega^{\varepsilon} (k)) < \gamma \};
\nonumber \\
& & \mbox{(e)} \ \ \eta^{\varepsilon}_2 (f^{\varepsilon} (t | x, k), \gamma) =
\overline{\mbox{mes}} \{ t \geq 0 \ | \ \rho^* (f^{\varepsilon} (t | x, k),
\omega^{\varepsilon} (k)) \geq \gamma \};
\nonumber \\
& & \mbox{(f)} \ \ \eta^{\varepsilon}_3 (f^{\varepsilon} (t | x, k), = \inf \{ t \geq 0 \ | \ \rho^*
(f^{\varepsilon} (t' | x, k),
\omega^{\varepsilon} (k)) < \gamma\ \mbox{for}\ t' > t\}.
\nonumber
\end{eqnarray}
     Here $\overline{\mbox{mes}} \{ \qquad \}$ is the external measure,
$\omega^{\varepsilon} (k) = \bigcup_{x \in X} \omega^{\varepsilon} (x,k)$.

There are another three important relaxation times. They  are bound up with the relaxation of a
$\varepsilon$-motion to its $\omega$-limit set. We do not consider them in this work.

\begin{proposition}\label{P4.16}
For  any $x \in X,\ k \in K, \ \varepsilon>0,\ \gamma > 0$    and $(k,x, \varepsilon)$-motion
$f^{\varepsilon} (t | x, k)$ the relaxation times (34a-f) are  defined (finite) and the inequalities
$\tau^{\varepsilon}_1 \leq \tau^{\varepsilon}_2 \leq \tau^{\varepsilon}_3,\ \eta^{\varepsilon}_1 \leq
\eta^{\varepsilon}_2 \leq \eta^{\varepsilon}_3, \ \tau^{\varepsilon}_i \geq \eta^{\varepsilon}_i\ (i
= 1, 2, 3)$ are valid.
\end{proposition}
{\bf Proof.} The validity of the inequalities is obvious due to the corresponding  inclusions
relations  between  the  sets  or  their complements from the right parts of (\ref{e34}). For the
same reason  it is   sufficient   to   prove    definiteness    (finiteness)    of
$\tau^{\varepsilon}_3 (f^{\varepsilon} (t | x, k), \gamma)$. Suppose the contrary: the set from the
right part of (34c) is empty  for  some $x \in X,\ k \in K,\ \gamma > 0$ and $(k, x,
\varepsilon)$-motion $f^{\varepsilon} (t | x, k)$.
  Then   there   is    such    sequence  $t_i \rightarrow \infty$  that
$\rho^* (f^{\varepsilon} (t_i |x,k), \omega^{\varepsilon} (x,k)) \geq \gamma$. Owing to the
compactness of $X$, from  the sequence $f^{\varepsilon} (t_i | x, k)$ can be chosen a  convergent
one.  Denote  its limit as $y$. Then $y$ satisfies the definition of $\omega$-limit  point  of $(k,x,
\varepsilon)$-motion  but  does  not  lie  in $\omega^{\varepsilon} (x,k)$.   The   obtained
contradiction    proves    the    existence    (finiteness)     of $\tau_3 (f^{\varepsilon} (t | x,
k), \gamma)$.

     In connection with the introduced relaxation times (34a-f) it
is possible to study many different  kinds  of  slow  relaxations: infiniteness of the relaxation
time for given $\varepsilon$, infiniteness  for any $\varepsilon$ small enough e.c. We  confine
ourselves to  one  variant only. The most attention will be paid to the times $\tau^{\varepsilon}_1$
and $\tau^{\varepsilon}_3$.

\begin{definition}\label{D4.4}  We  say  that  the  system  (\ref{e1})   possesses $\tau^0_i$-
($\eta^0_i$-)-{\it slow relaxations}, if there are  such  $\gamma >0$,  sequences  of numbers
$\varepsilon_j > 0,\ \varepsilon_j \rightarrow 0$, of points $(x_j, k_j) \in X \times K$, and    of
$(k_j, x_j, \varepsilon_j)$-motions $f^{\varepsilon_j} (t | x_j, k_j)$ that $\tau^{\varepsilon_j}_i
(f^{\varepsilon_j} (t | x_j, k_j), \gamma) \rightarrow \infty$ $(\eta^{\varepsilon_{j}}_i
(f^{\varepsilon_j} (t | x_j, k_j), \gamma) \rightarrow \infty)$ as $j \rightarrow \infty$.
\end{definition}

\begin{theorem}\label{T4.1}The system (\ref{e1}) possesses $\tau^0_3$-slow relaxations  if and only if it
possesses $\omega^0 (x,k)$-bifurcations.
\end{theorem}
{\bf Proof.} Suppose  that  the  system  (\ref{e1})  possesses    $\tau^3_0$-slow relaxations: there
are such $\gamma > 0$, sequences of numbers $\varepsilon_j > 0,\ \varepsilon_j \rightarrow 0$, of
points $(x_j, k_j) \in X \times K$ and of $(k_j, x_j, \varepsilon_j)$-motions  $f^{\varepsilon_j} (t
| x_j, k_j)$ that
\begin{equation}\label{e35}
\tau^{\varepsilon_j}_3
(f^{\varepsilon_j} (t | x,k), \gamma) \rightarrow \infty
\end{equation}
as $j \rightarrow \infty$.

     Using the  compactness  of $X \times K$,  choose  from  the  sequence
$(x_j, k_j)$
   a   convergent   one   (preserving   the   denotations):
$(x_j, k_j) \rightarrow (x^*, k^*)$.
 According to the  definition  of  the  relaxation
time $\tau^{\varepsilon}_3$  there is such sequence $t_j \rightarrow \infty$
that
\begin{equation}\label{e36}
\rho^* (f^{\varepsilon_j} (t_j | x_j, k_j), \omega^{\varepsilon_j}
(x_j, k_j)) \geq \gamma.
\end{equation}

     Choose  again  from $(x_j, k_j)$   a  sequence   (preserving   the
denotations) in such a manner, that the sequence $y_j = f^{\varepsilon_j} (t_j | x_j, k_j)$ would be
convergent: $y_j \rightarrow y^* \in X$. According  to  Proposition  \ref{P4.6} $(k^*, y^*)$-motion
is whole and $f^{\varepsilon_j} (t_j + t | x_j, k_j) \rightarrow f (t, y^*, k^*)$ uniformly over any
compact  segment  $t \in [ t_1, t_2]$. Two  cases  are  possible: $\omega^0 (y^*, k^*) \bigcap \alpha
(y^*, k^*) \neq \varnothing$
 or $\omega^0 (y^*, k^*) \bigcap \alpha (y^*, k^*) = \varnothing$.
We shall show that in the  first  case   there   are   $\omega^0 (x,k)$-bifurcations with   the
discontinuity $\tilde{\gamma} \geq \gamma / 2 \ (( y^*, k^*)$ is the point of bifurcation), in  the
second case there are  $\omega^0 (x,k)$-bifurcations  too $((p,k^*)$   is  the point of bifurcation,
where $p$ is any element from $\alpha (y^*, k^*))$, but the value of discontinuity can be less  than
$\gamma / 2$.  We  need  four lemmas.

\begin{lemma}\label{L4.2}Let $x \in X,\ k \in K,\  \varepsilon > 0,\
f^{\varepsilon} (t | x,k)$ be  $(k, x, k)$-motion, $t>0, y=f^{\varepsilon} (t |x,k)$.
 Then $\omega^0 (y,k) \subset \omega^{2 \varepsilon + \sigma} (x,k)$
for any $\sigma >0$.
\end{lemma}
     The proof is an obvious consequence of  the  definitions  and
Proposition \ref{P4.2}.

\begin{lemma}\label{L4.3} Let $x \in X,\ k \in K,\ t_0 > 0,\
y = f(t_0, x, k),\ , \delta > 0, \ \varepsilon = \varepsilon (\chi (\delta, t_0, T), T) + \delta$.
Then $\omega^{\delta} (x,k) \subset \omega^{\varepsilon} (y,k)$.
\end{lemma}
{\bf Proof.} Let $f^{\delta}(t|x,k)$ be $(k,x,\delta)$-motion. Then
\begin{equation}\label{e37}
f^* (t) = \left\{
\begin{array}{ll}
y, & \mbox{if}\ t =0;\\
f^{\delta} (t+t_0 |x,k), & \mbox {if}\ t > 0,
\end{array}
\right.
\end{equation}
is $(k,y, \varepsilon)$-motion, $ \omega (f^*) \subset \omega^{\varepsilon} (y,k) $, and $\omega
(f^*) = \omega (f^{\delta} (t|x,k))$.

Since $\varepsilon (\chi (\delta, t_0, T), T) \rightarrow 0$,
for $\delta \rightarrow 0$ we obtain

\begin{corollary}\label{C4.5}  Let  $x \in X,\ k \in K,\ t_0 >0, y = f(t_0, x,k)$. Then  $\omega^0 (x,k) =
\omega^0 (y,k)$.
\end{corollary}

\begin{lemma}\label{L4.4}Let $(k,y)$-motion be  whole  and
$\omega^0 (y,k) \bigcap \alpha (y,k) \neq \varnothing$. Then $y \in \omega^0 (y,k)$.
\end{lemma}
{\bf Proof}.  Let  $\varepsilon > 0,\ p \in \omega^0 (y,k) \bigcap \alpha (y,k)$.  Let  us construct
a periodical $(k,y, \varepsilon)$-motion. Suppose that $\delta = \frac{1}{2} \delta
(\frac{\varepsilon}{2}, T)$. There  is  such $t_1 > T$
 that for  some  $(k,y, \delta)$-motion  $f^{\delta} (t | y,k)\
\rho (f^{\delta} (t_1 | y,k), p) < \delta$. There is also such $t_2 < 0$ that $\rho(f(t_2, y,k), p) <
\delta$. Then it is possible to construct a periodical $(k,y, \varepsilon)$-motion, due to the
arbitrariness of $\varepsilon > 0 $ and $ y \in \omega^0 (y,k)$.

\begin{lemma}\label{L4.5}Let $y \in X,\ k \in K,\ (k,y)$-motion be whole. Then for  any
$p \in \alpha (y,k)$ $\omega^0 (p,k) \supset \alpha (y,k)$.
\end{lemma}
{\bf Proof.} Let $p \in \alpha (y,k),\ \varepsilon > 0$. Let  us  construct  a  periodical $(k,p,
\varepsilon)$-motion. Suppose that $\delta = \frac{1}{2} \delta (\varepsilon,T)$. There are two such
$t_1, t_2 < 0$ that $t_1 - t_2 > T$ and $\rho (f( t_{1, 2}, y,k), p) < \delta$.

Suppose
$$
f^* (t) = \left\{
\begin{array}{ll}
p, & \mbox{if}\ t = 0\ \mbox{or}\ t = t_1 - t_2;\\
f(t+t_2|y,k), & \mbox{if}\ 0 < t < t_1 - t_2,
\end{array}
\right. $$ where $f^* (t+n(t_1 - t_2)) = f^* (t)$.   Periodical $(k,p,\varepsilon)$-motion    is
constructed.  Since $\varepsilon > 0$   is  arbitrary, $p \in \omega^0 (p,k)$.    Using Proposition
\ref{P4.11} and the connectivity of  $\alpha(y,k)$,  we  obtain  the required: $\alpha (y,k) \subset
\omega^0 (p,k)$.

Let  us  return  to  the  proof  of  Theorem \ref{T4.1}. Note that according to Proposition
\ref{P4.15} if there are not $\omega^0 (x,k)$-bifurcations with the  discontinuity $\tilde{\gamma}
\geq \gamma / 2$, then from some  $\varepsilon_0 > 0$ (for $0 < \varepsilon \leq \varepsilon_0)\
d(\omega^{\varepsilon} (x,k), \omega^0 (x,k)) \leq \gamma / 2$   for  any $x \in X,\ k \in K$.
Suppose that the system has $ \tau_3^0$-slow relaxations and does not possess $\omega^0
(x,k)$-bifurcations with the discontinuity $\tilde{\gamma} \geq \gamma / 2$. Then  from  (\ref{e36})
it follows  that  for $0 < \varepsilon \leq \varepsilon_0$
\begin{equation}\label{e38}
p^* (f^{\varepsilon_j} (t_j | x_j, k_j), \omega^{\varepsilon} (x_j, k_j)) \geq \gamma / 2.
\end{equation}

According  to  Lemma  \ref{L4.2}  $\omega^0 (y_j, k_j) \subset \omega^{3 \varepsilon_j} (x_i, k_j)$.
Let $0 < \varkappa < \gamma / 2$. From some $j_0$ (for  $j > j_0$) $ 3 \varepsilon_j < \varepsilon_0
$ and $\rho (f^{\varepsilon_j} (t_j | x_j, k_j), y^*) < \gamma / 2 - \varkappa$. For $j
> j_0$  from (\ref{e38}) we obtain
\begin{equation}\label{e39}
\rho^* (y^*, \omega^0 (y_j, k_j)) > \varkappa.
\end{equation}
If $\omega^0 (y^*, k^*) \bigcap \alpha (y^*, k^*) \neq \varnothing$, then from (\ref{e39}) and Lemma
\ref{L4.4}  follows the  existence  of  $\omega^0(x,k)$-bifurcations  with  the discontinuity
$\tilde{\gamma} \geq \gamma / 2$.
 The obtained  contradiction  (if $\omega^0 (y^*, k^*) \bigcap \alpha
(y^*, k^*) \neq \varnothing$  and there are not $\omega^0 (x,k)$-bifurcations with the  discontinuity
$\tilde{\gamma} \geq \gamma / 2$, then  they  are)  proves   in   this   case   the   existence   of
$\omega^0 (x,k)$-bifurcations   with   the    discontinuity $\tilde{\gamma} \geq \gamma / 2$.    If
$\omega^0 (y^*, k^*) \bigcap \alpha (y^*, k^*) = \varnothing$,
 then there also exist  $\omega^0 (x,k)$-bifurcations.
Really, let  $p \in \alpha (y^*, k^*)$. Consider  such  a  sequence  $t_i \rightarrow - \infty$  that
$f(t_i, y^*, k^*) \rightarrow p$. According  to  Corollary   \ref{C4.5} $\omega^0 (f(t_i, y^*,k^*),
k^*) = \omega^0 (y^*, k^*)$, consequently,  according  to  Lemma  \ref{L4.5}, $d(\omega^0 (p,k^*),
\omega^0 (f (t_i, y^*, k^*)) \geq d(\alpha (y^*, k^*), \omega^0 (y^*, k^*)) > 0$ and there   are
$\omega^0 (x,k)$-bifurcations. The theorem is proved.

     Note that inverse to  Theorem \ref{T4.1}  is  not  true:  for
unconnected $X$ from the existence of
$\omega^0(x,k)$-bifurcations does  not
follow the existence of $\tau^0_3$-slow relaxations.

\begin{example}\label{E4.1} $(\omega^0 (x,k)$-bifurcations    without    $\tau^0_3$ -slow relaxations). Let
$X$ be a subset of  plane,  consisting  of  points with     coordinates $(\frac{1}{n}, 0)$    and
vertical      segment $J = \{ (x,y) | x=0, \ y \in [-1,1]\}$.
 Let us consider on $X$ a  trivial  dynamical
system $f(t, x) \equiv x$.  In  this  case $\omega^0_f (( \frac{1}{n}, 0 )) = \left\{ (\frac{1}{n},
0) \right\}$, $\omega^0_f ((0, y)) = J$. There are $\omega^0 (x,k)$ bifurcations: $(\frac{1}{n}, 0)
\rightarrow (0,0) \ \mbox{as} \ n \rightarrow \infty,\ \omega^0_f ((\frac{1}{n}, 0)) = \left\{
(\frac{1}{n}, 0) \right\},\ \omega^0_f ((0,0)) = J$.
   But   there   are    not    $\tau^3_0$ -slow    relaxations:
$\tau^{\varepsilon}_3 (f^{\varepsilon}(t |x), \gamma) = 0$
 for any $(x, \varepsilon)$-motion $f^{\varepsilon}(t |x)$ and
$\gamma > 0$. This  is  associated with the fact that for any $(x, \varepsilon)$-motion  and
arbitrary $t_0 \geq 0$ the following function $$ f^* (t) = \left\{ \begin{array}{ll} f^{\varepsilon}
(t|x), & \mbox{if}\ 0 \leq t \leq t_0;\\ f^{\varepsilon} (t_0 | x), & \mbox{if}\ t \geq t_0
\end{array}
\right. $$ is $(x, \varepsilon)$-motion too, consequently, each $(x, \varepsilon)$-trajectory
consists of the points of  $\omega^{\varepsilon}_f (x)$.
\end{example}

For connected  $X$  the existence  of  $\omega^0 (x,k)$-bifurcations  is equivalent to the existence
of  $\tau^0_3$-slow relaxations.

\begin{theorem}\label{T4.2}Let $X$ be connected. In this case the system  (\ref{e1}) possesses  $\tau^0_3$
-slow  relaxations  if  and  only  if  it   possesses $\omega^0 (x,k)$-bifurcations.
\end{theorem}
     One part of Theorem \ref{T4.2}  (only  if)  follows  from  the
Theorem \ref{T4.1}. Let us put off the proof of the  other  part  of  Theorem \ref{T4.2} till
Subsection \ref{SS4.4}, and  the  remained  part  of  the  present subsection devote to the study of
the set of singularities of  the relaxation time $\tau_2$  for perturbed motions.

\begin{theorem}\label{T4.3}
Let $\gamma > 0, \ \varepsilon_i > 0, \ \varepsilon_i \rightarrow 0, \ (x_i, k_i) \in X \times K,\
f^{\varepsilon_i} (t | x_i, k_i)$ be $(k_i, x_i, \varepsilon_i)$-motions, $\tau_2^{\varepsilon_i}
(f^{\varepsilon_i} (t | x_i, k_i), \gamma) \rightarrow \infty$. Then any limit point of the sequence
$\{ (x_i, k_i)\}$ is a point of $\omega^0 (x,k)$-bifurcation  with the discontinuity $\tilde{\gamma}
\geq \gamma$.
\end{theorem}
{\bf Proof.} Let $(x_0, k_0)$ be limit point of the sequence $\{ (x_i, k_i) \}$. Turning to
subsequence and  preserving  the  denotations,  let  us write down $(x_i, k_i) \rightarrow (x_0,
k_0)$. Let $X = \bigcup^n_{j=1} V_j$  be a finite open  covering of $X$. Note that
\begin{eqnarray*}
\lefteqn{\tau^{\varepsilon_i}_2 (f^{\varepsilon_i} (t | x_i, k_i) \gamma ) \leq} \\
& & \leq \sum^n_{j=1} \overline{\mbox{mes}} \{ t \geq 0 \ | \
f^{\varepsilon_j} (t | x_i, k_i) \in
V_j, \rho^* (f^{\varepsilon_i} (t |x_i, k_i), \omega^{\varepsilon_i}
(x_i, k_i)) \geq \gamma \}.
\end{eqnarray*}
Using this remark, consider a sequence of reducing coverings.  Let us find (similarly to the proof of
Theorem \ref{T3.1}) such $y_0 \in X$ and subsequence  in $\{ (x_i, k_i) \}$ (preserving  for   it the
previous denotation) that for any neighbourhood $V$ of the point $y_0$ $$ \overline{\mbox{mes}} \{ t
\geq 0 \ | \ f^{\varepsilon_i} (t | x_i, k_i) \in V, \ \rho(f^{\varepsilon_i} (t | x_i, k_i),
\omega^{\varepsilon_i} (x_i, k_i)) \geq \gamma \} \rightarrow \infty. $$ Let us show that $y_0 \in
\omega^0 (x_0, k_0)$. Let $\varepsilon > 0$. Construct such a $(k_0, x_0, \varepsilon)$-motion that
$y_0$ is its $\omega$-limit point. Suppose $\delta = \frac{1}{2} \delta (\frac{\varepsilon}{3}, T)$.
From some $i_0$  (for $i > i_0$)   the   following   inequalities   are   true: $\rho (x_i, x_0) <
\delta,\ \rho_K (k_i, k_0) < \delta,\ \varepsilon_i < \delta$, and $$ \overline{\mbox{mes}} \{ t \geq
0 \ | \ \rho (f^{\varepsilon_i} (t | x_i, k_i), y_0) < \delta,\ \rho^* (f^{\varepsilon_i} (t | x_i,
k_i),\ \omega^{\varepsilon_i} (x_i, k_i)) \geq \gamma \} > T. $$ On account of the last of these
inequalities for every $i > i_0$  there are such  $t_1, t_2 > 0$  that $t_2 - t_1 > T$ and
$\rho(f^{\varepsilon_i} (t_{1,2} | x_i, k_i), y_0) < \delta$. Let $i > i_0$. Suppose $$ f^* (t) =
\left\{ \begin{array}{ll} x_0, & \mbox{if}\ t=0;\\ f^{\varepsilon_i} (t | x_i, k_i), & \mbox{if}\ 0 <
t < t_2, \ t \neq t_1;\\ y_0, &  \mbox{if}\ t = t_1.
\end{array}
\right. $$ If   $t \geq t_1$,  then  $f^* (t+n(t_2 - t_1)) = f^* (t),\ n = 0,1, \ldots$. By  virtue
of the construction $f^*$ is $(k_0, x_0, \varepsilon)$-motion, $y_0 \in \omega (f^*)$.  Consequently
(due to the arbitrary choice of $\varepsilon >0),\ y_0 \in \omega^0 (x_0, k_0)$. Our choice  of the
point  $y_0$  guarantees  that $y_0 \not \in \omega^{\varepsilon_i} (x_i, k_i)$   from   some $i$.
Furthermore, for any $\varkappa >0$  exists  such $i = i(\varkappa)$   that  for $i > i(\varkappa)\
\rho^* (y_0, \omega^{\varepsilon_i} (x_i, k_i) > \gamma - \varkappa$.
 Consequently, $(x_0, k_0)$   is  the  point
of $\omega^0 (x,k)$-bifurcation with the discontinuity
$\tilde{\gamma} \geq \gamma$.

\begin{corollary}\label{C4.6} Let $\gamma > 0$. The set of all points $(x,k) \in X \times K$, for which there
are such sequences of numbers $\varepsilon_i > 0,\ \varepsilon \rightarrow 0$,  of  points $(x_i,
k_i) \rightarrow (x,k)$,
  and   of  $(k_i, x_i, \varepsilon_i)$-motions
$f^{\varepsilon_i} (t | x_i, k_i)$     that $\tau_2^{\varepsilon_i} (f^{\varepsilon_i} (t | x_i,
k_i), \gamma) \rightarrow \infty$, is nowhere dense in $X \times K$. The union  of  all $\gamma > 0$
these sets (for all $\gamma > 0$) is a set of first category in $X \times K$.
\end{corollary}

\subsection {Smale Order and Smale Diagram for General Dynamical Systems}\label{SS4.3}

Everywhere in this subsection one semiflow of homeomorphisms $f$ on $X$ is studied. We study here the
equivalence and preorder relations generated by semiflow.

\begin{definition}\label{D4.5} Let $x_1, x_2 \in X$. Say that points $x_1$ and $x_2$    are $f$-{\it
equivalent} (denotation  $x_1 \sim x_2$), if for any $\varepsilon > 0$   there  are  such $(x_1,
\varepsilon)$- and $(x_2, \varepsilon)$-motions $f^{\varepsilon} (t|x_1)$ and $f^{\varepsilon}
(t|x_2)$   that  for  some $t_1, t_2 > 0$ $$ f^{\varepsilon} (t_1 | x_1) = x_2,\ f^{\varepsilon} (t_2
| x_2) = x_1. $$
\end{definition}

\begin{proposition}\label{P4.17} The relation $\sim$  is  a  closed  $f$-invariant equivalence relation: the
set of pairs $(x_1, x_2)$, for which $x_1 \sim x_2$ is closed in $X \times K$; if $x_1 \sim x_2$ and
$x_1 \neq x_2$, then   $x_1$-  and   $x_2$-motions  are whole and for any $t \in (-\infty, \infty)\
f(t,x_1) \sim f (t, x_2)$. If  $x_1 \neq x_2$,  then $x_1 \sim x_2$ if and only if $\omega^0_f (x_1)
= \omega^0_f (x_2),\ x_1 \in \omega^0_f (x_1),\ x_2 \in \omega^0_f (x_2)$.\footnote{Compare  with
\cite{[52]}, Ch.6, Sec.~1, where analogous  propositions  are  proved  for equivalence relation
defined by action functional.}
\end{proposition}
{\bf Proof.} Symmetry  and  reflexivity  of  the  relation  $\sim$  are obvious. Let us prove its
transitivity.  Let $x_1 \sim x_2,\ x_2 \sim x_3, \ \varepsilon >0$. Construct $\varepsilon$-motions
which go from $x_1$  to $x_3$, and  from $x_3$  to $x_1$, gluing together $\delta$-motions, going
from $x_1$  to  $x_2$, from $x_2$  to  $x_3$  and from $x_3$  to  $x_2$, from $x_2$  to  $x_1$.
Suppose that $\delta = \varepsilon / 4$. Then,  according to Proposition \ref{P4.2}, as  a result of
the  gluing  we  obtain $\varepsilon$-motions  with  required  properties.  Therefore $x_1 \sim x_3$.

Let  us consider the closure of the relation $\sim$. Let $\varepsilon > 0,\ x_i, y_i \in X,\ x_i
\rightarrow x,\ y_i \rightarrow y,\ x_i \sim y_i$. Suppose $\delta = \frac{1}{2} \delta
(\frac{\varepsilon}{3}, T)$. There is such $i$ that $\rho (x_i, x) < \delta$ and $\rho (y_i, y) <
\delta$. Since $x_i \sim y_i$, there are binding them $\delta$-motions $f^{\delta} (t | y_i)$ and
$f^{\delta} (t | x_i):\ f^{\delta} (t_{1i} | x_i) = y_i,\ f^{\delta} (t_{2i} | y_i) = x_i$, for which
$t_{1i}, t_{2i} > 0$. Suppose that $$
 f^*_1 (t) = \left\{
\begin{array}{ll}
x, & \mbox{if}\ t=0;\\
y, & \mbox{if}\ t = t_{1i};\\
f^{\delta} (t | x_i), & \mbox{if}\ t \neq 0, t_{1i},
\end{array}
\right.
$$
$$
f^* (t) = \left\{
\begin{array}{ll}
y, & \mbox{if}\ t = 0;\\
x, & \mbox{if}\ t = t_{2i};\\
f^{\delta} (t|y_i), & \mbox{if}\ t \neq 0, t_{2i},
\end{array}
\right. $$ here $f^*_1$ and $f^*_2$  are  correspondingly $(x, \varepsilon)$- and  $(y,
\varepsilon)$-motions, $f^*_1 (t_{1i}) = y,\ f^*(t_{2i}) = x$. Since   was  chosen  arbitrarily,  it
is proved that $x \sim y$. Let  $x_{1,2} \in X,\ x_1 \sim x_2,\ x_1 \neq x_2$. Show that $\omega^0_f
(x_1) = \omega^0_f (x_2)$ and  $x_{1,2} \in \omega^0_f (x_1)$. Let $\varepsilon > 0,\ y \in
\omega^0_f (x_2)$. Prove that  $y \in \omega^{\varepsilon}_f (x_1)$. Really, let $f^{\varepsilon / 3}
(t | x_1)$ be $(x_1, \varepsilon / 3)$-motion, $f^{\varepsilon/3} (t_0 | x_1) = x_2,\ f^{\varepsilon
/3} (t | x_2)$     be $(x_2, \varepsilon / 3)$-motion, $y \in \omega (f^{\varepsilon/3}
(f^{\varepsilon / 3} (t | x_2))$. Suppose $$ f^* (t) = \left\{ \begin{array}{ll} f^{\varepsilon/3} (t
| x_1), & \mbox{if} \ 0 \leq t \leq t_0;\\ f^{\varepsilon/3} (t_0 | x_2), & \mbox{if} \ t > t_0,
\end{array}
\right. $$ here  $f^*$ is $(x_1, \varepsilon)$-motion (in  accordance  with  Proposition \ref{P4.2}),
$y \in \omega (f^*)$. Consequently, $y \in \omega^{\varepsilon} (x_1)$ and, due to arbitrary choice
of $\varepsilon > 0,\ y \in \omega^0 (x_1)$. Similarly $\omega^0 (x_1) \subset \omega^0 (x_2)$,
therefore $\omega^0 (x_1) = \omega^0 (x_2)$.  It  can  be  shown  that $x_1 \in \omega^0_f (x_1),\
x_2 \in \omega^0_f (x_2)$. According to Proposition \ref{P4.8}, the sets $\omega^0_f (x_{1,2})$ are
invariant and $x_{1,2}$-motions are whole.

Now, let us show that if $x_2 \in \omega^0_f (x_1)$  and $x_1 \in \omega^0_f (x_2)$ then $x_1 \sim
x_2$. Let $x_2 \in \omega^0_f (x_1),\ \varepsilon > 0$. Construct a  $\varepsilon$-motion  going
from $x_1$   to $x_2$. Suppose that $\delta = \delta =
 \frac{1}{2} \delta (\frac{\varepsilon}{2}, T)$.
 There is such a $(x_1, \delta)$-motion that $x_2$ is  its
$\omega$-limit point:  $f^{\delta} (t_1 | x_1) \rightarrow x_2,\ t_1 \rightarrow \infty$.  There  is
such $t_0 > 0$ that $\rho (f^{\delta} (t_0 | x_1), x_2) < \delta$. Suppose that $$ f^* (t) = \left\{
\begin{array}{ll} f^{\delta} (t | x_1), & \mbox{if}\ t \neq t_0;\\ x_2, & \mbox{if}\ t = t_0,
\end{array}
\right. $$ where $f^* (t)$   is $(x_1, \varepsilon)$-motion  and $f^* (t_0) = x_2$.   Similarly,   if
$x_1 \in \omega^0_f (x_2)$,
 then for any $\varepsilon > 0$ exists
$(x_2, \varepsilon)$-motion which  goes  from $x_2$
 to  $x_1$.
     Thus, if  $x_1 \neq x_2$, then $x_1 \sim x_2$ if  and  only  if
$x_1 \in \omega^0 (x_2)$   and $x_2 \in \omega^0_f (x_1)$. In this  case  $\omega^0_f(x_1) =
\omega^0_f(x_2)$. The  invariance  of  the relation $\sim$ follows now from the invariance of the
sets $\omega^0_f (x)$ and the  fact  that $\omega^0_f (x) = \omega^0_f (f (t,x))$ if $f(t,x)$   is
defined.   The proposition is proved.

Let us remind, that topological space is called {\it totally disconnected}
if there exist a base of topology, consisting of sets which are
simultaneously open and closed. Simple examples of such spaces are
discrete space and Cantor discontinuum.

\begin{proposition}\label{P4.18}
Factor space $\omega^0_f / \sim$ is  compact  and  totally
disconnected.
\end{proposition}
{\bf Proof.} This proposition follows  directly from Propositions \ref{P4.11}, \ref{P4.17} and
Corollary \ref{C4.4}.

\begin{definition}\label{D4.6}  (Preorder,  generated  by  semiflow). Let $x_1, x_2 \in X$. Let say
$x_1 \succsim x_2$ if for any $\varepsilon>0$ exists such a $(x_1, \varepsilon)$-motion
$f^{\varepsilon} (t | x_1)$ that $f^{\varepsilon} (t_0 | x_1) = x_2$ for some $t_0 \geq 0$.
\end{definition}

\begin{proposition}\label{P4.19} The  relation $\succsim$  is  a  closed  preorder relation on $X$.
\end{proposition}
{\bf Proof.} Transitivity of $\succsim$  easily follows from Proposition \ref{P4.2} about gluing of
$\varepsilon$-motions. The  reflexivity  is  obvious.  The closure can be proved similarly to the
proof of the  closure of $\sim$ (Proposition \ref{P4.17}, practically literal coincidence).

\begin{proposition}\label{P4.20} Let $x \in X$. Then $$ \omega^0_f (x) = \{ y \in \omega^0_f \ | \ x \succsim
y\}. $$
\end{proposition}
{\bf Proof}. Let $y \in \omega^0_f (x)$. Let us show that $x \succsim y$. Let $\varepsilon > 0$.
Construct a $\varepsilon$-motion  going  from $x$   to $y$.  Suppose $\delta = \frac{1}{2} \delta (
\frac{\varepsilon}{3}, T)$.   There   is such a $(x, \delta)$-motion $f^{\delta} (t|x)$ that $y$ is
its $\omega$-limit point: $f^{\delta} (t_j | x) \rightarrow$  for some sequence $t_j \rightarrow
\infty$. There  is  such $t_0 > 0$  that $\rho (f^{\delta} (t_0 | x), y) < \delta$. Suppose that $$
f^* (t) = \left\{ \begin{array}{ll} f^{\delta} (t | x), & \mbox{if}\ t \neq t_0;\\ y, & \mbox{if}\ t
= t_0,
\end{array}
\right. $$ here $f^* (t)$ is $(x, \varepsilon)$-motion and $f^* (t_0) = y$. Consequently, $x \succsim
y$.  Now suppose that $y \in \omega^0_f,\  x \succsim y$. Let  us  show  that  $y \in \omega^0_f
(x)$.  Let $\varepsilon > 0$. Construct such a $(x, \varepsilon)$-motion that $y$ is its
$\omega$-limit point.  To  do  this, use Proposition \ref{P4.9} and Corollary  \ref{C4.1}  and
construct a periodical $(y, \varepsilon / 3)$-motion $f^{\varepsilon / 3}:\ f^{\varepsilon / 3} (nt_0
| y) = y,\ n = 0,1, \ldots.$. Glue it together with $(x, \varepsilon / 3)$-motion going from $x$ to
$y (f^{\varepsilon / 3} (t_1 | x) = y)$: $$ f^* (t) = \left\{ \begin{array}{ll} f^{\varepsilon / 3}
(t|x), & \mbox{if}\ 0 \leq t \leq t_1;\\ f^{\varepsilon / 3} (t - t_1| y), & \mbox{if}\ t \geq t_1,
\end{array}
\right. $$ where $f^* (t)$ is $(x, \varepsilon)$-motion, $y \in \omega (f^*)$,  consequently
($\varepsilon > 0$  is arbitrary), $y \in \omega^0_f (x)$. The proposition is proved.

  We say that the set $Y \subset \omega^0_f$
is {\it saturated downwards},  if  for  any $y \in Y$ $$ \{ x \in \omega^0_f | y \succsim x \}
\subset Y. $$ It is obvious  that  every  saturated  downwards subset in $\omega^0_f$  is saturated
also for the equivalence relation $\sim$.

\begin{proposition}\label{P4.21}  Let $Y \subset \omega^0_f$  be  open  (in $\omega^0_f$)   saturated
downwards set. Then the set $At^0 (Y) = \{ x \in X \ | \ \omega^0_f (x) \subset Y \}$  is open in
$X$.
\end{proposition}
{\bf Proof.} Suppose the contrary. Let $x \in At^0 (Y),\ x_i \rightarrow x$ and for every $i = 1,2,
\ldots$ there is $y_i \in \omega^0_f (x_i) \setminus Y$. On account of the  compactness  of
$\omega^0_f \setminus Y$ there is a subsequence in $\{y_i \}$, which converges  to $y^* \in
\omega^0_f \setminus Y$. Let  us  turn  to  corresponding  subsequences  in $\{ x_i \},\ \{y_i \}$,
preserving the denotations:  $y_i \rightarrow y^*$. Let us show that $y \in \omega^0_f (x)$. Let
$\varepsilon > 0$. Construct a $\varepsilon$-motion going from $x$ to $y$. Suppose that $\delta =
\frac{1}{2} \delta ( \frac{\varepsilon}{3}, T)$. From  some $i_0\ \rho (x_i, x) < \delta$ and $\rho
(y_i, y^*) < \delta$.  Let  $i > i_0$.  There  is $(x_i, \delta)$-motion going from $x_i$ to $y_i:\
f^{\delta} (t_0 | x_i) = y_i$  (according to  Proposition \ref{P4.20}). Suppose that $$ f^* (t) =
\left\{ \begin{array}{ll} f^*, & \mbox{if}\ t = 0;\\ y^*, & \mbox{if}\ t = t_0;\\ f^{\delta} (t |
x_i), & \mbox{if}\ t \neq 0, t_0,
\end{array}
\right. $$ where $f^*$  is $(x, )$-motion  going  from  $x$  to  $y^*$. Since  $\varepsilon >0$  is
arbitrary, from this it follows  that $x \succsim y^*$  and,  according  to  Proposition \ref{P4.20},
$y^* \in \omega^0_f (x)$. The    obtained    contradiction $(y^* \in \omega^0_f (x) \setminus Y$, but
$\omega^0_f (x) subset Y$) proves the proposition.

\begin{theorem}\label{T4.4}
Let $x \in X$ be a point  of  $\omega^0_f (x)$-bifurcation.  Then there is such open in $\omega^0_f$
saturated downwards set $W$ that $x \in \partial At^0 (W)$.
\end{theorem}
{\bf Proof.} Let $x \in X$ be a point  of  $\omega^0_f$-bifurcation:  there  are such sequence $x_i
\rightarrow x$ and such $y^* \in \omega^0_f (x)$ that $\rho^* (y^*, \omega^0_f (x_i)) > \alpha > 0$
for all $i = 1, 2, \ldots$. Let us consider the set $\omega = \bigcup^{\infty}_{i=1} \omega^0_f
(x_i)$. The set  $\omega$  is saturated downwards (according to Proposition \ref{P4.20}).  We have to
prove that  it  possesses  open  (in $\omega^0_f$)  saturated  downwards neighbourhood which does not
contain  $y^*$. Beforehand let  us  prove the following lemma.

\begin{lemma}\label{L4.6}Let  $y_1, y_2 \in \omega^0_f,\ y_1 \not \in \omega^0_f (y_2)$. Then there exists
such an open saturated downwards set $Y$ that  $y_2 \in Y$,   $y_1 \notin Y$, and $Y \subset
\omega^{\varepsilon}_f (y_2)$ for some $\varepsilon > 0$.
\end{lemma}
{\bf Proof.} $\omega^0_f (y_2) = \bigcap_{\varepsilon>0} \omega^{\varepsilon} (y_2)$ (according to
Proposition  \ref{P4.7}). There are such $\varepsilon_0>0,\ \tau >0$ that if $0 < \varepsilon \leq
\varepsilon_0$ then $\rho^* (y_1, \omega^{\varepsilon} (y_2)) > \lambda$. This follows from the
compactness of $X$  and the so-called Shura-Bura lemma (\cite{[66]}, p.171-172): let a subset $V$ of
compact space  be  intersection of some family of closed sets. Then for  any  neighbourhood  of $V$
exists a finite collection of sets from that family,  intersection of which  contains  in  given
neighbourhood.  Note  now  that  if $\rho^* (z, \omega^0_f (y_2)) < \delta = \frac{1}{2} \delta
(\frac{\varepsilon}{3}, T)$
 and $z \in \omega^0_f$, then $z \in \omega^{\varepsilon}_f (y_2)$.
Really,  in  this case there are such $p \in \omega^0_f (y_2),\ (y_2, \delta)$-motion
$f^{\delta}(t|y_2)$, and  monotone sequence $t_i \rightarrow \infty$  that $\rho(z, p) < \delta,\
t_{j+1} - t_j > T$   and $\rho (f^{\delta}(t_i | y_2), p) < \delta$. Suppose $$ f^* (t) = \left\{
\begin{array}{ll} f^{\delta} (t | y_2), & \mbox{if}\ t \neq t_j;\\ z, & \mbox{if}\ t = t_j,
\end{array}
\right. $$ here $f^* (t)$  is $(y_2, \varepsilon)$-motion  and $z \in \omega (f^*) \subset
\omega^{\varepsilon} (y_2)$.   Strengthen somewhat this statement. Let $z \in \omega^0_f$  and  for
some $n>0$  exist  such chain $\{z_1, z_2, \ldots, z_n \} \in \omega^0_f$ that $y_2 = z_1,\ z = z_n$
and for any $i = 1, 2, \ldots, n-1$ either $z_i \succsim z_{i+1}$   or  $\rho (z_i, z_{i+1}) < \delta
= \frac{1}{2} \delta (\frac{\varepsilon}{7}, T)$. Then $z \in \omega^{\varepsilon} (y_2)$ and such a
$(y_2, \varepsilon)$-motion that $z$  is  its $\omega$-limit  point  is  constructed  as follows. If
$z_i \succsim z_{i+1}$, then find $(z_i, \delta)$-motion  going  from  $z_i$    to $z_{i+1}$, and for
every $i=1, \ldots,n$ find a periodical $(z_i, \delta)$-motion.  If $z_1 \succsim z_2$, then suppose
that $f^*_1$ is $(z_1, \delta)$-motion going from $z_1$  to  $z_2$, $f_1^* (t_1) = z_2$; and if $\rho
(z_1, z_2) < \delta,\ z_1 \succsim z_2$, then suppose  that  $f^*_1$   is  a periodical $(z_2,
\delta)$-motion and $t_1 > 0$ is such a number that $t_1 > T$  and $f^*_1 (t_1) = z_2$.
 Let $f^*_1, \ldots, f^*_k,\ t_1, \ldots, t_k$ be  already  determined.
Determine    $f^*_{k+1}$. Four variants are possible:

1) $f^*$  is periodical $(z_i, \delta)$-motion,
$i < n,\ z_i \succsim z_{i+1}$, then  $f^*_{k+1}$   is
$(z_i, \delta)$-motion going from $z_i$ to    $z_{i+1}$
$f^*_{k+1}(t_{k+1}) = z_{i+1}$;

2) $f^*_k$  is periodical $(z_i, \delta)$-motion,
$i < n,\ \rho(z_i,z_{i+1})< \delta$, then $f^*_{k+1}$ is
periodical $(z_{i+1}, \delta)$-motion, $f^*_{k+1} (t_{k+1} ) = z_{i+1},\
t_{k+1} > T$;

3) $f^*_k$ is $(z_i, \delta)$-motion going from $z_i$ to   $z_{i+1}$,
then  $f^*_{k+1}$ is
periodical $(z_{i+1}, \delta)$-motion, $f^*_{k+1} (t_{k+1}) =
z_{i+1}, \ t_{k+1} > T$;

4)  $f^*_k$ is periodical $(z_n, \delta)$-motion, then the  constructing  is
finished.

After constructing  the whole chain of $\delta$-motions $f^*_k$   and  time moments  $t_k$, let us
denote the number of its elements  by  $q$  and  assume that $$ f^* (t) = \left\{
\begin{array}{ll}
z_1, & \mbox{if}\ t = 0;\\ f^*_1 (t), & \mbox{if}\ 0 < t \leq t_1;\\ f^*_k \left( t- \sum^{k-1}_{j=1}
t_j \right), & \mbox{if}\ \sum^{k-1}_{j=1} t_j < t \leq \sum^k_{j=1} t_j (k<q);\\ f^*_q \left( t -
\sum^{q-1}_{j=1} t_j \right), & \mbox{if}\ t > \sum^{q-1}_{j=1} t_j.
\end{array}
\right. $$ Here $f^*(t)$ is $(y_2, \varepsilon)$-motion, and $z_n = z$ is its $\omega$-limit point.
The set  of  those $z \in \omega^0_f$ for  which  exist  such   chains $z_1, \ldots, z_n\ (n=1, 2,
\ldots)$ is an openly-closed (in $\omega^0_f$) subset  of $\omega^0_f$,  saturated downwards.
Supposing $0 < \varepsilon \leq \varepsilon_0$, we obtain  the  needed  result.  Even more strong
statement was proved: $Y$ can  be  chosen  openly-closed (in  $\omega^0_f$), not only open.

Let us  return  to  the  proof  of  Theorem \ref{T4.4}.  Since $\omega = \bigcup^{\infty}_{i=1}
\omega^0_f (x_i)$ and each $z \in \omega^0_f (x_i)$ has  an  open  (in $\omega^0_f$) saturated
downwards neighbourhood $W_z$ which does not contain $y^*$, then the  union of these neighborhoods is
an open (in $\omega^0_f$) saturated downwards set which includes $\omega$ but does not contain $y^*$.
Denote this  set  by $W$: $W = \bigcup_{z \in \omega} W_z$. Since $x_i \in At^0 (W),\ x \not \in At^0
(W)$ and $x_i \rightarrow x$, then $x \in \partial At^0 (W)$. The theorem is proved.

The following proposition will be used in Subsection \ref{SS4.4}  when  studying slow relaxations of
one perturbed system.

\begin{proposition}\label{P4.22} Let $X$ be  connected, $\omega^0_f$   be  disconnected.
Then there is such $x
\in X$ that $x$-motion is whole and $x \not \in \omega^0_f$.  There  is also such partition of
$\omega^0_f$  into openly-closed (in $\omega^0_f$)  subsets: $$ \omega^0_f = W_1 \bigcup W_2,\ W
\bigcap W_2 = \varnothing, \alpha_f (x) \subset W_1 \mbox{but} \omega^0_f (x) \subset W_2. $$
\end{proposition}
{\bf Proof.} Repeating the proof of Lemma \ref{L3.3}  (the  repetition is practically literal,
$\omega^0_f$ should be substituted instead  of $\overline{\omega_f}$), we obtain that $\omega^0_f$ is
not Lyapunov stable. Then, according to Lemma \ref{L3.2}, there is such $x \in X$ that $x$-motion is
whole and $x \not \in \omega^0_f$. Note now that the set $\alpha_f(x)$ lies in equivalence class by
the  relation $\sim$, and the set $\omega^0_f$ is saturated  by  the  relation $\sim$ (Proposition
\ref{P4.17}, Lemma  \ref{L4.5}).  $\alpha_f (x) \bigcap \omega^0_f (x) = \varnothing$, otherwise,
according to Proposition \ref{P4.17} and Lemma \ref{L4.4},  $x \in \omega^0_f$. Since $\omega^0_f /
\sim$   is totally disconnected space (Proposition \ref{P4.18}), there exists partition of it into
openly-closed subsets, one of  which  contains  image  of $\alpha_f (x)$
 and  the  other  contains  image  of  $\omega^0_f (x)$  (under  natural
projection $\omega^0_f \rightarrow \omega^0_f / \sim$). Prototypes of these openly-closed  sets  form
the needed partition of $\omega^0_f$. The proposition is proved.

\subsection{Slow Relaxations in One Perturbed System}\label{SS4.4}

In this subsection, as in the preceding one, we investigate one semiflow of homeomorphisms $f$ over a
compact space $X$.

\begin{theorem}\label{T4.5}
$\eta^0_1$- and $\eta_2^0$-slow relaxations are  impossible  for
one semiflow.
\end{theorem}
{\bf Proof.} It is enough to  show  that  $\eta^0_2$-slow  relaxations  are impossible. Suppose the
contrary: there  are  such $\gamma>0$  and  such sequences  of  numbers $\varepsilon_i > 0\
\varepsilon_i \rightarrow 0$,  of  points $x_i \in X$   and   of $(x_i, \varepsilon_i)$-motions
$f^{\varepsilon_i} (t | x_i)$ that $\eta^{\varepsilon_i}_2 (f^{\varepsilon_i} (t | x_i), \gamma)
\rightarrow \infty$.  Similarly  to the proofs of the theorems \ref{T4.3} and  \ref{T3.1},  find  a
subsequence in $\{ f^{\varepsilon_i} (t | x_i) \}$ and such $y^* \in X$ that $\rho^* (y^*,
\omega^0_f) \geq \gamma$ and,  whatever  be  the neighbourhood $V$  of the point $y^*$ in $X$,
$\overline{\mbox{mes}} \{ t \geq 0 \ | \ f^{\varepsilon_i} (t | x_i) \in V \} \rightarrow \infty\ (i
\rightarrow \infty, f^{\varepsilon_i} (t | x_i)$
 belongs to the chosen subsequence). As in the  proof  of
Theorem \ref{T4.3}, we  have  $y^* \in \omega^0_f (y^*) \subset \omega^0_f$.  But,  according  to the
constructing, $\rho^* (y^*, \omega^0_f) \geq \gamma > 0$. The obtained contradiction proves the
absence of $\eta^0_2$-slow relaxations.

\begin{theorem}\label{T4.6}
Let $X$ be connected. Then, if $\omega^0_f$  is  connected
then the semiflow $f$ has not    $\tau^0_{1,2,3}$- and
$\eta^0_3$-slow relaxations. If $\omega^0_f$
is  disconnected,  then  $f$  possesses  $\tau^0_{1,2,3}$-   and
$\eta^0_3$-slow
relaxations.
\end{theorem}
{\bf Proof.} Let $X$ and $\omega^0_f$ be  connected.  Then,  according  to  the propositions
\ref{P4.17} and \ref{P4.18}, $\omega^0_f (x) = \omega^0_f$  for  any $x \in X$.  Consequently,
$\omega^0 (x)$-bifurcations are absent.  Therefore  (Theorem \ref{T4.1}) $\tau_3$-slow relaxations
are absent. Consequently, there are not other $\tau^0_i$- and $\eta^0_i$-slow relaxations due to the
inequalities $\tau^{\varepsilon}_i \leq \tau^{\varepsilon}_3$   and $\eta^{\varepsilon}_i \leq
\tau^{\varepsilon}_3\ (i = 1,2,3)$ 1,2,3) (see Proposition \ref{P4.16}). The first part of  the
theorem  is proved.

Suppose now that $X$ is connected and $\omega^0_f$ is  disconnected.  Let us use Proposition
\ref{P4.22}. Find such $x \in X$ that $x$-motion is whole, $x \not \in \omega^0_f$,
  and  such  partition  of  $\omega^0_f$   into  openly-closed   subsets
$\omega^0_f = W_1 \bigcup W_2,\ W_1 \bigcap W_2 = \varnothing$ that $\alpha_f (x) \subset W_1,\
\omega^0_f (x) \subset W_2$.  Suppose $\gamma = \frac{1}{3} r (W_1, W_2)$. There is such $t_0$  that
for  $t < t_0\ \rho^* (f(t,x), W_2) > 2 \gamma$. Let $p \in \alpha_f (x),\ t_j < t_0,\ t_j
\rightarrow - \infty,\ f(t_j, x) \rightarrow p$. For  each $j = 1,2, \ldots$ exists such $\delta_j >
0$ that for  $\varepsilon < \delta_j\ d(\omega^{\varepsilon}_f (f (t_j, x)), \omega^0_f (f (t_i, x)))
< \gamma$    (this   follows   from the Shura-Bura   lemma   and   Proposition    \ref{P4.8}). Since
$\omega^0_f (f (t_j, x)) = \omega^0_f (x)$ (Corollary \ref{C4.5}), for $\varepsilon < \delta_j\
d(\omega^{\varepsilon}_f (f(t_j, x)), W_2) < \gamma$. Therefore $\rho^* (f (t,x),
\omega^{\varepsilon}_f (f (t_j, x))) > \gamma$ if $t \in [t_j, t_0],\ \varepsilon > \delta_j$.
Suppose $x_i = f(t_j,\ x),\ \varepsilon_j > 0,\ \varepsilon_j < \delta_j,\ \varepsilon_j \rightarrow
0,\ f^{\varepsilon_j} (t | x_j) = f(t,x_j)$.
 Then $\tau^{\varepsilon_j}_1 (f^{\varepsilon_j} (t | x_j), \gamma) \geq t_0
- t_i \rightarrow \infty$. The existence of  $\tau_1$-  (and  consequently of  $\tau_{2,3}$-)-slow
relaxations is proved. To prove the  existence  of $\eta_3$-slow relaxations we need the following
lemma.

\begin{lemma}\label{L4.7}For any $\varepsilon > 0,\ \varkappa > 0$
$$
\overline{\omega^{\varepsilon}_f} \subset \omega^{\varepsilon+ \varkappa}_f.
$$
\end{lemma}
{\bf Proof.} Let  $y \in \overline{\omega^{\varepsilon}_f}$: there are such sequences  of  points
$x_i \in X,\ y_i \in \omega^{\varepsilon}_f$,
 of $(x_i, \varepsilon)$-motions $f^{\varepsilon} (t | x_i)$,
of numbers  $t^i_j > 0,\ t^i_j \rightarrow \infty \ \mbox{as} \ j \rightarrow \infty $  that $y_i
\rightarrow y,\ f^{\varepsilon} (t^i_j | x_i) \rightarrow y_i \ \mbox{as} \ j \rightarrow \infty$.
Suppose that $\delta = \frac{1}{2} \delta (\frac{\varkappa}{3}, T)$.
 There  is  such  $y_i$
that $\rho (y_i, y) < \delta$. For this $y_i$  there is such monotone  sequence $t_j \rightarrow
\infty$ that $t_j - t_{j-1} > T$ and $\rho (y_i, f^{\varepsilon} (t_j | x_i )) < \delta$. Suppose $$
f^* (t) = \left\{ \begin{array}{ll} f^{\varepsilon} (t | x_i), & \mbox{if}\ t \neq t_j;\\ y, &
\mbox{if}\ t = t_j\ (j = 1, 2, \ldots),
\end{array}
\right. $$ where $f^* (t)$ is $(x_i, \varkappa + \varepsilon)$-motion and $y \in \omega (f^*)$.
Consequently, $y \in \omega^{\varepsilon + \varkappa}_f$. The lemma is proved.

\begin{corollary}\label{C4.7} $\omega^0_f = \bigcap_{\varepsilon > 0} \overline{\omega^{\varepsilon}_f}$.
\end{corollary}
     Let us return to the proof of Theorem \ref{T4.6}  and  show  the
existence of $\eta^0_3$-slow relaxations if $X$ is connected and $\omega^0_f$  is  not. Suppose that
$\gamma = \frac{1}{5} r (W_1, W_2)$. Find such $\varepsilon_0 > 0$ that for $\varepsilon <
\varepsilon_0\ d(\omega^{\varepsilon}_f, \omega^0_f) > \gamma$ (it exists according to Corollary
\ref{C4.7} and the Shura-Bura lemma). There  is  $t_1$   for  which $d(f(t_1, x), \omega^0_f) > 2
\gamma$. Let $t_j < t_1,\ t_j \rightarrow - \infty,\ x_j = f(x_j, x)$.
 As $(x_j, \varepsilon)$-motions let  choose  true  motions
$f(t,x_j)$. Suppose  that  $\varepsilon_j \rightarrow 0,\ 0 < \varepsilon_j < \varepsilon_0$.  Then
$\eta^{\varepsilon_j} (f(t,x_j), \gamma) > t_1 - t_j \rightarrow \infty$  and, consequently,
$\eta^0_3$-slow relaxations exist. The theorem is proved.

In conclusion of this subsection let us give {\it the proof of  Theorem \ref{T4.2}.} We consider
again the family of  parameter  depending semiflows.

{\bf Proof.} Let $X$ be  connected  and $\omega^0(x,k)$-bifurcations  exist. Even if for one $k \in
K\ \omega^0 (k)$ is disconnected, then, according to  Theorem \ref{T4.6}, $\tau^0_3$-slow relaxations
exist. Let $\omega^0 (k)$ be connected for any $k \in K$. Then $\omega^0 (x,k) = \omega^0 (k)$ for
any $x \in X,\ k \in K$. Therefore  from  the existence  of $\omega^0 (x,k)$-bifurcations follows  in
this  case   the existence of $\omega^0 (k)$-bifurcations. Thus, Theorem \ref{T4.2} follows from the
following lemma which is of interest by itself too.

\begin{lemma}\label{L4.8}If the system  (\ref{e1})  possesses  $\omega^0 (k)$-bifurcations, then it possesses
$\tau^0_3$- and $\eta^0_3$-slow relaxations.
\end{lemma}
{\bf Proof.} Let $k^*$ be a point of $\omega^0 (k)$-bifurcation:  and there are such $\alpha > 0,\
y^* \in \omega^0 (k^*)$  that $\rho^* (y^*, \omega^0 (k_i)) > \alpha > 0$  for  any $i = 1, 2,
\ldots$.
 According to Corollary \ref{C4.7} and the Shura-Bura  lemma,
for every $i$ exists $\delta_i > 0$ for  which $\rho^*(y^*, \omega^{\delta_i} (k_i)) > 2 \alpha / 3$.
Suppose that $0 < \varepsilon_i \leq \delta_i$, $\varepsilon_i \rightarrow 0$. As the
$\varepsilon$-motions appearing in  the  definition  of slow   relaxations   take   the   real $(k_i,
y_i)$-motions,    where $y_i = f(-t_i, y^*, k^*)$, and $t_i$ are determined as follows: $$ t_i = \sup
\{ t>0 \ | \ \rho (f(t', x, k), f (t' x, k')) < \alpha / 3 \} $$
 under the conditions
$t' \in [0,t],\ x \in X,\ \rho_K (k, k') < \rho_K (k^*, k_i) \}$.

Note    that  $\rho^*(f(t_i, y_i, k_i), \omega^{\varepsilon_i}
(k_i)) \geq \alpha / 3$,    consequently, $\eta^{\varepsilon_i}_3
(f(t,y_i, k_i), \alpha / 4) > t_i$
 and $t_i \rightarrow \infty$ as $i \rightarrow \infty$.
The last follows from  the compactness of $X$  and $K$ (see the proof  of  Proposition \ref{P4.1}).
Thus,  $\eta^0_3$-slow relaxations exist and then $\tau^0_3$-slow relaxations exist too. Lemma
\ref{L4.8} and Theorem \ref{T4.2} are proved. \vspace{3mm} $$
\begin{array}{lcr}
* \ \ \ & & \ \ \ * \\
& \  & \\
& * &
\end{array}
$$
\vspace{3mm}

In Sections \ref{S1}-\ref{S4} the basic notions of the theory  of transition  processes  and slow
relaxations are   stated.   Two directions of further development  of  the  theory  are possible:
introduction of new  relaxation  times  and  performing  the  same studies for them or widening the
circle  of solved  problems  and supplementing the  obtained  existence  theorems  with analytical
results.

Among  interesting  but  unsufficiently  explored  relaxation times let us mention the approximation
time $$ \tau_{ }(x,k,\varepsilon) = \inf \{ t \geq 0 \ | \ d(\omega (x,k), f ([0,t], x,k)) <
\varepsilon \} $$ and the averaging time $$ \tau_{ v} (x,k,\varepsilon, \varphi) = \inf \left\{ t
\geq 0 \ \biggm| \ \left| \frac{1}{t'} \int^{t'}_0 \ \varphi (f(\tau, x,k)) d \tau - \langle \varphi
\rangle _{x,k} \right| < \varepsilon\ \mbox{for}\ t' > t \right\}, $$ here $\varepsilon>0$, $\varphi$
is a continuous  function  over  the  phase  space  $X$, $$ \langle \varphi \rangle _{x,k} = \lim_{t
\rightarrow \infty} \frac{1}{t} \int^t_0 \varphi (f(t, x,k)) d \tau$$ (if the limit exists).

The approximation time is the time necessary  for  the motion to visit the
$\varepsilon$-neighbourhood of each  its $\omega$-limit  point.  The averaging time depends on
continuous function $\varphi$ and shows the time necessary for establishing the average value of
$\varphi$ with accuracy $\varepsilon$ along the trajectory.

As the most important problem of analytical  research, one  should consider the problem of studying
the  asymptotical  behaviour under $T \rightarrow \infty$ of ``domains of  delay", that is,  the sets
of those pairs $(x,k)$ (the initial  condition,  parameter)  for  which $\tau_i (x,k, \varepsilon) >
T$ (or $\eta_i (x,k, \varepsilon) > T)$. Such estimations for particular  two-dimensional  system are
given in the work \cite{[67]}.

``Structurally stable systems are not dense". It  would  not  be exaggeration to say that the so
titled work by Smale \cite{[39]} opened a new era in the understanding of dynamics. Structurally
stable (rough) systems  are  those  whose   phase   portraits   do   not change qualitatively under
small perturbations (accurate definitions with detailed  motivation  see in  \cite{[5]}).   Smale
constructed   such structurally unstable system that any other system close enough to it is also
structurally unstable. This result broke the  hopes  to classify if not all then ``almost  all"
dynamical  systems.  Such hopes were associated with  the  successes  of classification  of
two-dimensional dynamical systems \cite{[13],[68]} among which structurally stable ones are dense.

There  are  quite  a  number  of  attempts  to  correct   the catastrophic situation with structural
stability: to  invent  such natural notion of stability, for which almost all systems would be
stable.  The weakened definition of structural stability is proposed in the works
\cite{[69],[70],[71]}: the system is stable if almost all  trajectories  change little  under small
perturbations. This stability is already typical, almost all systems are stable in this sense.

The other way to  get  rid  of  the  ``Smale  nightmare"  (the existence of domains  of structurally
unstable  systems)  is  to consider the $\varepsilon$-motions, subsequently considering (or not)  the
limit $\varepsilon \rightarrow 0$. The picture  obtained  (even in limit $\varepsilon \rightarrow 0$)
is more stable  than the phase portrait (the accurate formulation  see  above  in Section \ref{S4}).
It seems to be obvious that one should first study those (more rough) details of dynamics, which do
not disappear under small perturbations.

The  approach  based  on  consideration  of  limit  sets   of $\varepsilon$-motions, in the form
stated here was proposed  in  the  paper \cite{[22]}. It  is necessary  to  note the  conceptual
proximity  of this approach to the method of quasi-averages in statistical physics \cite{[72]}. By
analogy, the stated approach could  be  called  the  method  of ``quasi-limit" sets.

Unfortunately, elaborated analytical or numerical methods  of studying  (constructing  or,  wider,
localizing)  limit  sets  of $\varepsilon$-motions for dynamical systems of general type are
currently absent. However, the author does not give up the hope for the possibility of elaboration of
such methods.

Is the subject of this work in the ``mainstream" of Dynamics? I don't know, but let us imagine an
experimental situation: we observe a dynamic of a system. It might be a physical or a chemical
system, or just a computational model, the precise nature of the system is not important. How long
should we monitor the system in order to study the limit behaviour? When does the transition process
turn into the limit dynamics? This work tries to state these problems mathematically and to answer
them, at least partially.

\end{document}